\documentclass[review,12pt]{elsarticle}
\usepackage[a4paper, total={8in, 11in}, margin=0.5in]{geometry}
\usepackage{setspace}
\usepackage{subcaption}
\usepackage{multirow}
\usepackage{multicol}
\usepackage{graphicx}
\usepackage{siunitx}
\RequirePackage{amsmath}
\usepackage{bm}
\usepackage{xfrac}
\usepackage{float}
\usepackage{arydshln} 
\usepackage{gensymb}

\usepackage{amsmath,esint,amsfonts}

\usepackage[most]{tcolorbox}
\usepackage[colorinlistoftodos]{todonotes}
\usepackage{cleveref} 

\journal{Journal}

\biboptions{sort&compress}

\begin{document}

	\title{Taylor-Couette flow in an elliptical enclosure generated by an inner rotating circular cylinder}

	\author{Akash Unnikrishnan$^{a,}$\fnref{Corresponding Author}}
	\author{Surya Pratap Vanka$^{b}$}
	\author{Vinod Narayanan$^{a}$}
	\address{a Department of Mechanical Engineering,\\
		Indian Institute of Technology Gandhinagar,		Gandhinagar, Gujarat 382355, India}
	\address{b Department of Mechanical Science and Engineering,\\
		University of Illinois at Urbana-Champaign, Urbana, Illinois 61801, USA}
	\fntext[Author to whom correspondence should be addressed]{\vspace{0.3cm}Corresponding Author Email Address: \url{akash.unnikrishnan@iitgn.ac.in}}

\begin{abstract}
    Taylor-Couette flow between rotating cylinders is a classical problem in fluid mechanics and has been extensively studied in the case of two concentric circular cylinders. There have been relatively small number of studies in complex-shaped cylinders with one or both cylinders rotating. In this paper, we study the characteristics of Taylor cells in an elliptical outer cylinder with a rotating concentric inner circular cylinder. We numerically solve the three-dimensional unsteady Navier-Stokes equations assuming periodicity in the axial direction. We use a Fourier-spectral meshless discretization by interpolating variables at scattered points using polyharmonic splines and appended polynomials. A pressure-projection algorithm is used to advance the flow equations in time. Results are presented for an ellipse of aspect ratio two and for several flow Reynolds numbers ($Re = \omega r_i (b-r_i))/\nu$, where $\omega$ = angular velocity [rad/s], $r_i$ = radius of inner cylinder, $b$ = semi-minor axis, and $\nu$ = kinematic viscosity) from subcritical to 300. Streamlines, contours of axial velocity, pressure, vorticity, and temperature are presented along with surfaces of Q criterion. The flow is observed to be steady until $Re = 300$ and unsteady at $Re = 350$. 
    
\end{abstract}
\maketitle

\section{Introduction}

    The Taylor-Couette flow \cite{mallock1889iv,mallock1896iii,couette1890distinction,taylor1923viii} between two rotating concentric circular cylinders has been extensively studied in literature since the early works of \citet{mallock1889iv,mallock1896iii,couette1890distinction}, and \citet{taylor1923viii}. Numerous experimental and computational studies have been since conducted at Reynolds numbers that range from subcritical to supercritical regimes, and to the turbulent regime. It has been well documented that at modest Reynolds numbers, the initially two-dimensional flow between the concentric circular cylinders bifurcates to a three-dimensional flow with pairs of toroidal vortices known as Taylor cells. As the Reynolds number is increased, the steady Taylor cells transition to spiral vortices, unsteady chaotic flow, and eventually to turbulence \cite{coles_1965,andereck_liu_swinney_1986,fenstermacher_swinney_gollub_1979,snyder1968stability,snyder1969change}. The widely studied geometry has been the two-dimensional concentric cylinders configuration with narrow and wide gaps \cite{diprima1972flow,diprima1972non,diprima1975nonlinear,diprima1977amplification,stuart1980mathematics,smith1982turbulent,davey1968instability} in tall as well as short aspect ratio geometries \cite{lorenzen1985anomalous,mullin2002symmetry,czarny2002spiral,schulz2003effect}. Numerical simulations \cite{bilson2007direct,pirro2008direct,brauckmann2013direct,dong2007direct,ostilla2013optimal} and Particle Image Velocimetry studies \cite{wereley1998spatio,huisman2012ultimate} have provided detailed structure and transitions of these cells. A collage of fascinating flow patterns from steady toroidal vortices to travelling waves, spiral wavy vortices, and mixed laminar/turbulent patches have been observed. \citet{coles_1965} reported several different flow patterns for gap Reynolds number between 114 and 1600 including patches of laminar and turbulent regions in an alternating spiral pattern. \citet{andereck_liu_swinney_1986} conducted experiments with varying combinations of both inner and outer cylinders rotating and observed that the number of azimuthal wave modes increases for a counter-rotating outer cylinder compared to a stationary outer cylinder. By increasing the rotational speeds of the inner and outer cylinders, many flow regimes ranging from modulated wavy vortices to intermittent and spiral turbulence, and a featureless turbulent flow at Re of 1000 were observed. Numerous practical applications of the Taylor-Couette flow have been developed for particle separation \cite{wereley1999inertial,wang2005cfd,majji2018inertial,dash2020particle}, membrane filtration \cite{holeschovsky1991quantitative,wereley1999inertial,lee2001rotating,serre2008stability,tilton2010pressure}, desalination \cite{tu2018scale}, and devices for chemical reaction engineering \cite{cohen1983experimental}. For continuous operation of separation devices, the extended Taylor-Couette-Poiseuille flow \cite{poncet2011numerical,campero1999flow,wereley1999velocity} has been studied extensively. Studies of heat transfer in Taylor-Couette flow have also been conducted with constant temperature and constant heat flux boundary conditions \cite{ball1988bifurcation,teng2015direct,guillerm2015flow,leng2021flow,khawar2022counter}.  
    
    While the most widely studied configuration has been the concentric cylinder geometry, geometrical variations in the shapes of the cylinders provide a large and uncharted opportunity for exploring the rich fluid physics and exploiting them for new device development. Geometrical variations to the coaxial configuration may include change of the cross-sectional shapes of the inner and outer cylinders (with uniformity in the axial direction), axial changes to the cross-sectional dimensions, and creation of periodic disturbances to name a few. The widely studied cross-sectional shape is an eccentrically placed inner circular cylinder which produces a non-axisymmetric base flow whose stability characteristics are different from the concentric case. Such a configuration not only alters the base sub-critical flow but also the super-critical states of the Taylor cells \cite{eagles1978effects,oikawa1989,shu2004numerical,krueger1966relative}. \citet{krueger1966relative} performed experiments to determine the critical rotation rates for different values of eccentricity. As the eccentricity is increased, it is found that the flow became more unstable, forming Taylor cells at earlier Reynolds numbers. A linear stability analysis performed by \citet{oikawa1989} indicates that the subcritical flow becomes unstable at lower Reynolds numbers as the eccentricity is increased. \citet{RUIXIU19921323} used the approach of  \citet{stuart1980mathematics,diprima1972flow,diprima1972non,diprima1975nonlinear,diprima1977amplification} to investigate the linear stability of flow between the eccentric cylinders and found results in agreement with previous studies. Stability of eccentric Taylor-Couette-Poiseuille flow was studied by \citet{leclercq2013temporal,leclercq2014absolute} using a bipolar coordinate system. 

    In contrast to the amount of research conducted on the basic Taylor-Couette geometry, there have been only a small number of studies of Taylor cells in other configurations. \citet{sprague2008tailored,sprague2009continuously} considered a geometry in which the Taylor vortices are continuously tailored by changing both the inner and outer cylinder radii. In concentric spheres, with the inner sphere rotating, \citet{wimmer1976experiments,nakabayashi1988modulated,nakabayashi1988spectral} showed the occurrence of Taylor vortices. Subsequently \citet{wimmer1995experimental} extended the study to ellipsoids and coaxial cones. \citet{denne1999travelling} observed that the Taylor vortices in coaxial cones have a different structure than the cylindrical Taylor-Couette flow. \citet{soos2007taylor} simulated the flow development in a mixer with a rotating lobed inner cylinder inside a circular outer cylinder. They observed enhanced mixing in three- and four-lobed configurations resulting from the cyclic squeezing and expansion of the Taylor cells. \citet{snyder1968experiments,snyder1968stability,snyder1969change} considered a square outer cylinder with a smooth inner circular cylinder and experimentally characterized the Taylor cells. A similar experiment with short cylinders was performed by \citet{mullin1985bifurcation} who found anomalous modes when the end wall boundary layers interacted with the Taylor cells. Such anomalous cells were also observed by \citet{benjamin1981anomalous}, and \citet{lorenzen1985anomalous} at higher Reynolds numbers. One configuration where the inner rotating cylinder is ribbed was considered by \citet{greidanus2015turbulent}. They measured the drag and torque on the rotating cylinder and found a reduction of $5.3\%$ in drag at a turbulent Reynolds number of $4.7 \times 10^4$. The ribs were placed vertically along the circumference of the inner cylinder. The riblets were of small dimension and did not alter the macroscopic features of the main Taylor vortices but contributed to reduction in the drag by the turbulent boundary layer. \citet{razzak2020numerical} considered axial bellows where the outer cylinders contained waves of a certain amplitude and wavelength. The numerical study characterized the flow patterns, distributions of velocities and pressure within the bellow cavities at different Reynolds numbers. Recently, \citet{mayank_2021} studied a geometry with the outer wall diffusing at a fixed angle. The three-dimensional flow fields were computed for different diffuser angles and different Reynolds numbers. The structure of the Taylor cells was documented in both steady and unsteady regimes. Other geometries where the cross-section varies in the axial direction are cones and spheres \cite{wimmer1976experiments,wimmer1995experimental,denne1999travelling,nakabayashi1988modulated,nakabayashi1988spectral}. A stability analysis of flow between an outer non-circular cylinder and a coaxial inner cylinder was reported by \citet{eagles1997stability}. Another non-traditional geometry in which Taylor vortices have been investigated is a “stadium" where two semi-circular end chambers are connected to a rectangular enclosure \cite{kobine1995dynamics}.

    The present paper deals with a novel geometry in which we have not found any studies of the structure and characteristics of Taylor vortices. The geometry is an elliptical cylinder inside which a circular cylinder is placed concentrically. The inner cylinder is rotated while the outer one is stationary. Such a geometry has applications to lubrication inside elliptical bearings \cite{pinkus1956analysis,hashimoto1984performance,hashimoto2001improvement,ebrahimi2020geometrical,singh1989elastothermohydrodynamic}. The base flow in an elliptical cylinder with a rotating inner cylinder is more complex than that in an eccentric circular cylinder configuration because of the formation of corner vortices. Because of the larger gap along the major axis of the ellipse, the circular motion created by the inner rotating cylinder drives another set of counter-rotating vortices akin to the Moffatt vortices in shear driven cavities \cite{vanka2008immersed,jyotsna1995multigrid,darr1991separated,erturk2007fine,an_bergada_mellibovsky_2019,mcquain1994steady,moffatt1964viscous,ren_guo_2017}. The two-dimensional flow inside an ellipse with a rotating cylinder was recently investigated by us \cite{unnikrishnan2022shear} using a high accuracy meshless solution procedure \cite{Shantanu2016,SHAHANE2021110623,shahane2021high,bartwal2021application}. In \cite{unnikrishnan2022shear}, we reported the effects of Reynolds number, aspect ratio of ellipse, and the eccentricity of placement of the inner cylinder on the flow and pressure fields. The numerical algorithm and the flow were two-dimensional and did not capture any three-dimensional vortices generated by instabilities of the base flow.

    In this paper, we present results of a three-dimensional study performed with a recently developed Fourier spectral meshless method. The Fourier spectral method assumes periodicity in the axial direction and solves the Navier-Stokes equations in Fourier space at scattered locations in the cross-stream plane. We show the formation of Taylor vortices similar to those in circular cylinders with differences arising due to the squeezing and expansion of the vortices as the flow passes through the major and minor axes. Section 2 first briefly describes the numerical algorithm. Section 3 presents results of validation and grid-independency tests. Results for concentric placement of the inner cylinder are given in Section 4. A summary of major findings is given in Section 5.

\section{Methodology}
\subsection{Flow Geometry and Equations Solved}
The flow geometry studied is shown in Figure 1. An inner cylinder of radius $r_i$ rotates inside an ellipse of the semi major axis $a$ and semi minor axis $b$. In this study, the aspect ratio $a/b$ is taken as two, and the gap $(b-r_i)$ is taken to be equal to $r_i$. Thus, the semi minor axis of the ellipse is twice the inner cylinder radius. Non-dimensionally, we have taken the radius of the inner cylinder to be $0.5$, and the gap to be $0.5$. The flow is considered periodic in the axial ($z$) direction, and the number of wave numbers resolved in Fourier space is decided based on the energy decay with wave number.  The flow Reynolds number is defined as $Re = \omega r_i (b-r_i)/\nu$, where $\omega$ is the speed of rotation (rad/s) of the inner cylinder and $\nu$ is the fluid kinematic viscosity. The fluid is considered Newtonian, and the density and viscosity are considered constant. The flow is considered to be incompressible. 

\begin{figure}[H]
    \centering
    \includegraphics[width=0.7\textwidth]{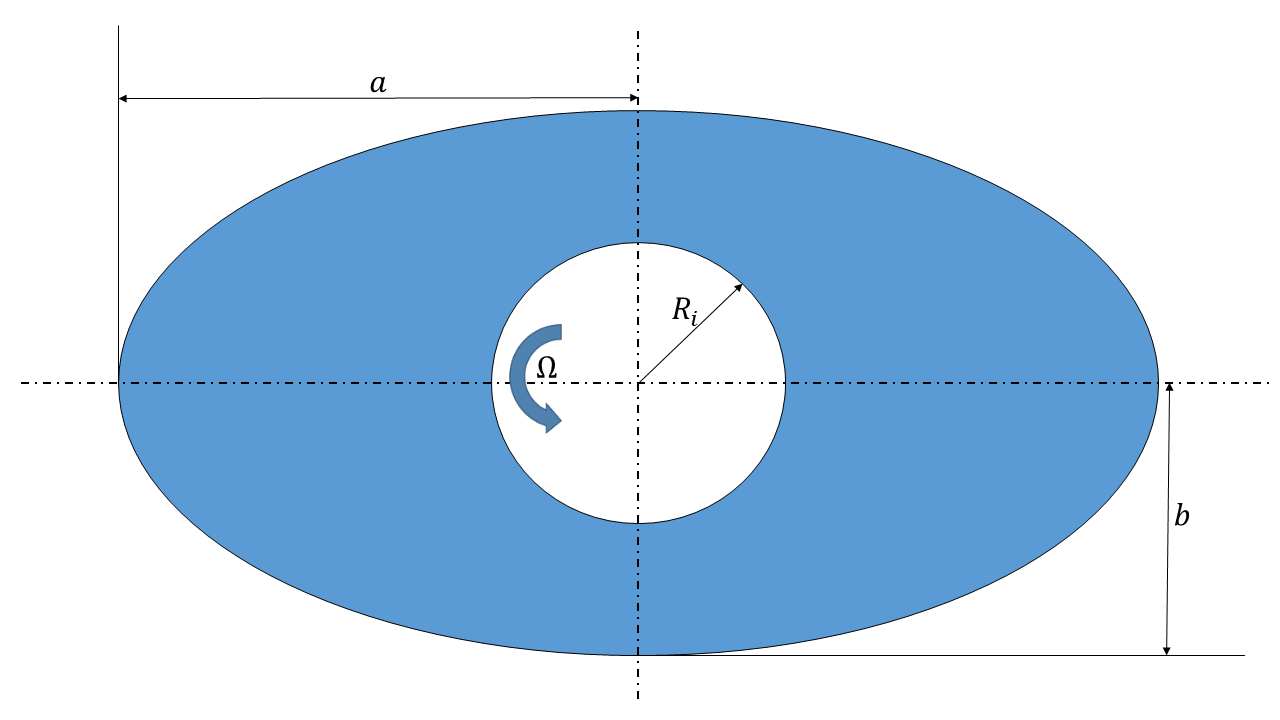}
    \caption{The cross section of the model analysed in this study}
    \label{fig:model}
\end{figure}

The walls of the cylinders are taken as non-penetrating and have no slip. For the heat transfer computations, the inner wall is heated at a non-dimensional temperature of one, and the outer wall is held at a non-dimensional temperature of zero. The three-dimensional time-dependent equations are solved marching in time to determine the onset of unsteadiness of the flow with the Reynolds number.  The equations solved are
\begin{align}
    \frac{\partial u_i}{\partial x_i} &= 0;\\
    \rho\left(\frac{\partial u_i}{\partial t} +u_j\frac{\partial u_i}{\partial x_j}\right) &= -\frac{\partial p}{\partial x_i} +\mu \nabla^2u_i \\
    \rho c_p \left(\frac{\partial T}{\partial t} + u_j\frac{\partial T}{\partial x_j}\right) &= k\nabla^2T + S
\end{align}
where, $\mu$ is dynamic viscosity of fluid, $\rho$ is the density of the fluid, $T$ is the temperature, $u_i$ are components of velocity, $k$ is the thermal conductivity and $S$ denotes the source term.
Here the tensor summation notation is used for the continuity equation and the advection terms. The steady state is assumed when the differences in values at adjacent time steps have decreased below $10^{-5}$ non-dimensional units. The torque and Nusselt number are also computed in time and compared between time steps. The non-dimensional torque is computed as
\begin{equation}
    \tau_c = \mu r_iu_{\theta}L
    \label{eqn:non-dimensional_torque}
\end{equation}
where, $u_{\theta}$ is the tangential velocity of the inner cylinder, $r_i$ is the radius of inner cylinder and $L$ is the length in the axial direction.
The Nusselt number is computed using the relation
\begin{equation}
    Nu = \frac{2 r_i}{k_t}\left(\frac{1}{2 \pi r_i L \Delta T} \int_{\theta=0}^{\theta=2 \pi} \int_{z=0}^{z=L}\left[k_t \frac{\partial T}{\partial r} r_i\right] d \theta d z\right)
    \label{eqn:Nusselt_number}
\end{equation}
where $k_t$ is the thermal conductivity, and $\Delta T$ is the temperature difference between the inner and outer cylinders.
 
\subsection{Numerical Method}
\label{sec:numerical_method}
\subsubsection{Spectral Decomposition}
    To implement the third dimention of the Cartesian space, a Fourier spectral method is used. Because of periodicity, the governing variables, $\psi(x, y, z, t)$, are first expressed in Fourier space by the expansion (\citet{fornberg1998practical,canuto2007spectral}):  
    \begin{equation}
        \psi\left(x, y, z_l, t\right)=\sum_{k =-N}^{k=N} \widehat{\psi_k}(x, y, t) \exp \left(\mathrm{i} k z_l\right), \quad l = 0,1,...,2N
        \label{Eq:dft_field_variable}
    \end{equation}
    where, $\widehat{\psi_k}$ is a complex valued function corresponding to wave number $k$ ranging from $-N$ to $N$ and $\mathfrak{i}^2=-1$. $\widehat{\psi_k}$ can be represented with an inverse Fourier transform given as in \cref{Eq:ifft_field_variable}. $\psi$ being a real valued function (the field variables), the transform coefficient functions $\widehat{\psi_k}(x,y,t)$ are complex conjugates for positive and negative values of $k$. 
    \begin{equation}
        \widehat{\psi_k}(x, y, t)=\frac{1}{2 N+1} \sum_{l=0}^{l=2N} \psi\left(x, y, z_l, t\right) \exp \left(-\mathfrak{i} k z_l\right), \quad k = -N,...,0,...,N
        \label{Eq:ifft_field_variable}
    \end{equation}
    
    The partial differential equations can be transformed to wave number space by computing each derivative in the transformed space. For example, the partial differential equation 
    \begin{equation}
        \frac{\partial \psi}{\partial t}+\left[\frac{\partial(u \psi)}{\partial x}+\frac{\partial(v \psi)}{\partial y}+\frac{\partial(w \psi)}{\partial z}\right]=\Gamma\left[\frac{\partial^2 \psi}{\partial x^2}+\frac{\partial^2 \psi}{\partial y^2}+\frac{\partial^2 \psi}{\partial z^2}\right]+S
        \label{Eq:transport_eqn}
    \end{equation}
    in transformed space becomes
    \begin{equation}
        \frac{\partial \widehat{\psi_k}}{\partial t}+\left[\frac{\partial \widehat{(u \psi)_k}}{\partial x}+\frac{\partial \widehat{(v \psi)_k}}{\partial y}+\mathfrak{i} \widehat{k(w \psi)_k}\right]=\Gamma\left[\frac{\partial^2 \widehat{\psi_k}}{\partial x^2}+\frac{\partial^2 \widehat{\psi_k}}{\partial y^2}-k^2 \widehat{\psi_k}\right]+\widehat{S_k}
        \label{Eq:transport_eqn_ft}
    \end{equation}
    where, $[u, v, w]$ are components of velocity, $\Gamma$ is diffusion coefficient and $S$ is the source term.
    
    For integrating the time-dependent Navier-Stokes equations, we use the popular fractional step procedure \cite{kim1985application} which consists of a momentum step, followed by a pressure computation step which projects the momentum velocities to a divergence-free state. In our current work, we use an explicit procedure to compute advection and diffusion fluxes. However, the pressure-projection step requires solution of an implicit equation, and thus an iterative algorithm.  The momentum step computes intermediate values for the velocities by explicitly updating them in wave number space. Thus,
    \begin{align}
        \frac{\widehat{\tilde{u}_k}-\widehat{u_k^n}}{\Delta t}&=-\left[\frac{\partial\widehat{(\left.u^n u^n\right)_k}}{\partial x}+\frac{\partial \widehat{(\left.v^n u^n\right)_k}}{\partial y}+\mathfrak{i} k \widehat{(\left.w^n u^n\right)_k}\right]+\nu\left[\frac{\partial^2 \widehat{u_k^n}}{\partial x^2}+\frac{\partial^2 \widehat{u_k^n}}{\partial y^2}-k^2 \widehat{u_k^n}\right]\\
        \frac{\widehat{\tilde{v}_k}-\widehat{v_k^n}}{\Delta t}&=-\left[\frac{\partial\widehat{(\left.u^n v^n\right)_k}}{\partial x}+\frac{\partial \widehat{(\left.v^n v^n\right)_k}}{\partial y}+\mathfrak{i} k \widehat{(\left.w^n v^n\right)_k}\right]+\nu\left[\frac{\partial^2 \widehat{v_k^n}}{\partial x^2}+\frac{\partial^2 \widehat{v_k^n}}{\partial y^2}-k^2 \widehat{v_k^n}\right]\\
        \frac{\widehat{\tilde{w}_k}-\widehat{w_k^n}}{\Delta t}&=-\left[\frac{\partial\widehat{(\left.u^n w^n\right)_k}}{\partial x}+\frac{\partial \widehat{(\left.v^n w^n\right)_k}}{\partial y}+\mathfrak{i} k \widehat{(\left.w^n w^n\right)_k}\right]+\nu\left[\frac{\partial^2 \widehat{w_k^n}}{\partial x^2}+\frac{\partial^2 \widehat{w_k^n}}{\partial y^2}-k^2 \widehat{w_k^n}\right]
        \label{Eq:fractional_timestep_initial}
    \end{align}
    The superscript $n$ denotes values at previous time step. Note that the explicit time step is usually much more restrictive than that of a semi-implicit or fully implicit algorithm.  However, the explicit algorithm is simpler to implement. Since these momentum velocities do not satisfy the continuity equation, a pressure Poisson equation is solved to project the intermediate velocities to a divergence-free state. The pressure Poisson equation in wave space is given by
    \begin{equation}
        \frac{\partial^2 \widehat{p_k^{n+1}}}{\partial x^2}+\frac{\partial^2 \widehat{p_k^{n+1}}}{\partial y^2}-k^2 \widehat{p_k^{n+1}}=\frac{\rho}{\Delta t}\left[\frac{\partial \widehat{\tilde{u}_k}}{\partial x}+\frac{\partial \widehat{\tilde{v}_k}}{\partial y}+\mathrm{i} k \widehat{\tilde{w}_k}\right]
        \label{Eq:fractional_timestep_pressure_poisson}
    \end{equation}
    After the pressure Poisson equation is solved, the momentum velocities are corrected in wave space as 
    \begin{equation}
        \begin{aligned}
        &\widehat{u_k^{n+1}}=\widehat{\tilde{u_k}}-\frac{\Delta t}{\rho} \frac{\partial \widehat{p_k^{n+1}}}{\partial x} \\
        &\widehat{v_k^{n+1}}=\widehat{\tilde{v}_k}-\frac{\Delta t}{\rho} \frac{\partial \widehat{p_k^{n+1}}}{\partial y} \\
        &\widehat{w_k^{n+1}}=\widehat{\tilde{w}_k}-\frac{\Delta t}{\rho} \mathrm{i} k \widehat{p_k^{n+1}}
        \end{aligned}
        \label{Eq:fractional_timestep_correction}
    \end{equation}
    Once the velocities are updated in wave space, they are back-transformed to physical space for computation of the next time step values. The temperature equation is subsequently solved with explicit computation of the advection and diffusion terms. 
    To compute the cross-stream derivatives $\frac{\partial \psi}{\partial x}$,$\frac{\partial \psi}{\partial y}$, etc. we use a meshless method that interpolates variables at scattered data using polyharmonic radial basis functions (PHS-RBF). For complex geometries, it is common to use unstructured finite volume methods \cite{shahane2019finite} or finite element methods. Here we use the meshless method for the flexibility and accuracy it provides. We discretize the cross-section by a set of scattered points generated here as vertices of a finite element grid, using the GMSH software \cite{gmsh}. The set of scattered points do not use any underlying grid and are not connected by edges and elements.
    
    The first step in the meshless method is to interpolate a variable between the scattered points. We use a cloud based interpolation in which any variable at a base point is interpolated over a cloud of neighbor points using the PHS-RBF. An arbitrary variable $s(\textbf{x})$ is interpolated as 
    \begin{equation}
	   s(\textbf{x}) = \sum_{i=1}^{q} \lambda_i \phi (||\bm{x} - \bm{x_i}||_2) + \sum_{i=1}^{m} \gamma_i P_i (\bm{x})
	   \label{Eq:RBF_interp}
    \end{equation}
    where, $\phi(r)=r^{2a_1+1},\hspace{0.1cm} a_1 \in \mathbb{N}$ is the PHS-RBF, $m$ is the number of monomials ($P_i$) upto a maximum degree of $l$ and $(\lambda_i, \gamma_i)$ are $q+m$ coefficients. We use $a_1 =1$ in this study. $q$ equations are obtained by collocating \cref{Eq:RBF_interp} over the $q$ cloud points. $m$ additional equations required to close the linear system are imposed as constraints on the polynomials \cite{flyer2016onrole_I}:
    \begin{equation}
        \sum_{i=1}^{q} \lambda_i P_j(\bm{x_i}) =0 \hspace{0.5cm} \text{for } 1 \leq j \leq m
    \label{Eq:RBF_constraint}
    \end{equation}

    In a matrix vector form, we can write these equations as
    
    \begin{equation}
        \begin{bmatrix}
        \bm{\Phi} & \bm{P}  \\
        \bm{P}^T & \bm{0} \\
        \end{bmatrix}
        \begin{bmatrix}
        \bm{\lambda}  \\
        \bm{\gamma} \\
        \end{bmatrix} =
        \begin{bmatrix}
        \bm{A}
        \end{bmatrix}
        \begin{bmatrix}
        \bm{\lambda}  \\
        \bm{\gamma} \\
        \end{bmatrix} =
        \begin{bmatrix}
        \bm{s}  \\
        \bm{0} \\
        \end{bmatrix}
        \label{Eq:RBF_interp_mat_vec}
    \end{equation}
    
    where, transpose is denoted by the superscript $T$, $\bm{\lambda} = [\lambda_1,...,\lambda_q]^T$, $\bm{\gamma} = [\gamma_1,...,\gamma_m]^T$, $\bm{s} = [s(\bm{x_1}),...,s(\bm{x_q})]^T$ and $\bm{0}$ is the vector of zeros. Dimensions of the submatrices $\bm{P}$ and $\bm{\Phi}$ are $q\times m$ and $q\times q$ respectively.

    Let ($l=2$) be the degree of appended polynomial, and let the dimension of the problem ($d=2$), there are $m=\binom{l+d}{l}=6$ polynomial terms, given as $[1, x, y, x^2, xy, y^2]$. The differential operators can be obtained by differentiating the RBF and the polynomials. 
    \begin{equation}
        \mathcal{L} [s(\textbf{x})] = \sum_{i=1}^{q} \lambda_i \mathcal{L} [\phi (\bm{||\bm{x} - \bm{x_i}||_2})] + \sum_{i=1}^{m} \gamma_i \mathcal{L}[P_i (\bm{x})]
        \label{Eq:RBF_interp_L}
    \end{equation}
    \Cref{Eq:RBF_interp_L} applied to all the points in the domain leads to a rectangular matrix vector system given by \cref{Eq:RBF_interp_mat_vec_L}.
    \begin{equation}
        \mathcal{L}[\bm{s}] =
        \begin{bmatrix}
        \mathcal{L}[\bm{\Phi}] & \mathcal{L}[\bm{P}]  \\
        \end{bmatrix}
        \begin{bmatrix}
        \bm{\lambda}  \\
        \bm{\gamma} \\
        \end{bmatrix}
        \label{Eq:RBF_interp_mat_vec_L}
    \end{equation}
    where, $\mathcal{L}[\bm{P}]$ and $\mathcal{L}[\bm{\Phi}]$ are matrices of sizes $q\times m$ and $q\times q$ respectively. Substituting \cref{Eq:RBF_interp_mat_vec} in \cref{Eq:RBF_interp_mat_vec_L} results in:
    \begin{equation}
        \begin{aligned}
        \mathcal{L}[\bm{s}] &=
        \left(\begin{bmatrix}
        \mathcal{L}[\bm{\Phi}] & \mathcal{L}[\bm{P}]  \\
        \end{bmatrix}
        \begin{bmatrix}
        \bm{A}
        \end{bmatrix} ^{-1}\right)
        \begin{bmatrix}
        \bm{s}  \\
        \bm{0} \\
        \end{bmatrix}
        =
        \begin{bmatrix}
        \bm{B}
        \end{bmatrix}
        \begin{bmatrix}
        \bm{s}  \\
        \bm{0} \\
        \end{bmatrix}\\
        &=
        \begin{bmatrix}
        \bm{B_1} & \bm{B_2}
        \end{bmatrix}
        \begin{bmatrix}
        \bm{s}  \\
        \bm{0} \\
        \end{bmatrix}
        = [\bm{B_1}] [\bm{s}] + [\bm{B_2}] [\bm{0}]
        = [\bm{B_1}] [\bm{s}]
        \end{aligned}
        \label{Eq:RBF_interp_mat_vec_L_solve}
    \end{equation}
    If the discrete values are known, the coefficients $(\lambda, \gamma)$ can be evaluated and hence the function can be evaluated at any arbitrary location within the interpolation zone. Or, if $\mathcal{L}[\bm{s}]$ satisfies a governing equation, the values of $s(x)$ at any arbitrary location can be evaluated.
    In the present algorithm, the intermediate momentum velocities are first computed in wave number space for all the cross-sectional points and for all wave numbers. Since this is an explicit step, only updates are needed. The pressure Poisson equation is then solved implicitly, also in wave number space. We use the BiCGSTAB algorithm from the eigen library \cite{eigenweb} to solve the implicit discrete pressure-Poisson equations for the required wave numbers.  After the velocities are updated to be divergence-free, they are transformed to physical space, and a new time step is initiated. The time marching is continued until steady state or until the prescribed time is reached. Application of this algorithm to the Taylor-Couette problem in an elliptic outer cylinder are described in the next sections below. Further details of the meshless method for two-dimensional problems can be found in \cite{Shantanu2016,SHAHANE2021110623,bartwal2021application,radhakrishnan2021non}.

\section{Validation and Grid Independency}
\begin{figure}[H]
    \centering
    \begin{subfigure}{0.45\textwidth}
        \centering
        \includegraphics[height=60mm]{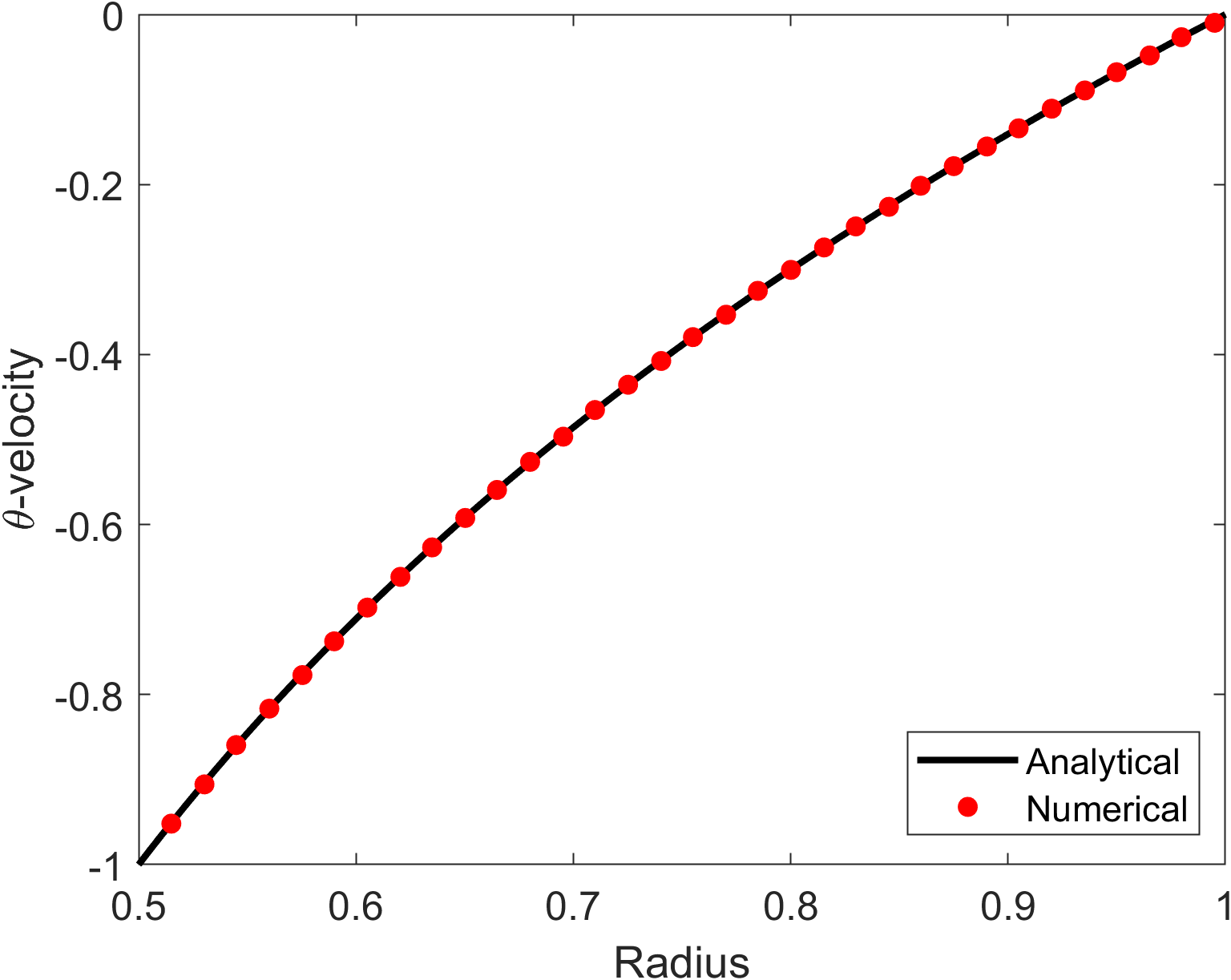}
        \caption{}
        \label{fig:code_validation_a}
    \end{subfigure}
    \hfill
    \begin{subfigure}{0.45\textwidth}
        \centering
        \includegraphics[height=60mm]{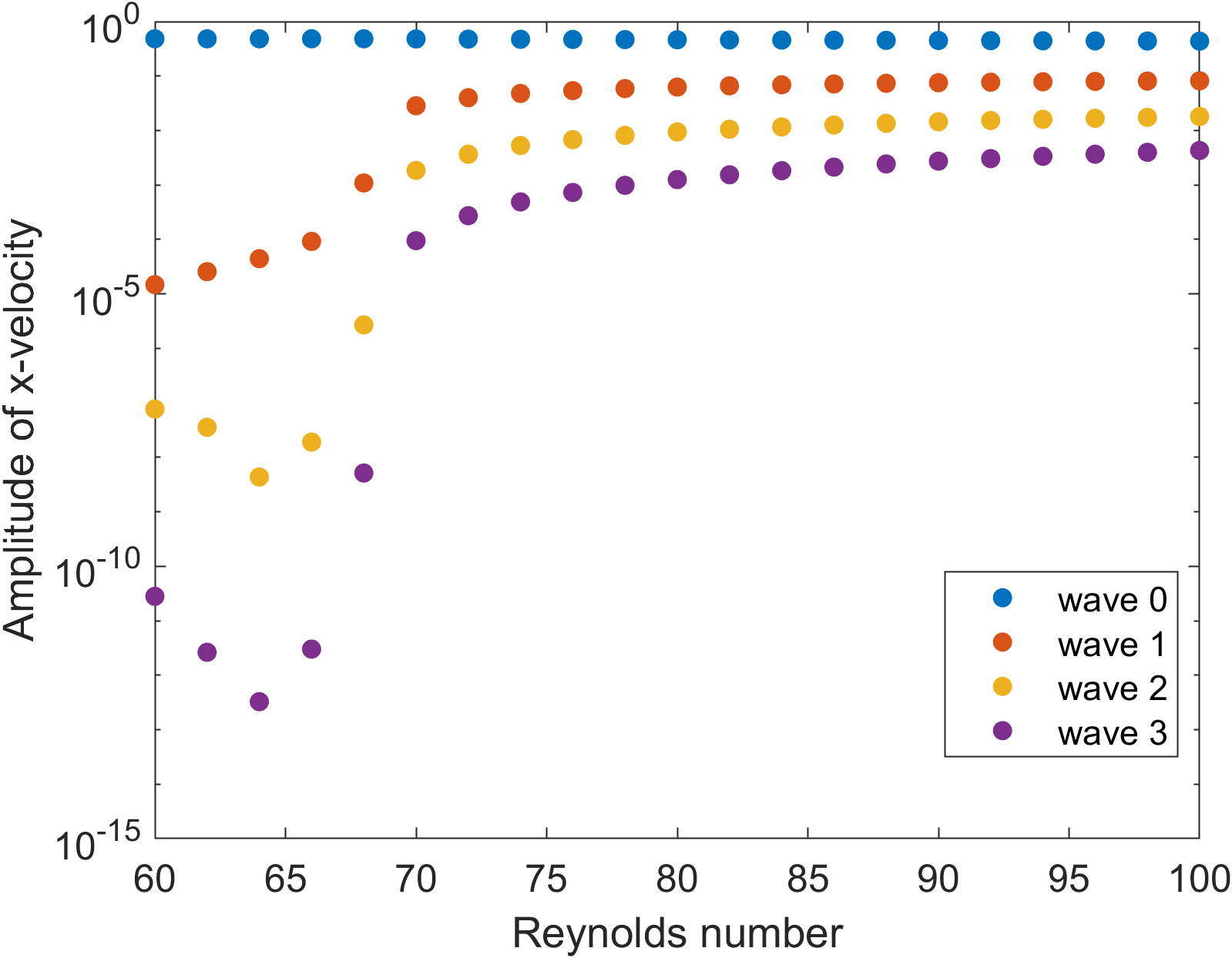}
        \caption{}
        \label{fig:code_validation_b}
    \end{subfigure}
    \hfill
    \par \bigskip
    \begin{subfigure}{0.55\textwidth}
        \centering
        \includegraphics[width=\textwidth]{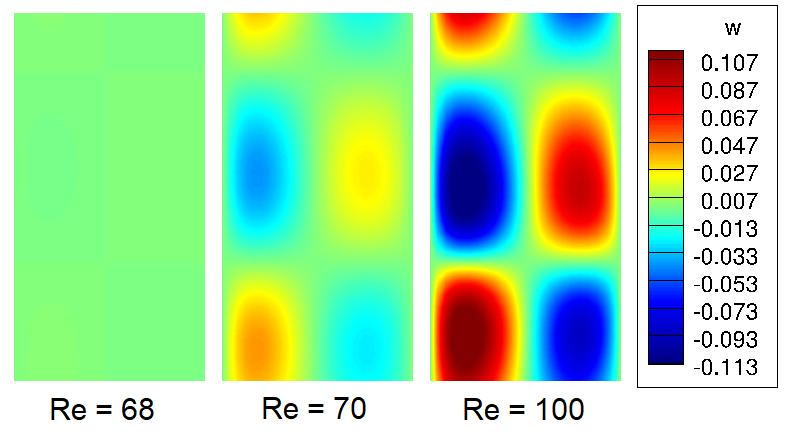}
        \caption{}
        \label{fig:code_validation_c}
    \end{subfigure}
    \hfill
    \par \bigskip 
    \begin{subfigure}{0.45\textwidth}
        \centering
        \includegraphics[height=60mm]{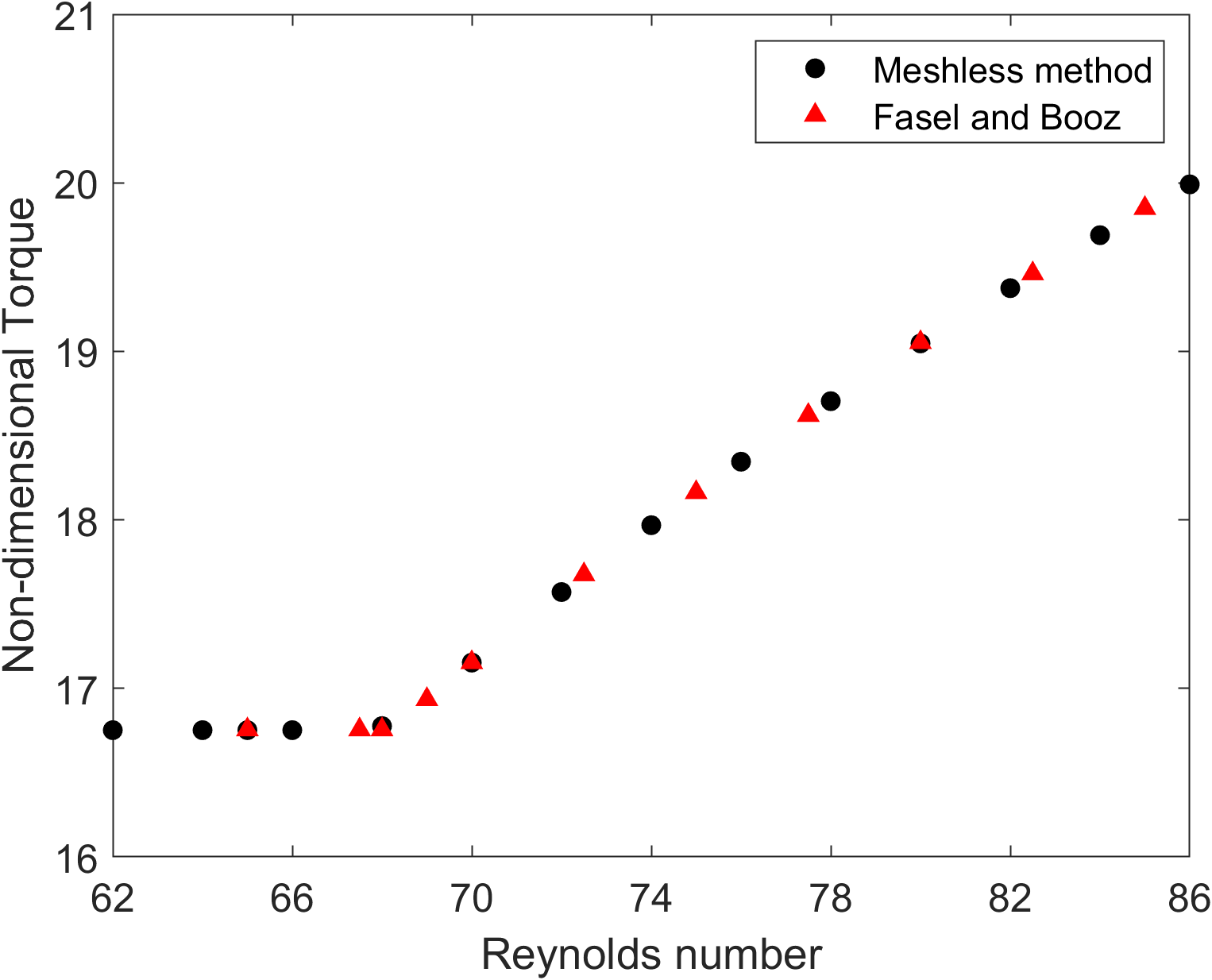}
        \caption{}
        \label{fig:code_validation_d}
    \end{subfigure}
    \caption{Plots and contours for validating the code for subcritical and supercritical stages of Taylor Couette setup: a) Comparison of analytical and numerical solution of cylindrical Couette flow data (subcritical stage) taken on a radius length at an axial location of $z = 0.5$; b) Plots of four domain averaged amplitudes of x-velocity at several Reynolds numbers. Transition of cylindrical Couette flow to Taylor Couette flow is seen to occur between Reynolds number of 68 and 70; c) Contours of axial velocity in the $r-z$ plane near supercritical Reynolds number; and d) Non-dimensional torque vs Reynolds number plot for circular Taylor Couette flow validated with \citet{fasel_booz_1984}}
    \label{fig:code_validation}
\end{figure}

The above-described numerical algorithm has been implemented in an open-source C++ code MemPhys \cite{memphys} and checked for correctness. The computer code was validated in a number of two-dimensional flows which have only the zeroth wave number in the axial direction including the subcritical cylindrical Couette flow as presented in  \cref{fig:code_validation_a}. Subsequently, the code was verified to reproduce the axisymmetric Taylor-Couette flow in a concentric annular gap. Comparisons of flow patterns and the torque on the inner cylinder were made with data of \citet{fasel_booz_1984}. The critical Reynolds number for the onset of the Taylor vortices was also narrowed to the range of 68-70 (\cref{fig:code_validation_b,fig:code_validation_c,fig:code_validation_d}), which is in good agreement with the value of 69.19 quoted by \citet{fasel_booz_1984}. \Cref{fig:code_validation_b} plots the averaged amplitudes of x-velocity over the domain for the first four Fourier modes against Reynolds numbers between 60 to 100.

\begin{figure}[H]
    \centering
    \includegraphics[width = 0.6\textwidth]{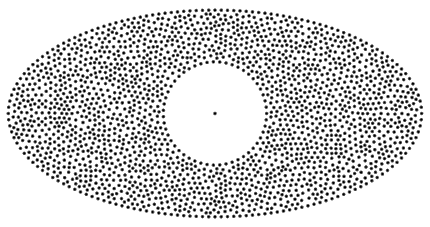}
    \caption{Sample point distribution of 1874 points}
    \label{fig:scattered_points}
\end{figure}

\begin{figure}[H]
     \centering
     \begin{subfigure}[b]{0.45\textwidth}
         \centering
         \includegraphics[height=60mm]{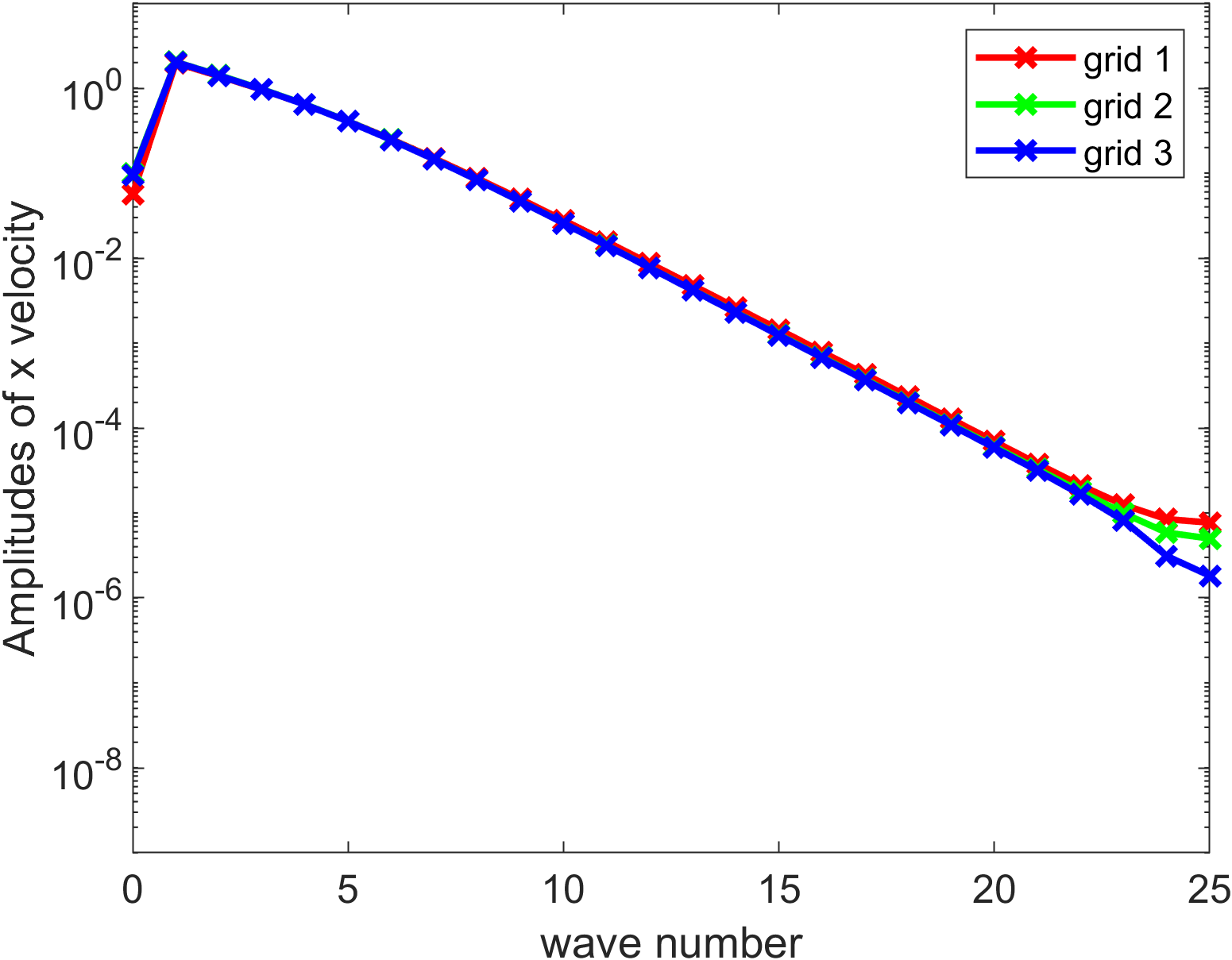}
         \caption{}
         \label{}
     \end{subfigure}
     \hfill
     \begin{subfigure}[b]{0.45\textwidth}
         \centering
         \includegraphics[height=60mm]{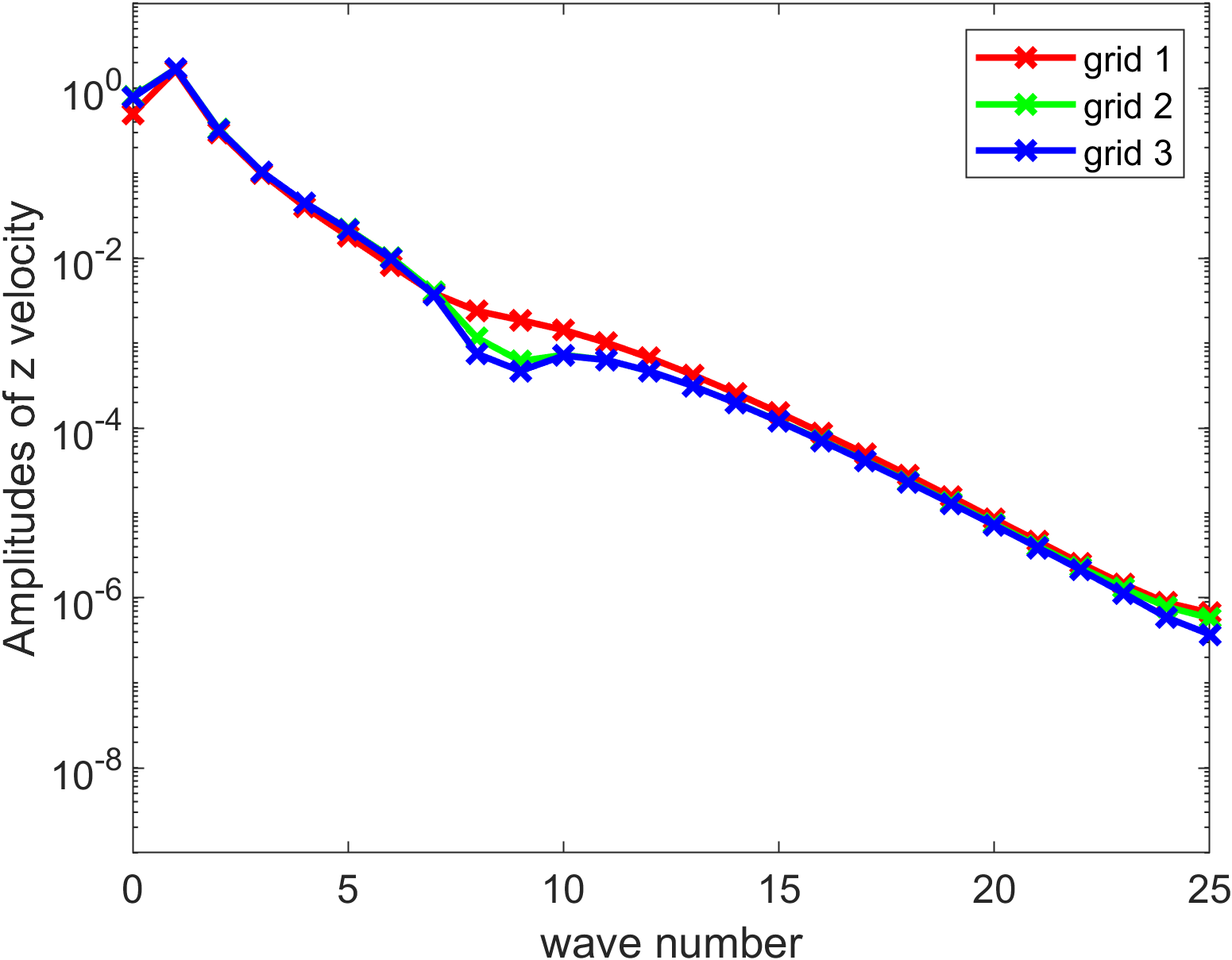}
         \caption{}
         \label{}
     \end{subfigure}
     \begin{subfigure}[b]{0.45\textwidth}
         \centering
         \includegraphics[height=60mm]{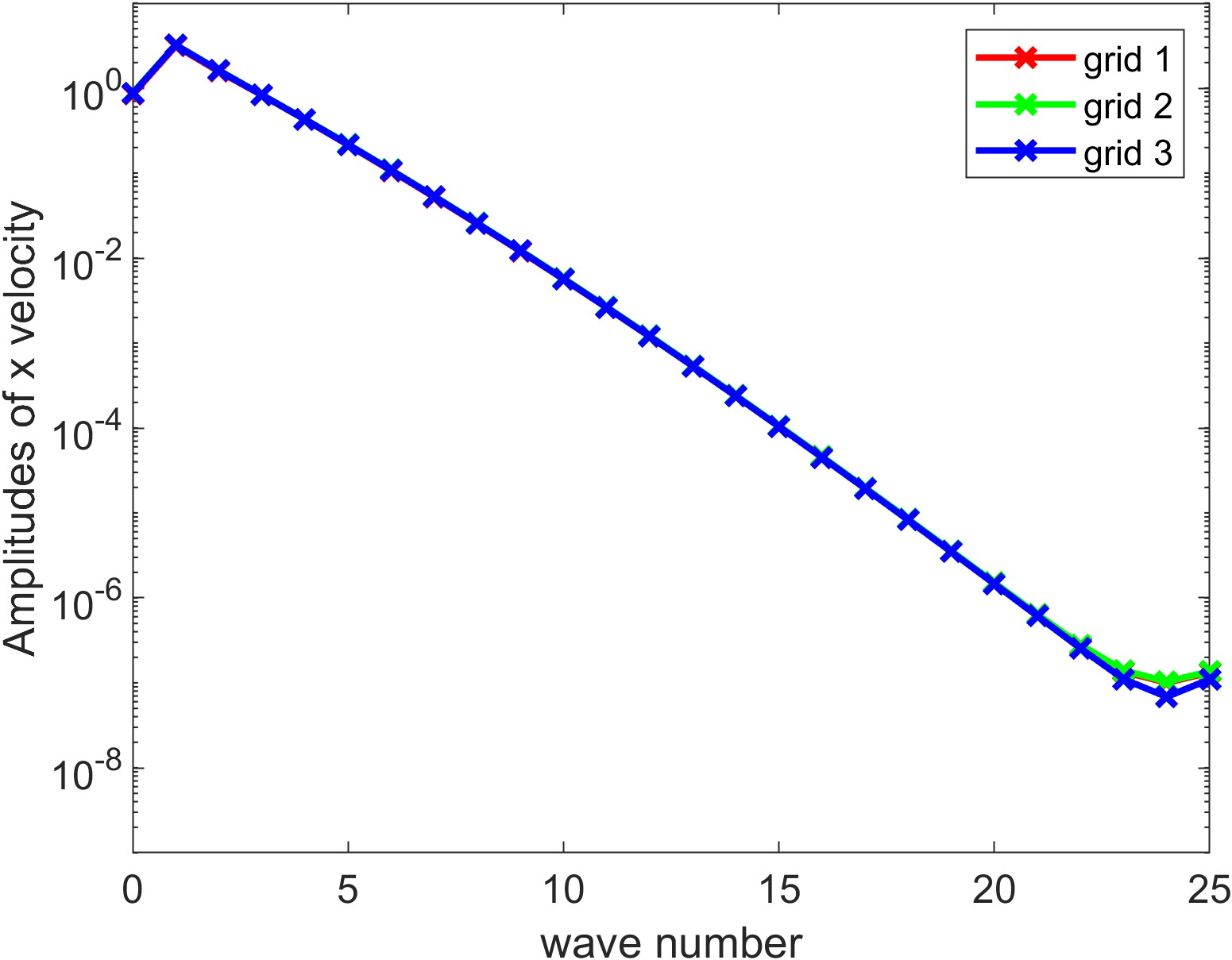}
         \caption{}
         \label{}
     \end{subfigure}
     \hfill
     \begin{subfigure}[b]{0.45\textwidth}
         \centering
         \includegraphics[height=60mm]{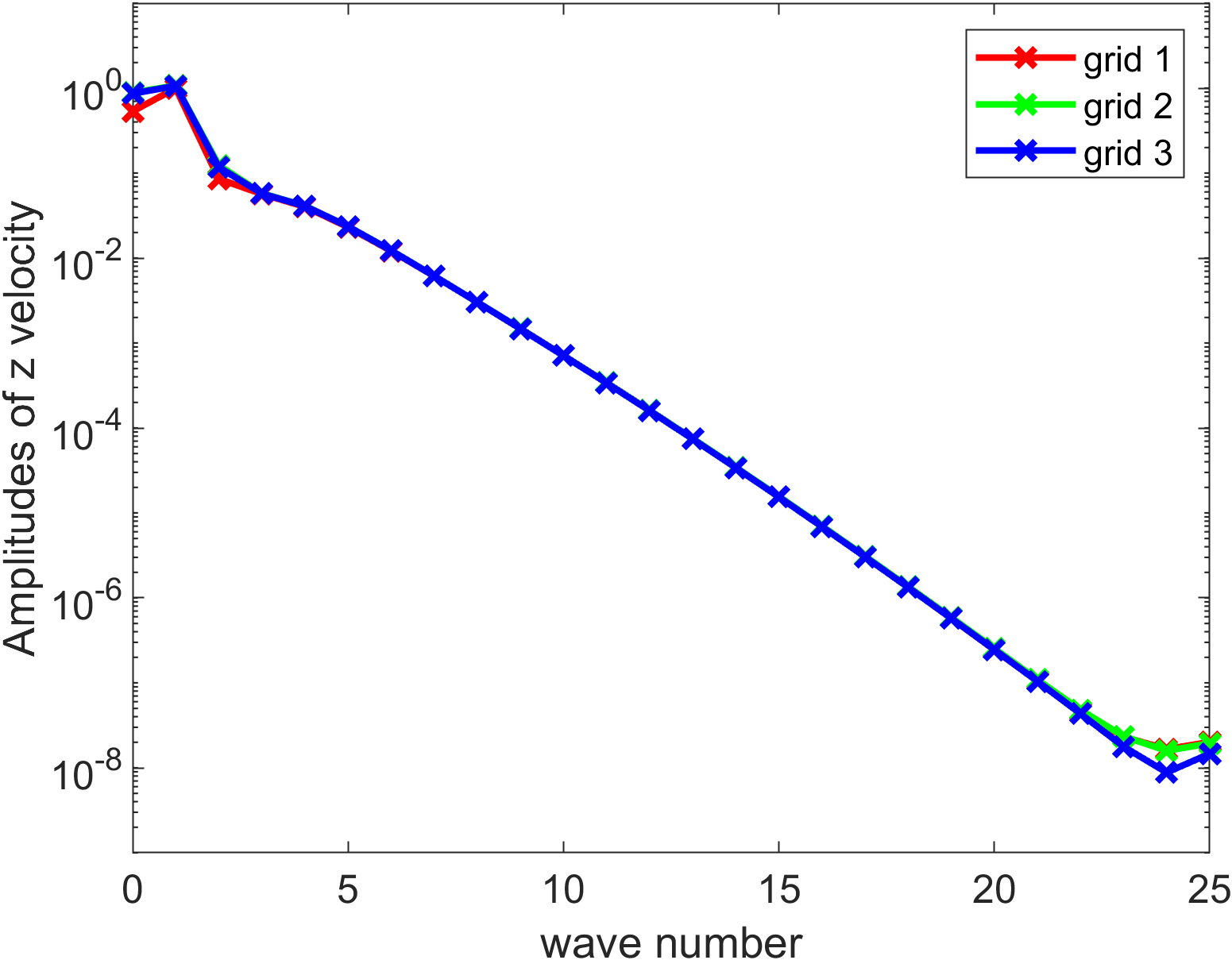}
         \caption{}
         \label{}
     \end{subfigure}
        \caption{a) Amplitude of x velocity vs wave number at $\theta = 0\degree$ $r = 0.7$; b) Amplitude of z velocity vs wave number at $\theta = 0\degree$ $r = 0.7$; c) Amplitude of x velocity vs wave number at $\theta = 0\degree$ $r = 0.9$; and d) Amplitude of z velocity vs wave number at $\theta = 0\degree$ $r = 0.9$;}
        \label{fig:amp_vs_wave_elliptical}
\end{figure}

Calculations of three-dimensional flow in an elliptical enclosure with a rotating inner cylinder were then initiated by first determining the number of scattered points needed for accuracy, and the number of wave modes needed in the axial direction to resolve the streamwise variations of field variables. These grid-independency calculations were performed at $Re = 300$, which is slightly below the Reynolds number at which the flow is observed to become unsteady. We use three different sets of scattered points in the cross-sectional plane. In the periodic direction, we consider twenty five modes ($N = 25$) and study the energy content as a function of the wave number. The cross-sectional distributions of scattered points (\cref{fig:scattered_points}) are characterized in terms of average point spacing between adajacent scattered points. The energy content in each wave number is then plotted for the three velocities at eight discrete locations ($r = 0.6$, $0.7$, $0.8$, $0.9$ and $\theta=0\degree$, $90\degree$). In \cref{fig:amp_vs_wave_elliptical} we show here four representative plots of the amplitudes of the velocities versus the streamwise wave number. It is seen that the medium point spacing of   $0.0158$ gives values very close to those of $\approx 0.008$.  Also, the amplitude of the velocities drops below 1.0e-5 for wave numbers greater than $15$.  We therefore conclude that the results with point spacing of $\approx 0.0158$ and $15$ wave numbers are sufficiently accurate while computationally economical. The flow is considered steady and converged when the time variations in the variables decrease and stay below 1.0e-5. 

\section{Results for Concentric Inner Cylinder}

\subsection{Torque and heat transfer rates}
\begin{figure}[H]
    \centering
    \begin{subfigure}{0.45\textwidth}
         \centering
         \includegraphics[width=\textwidth]{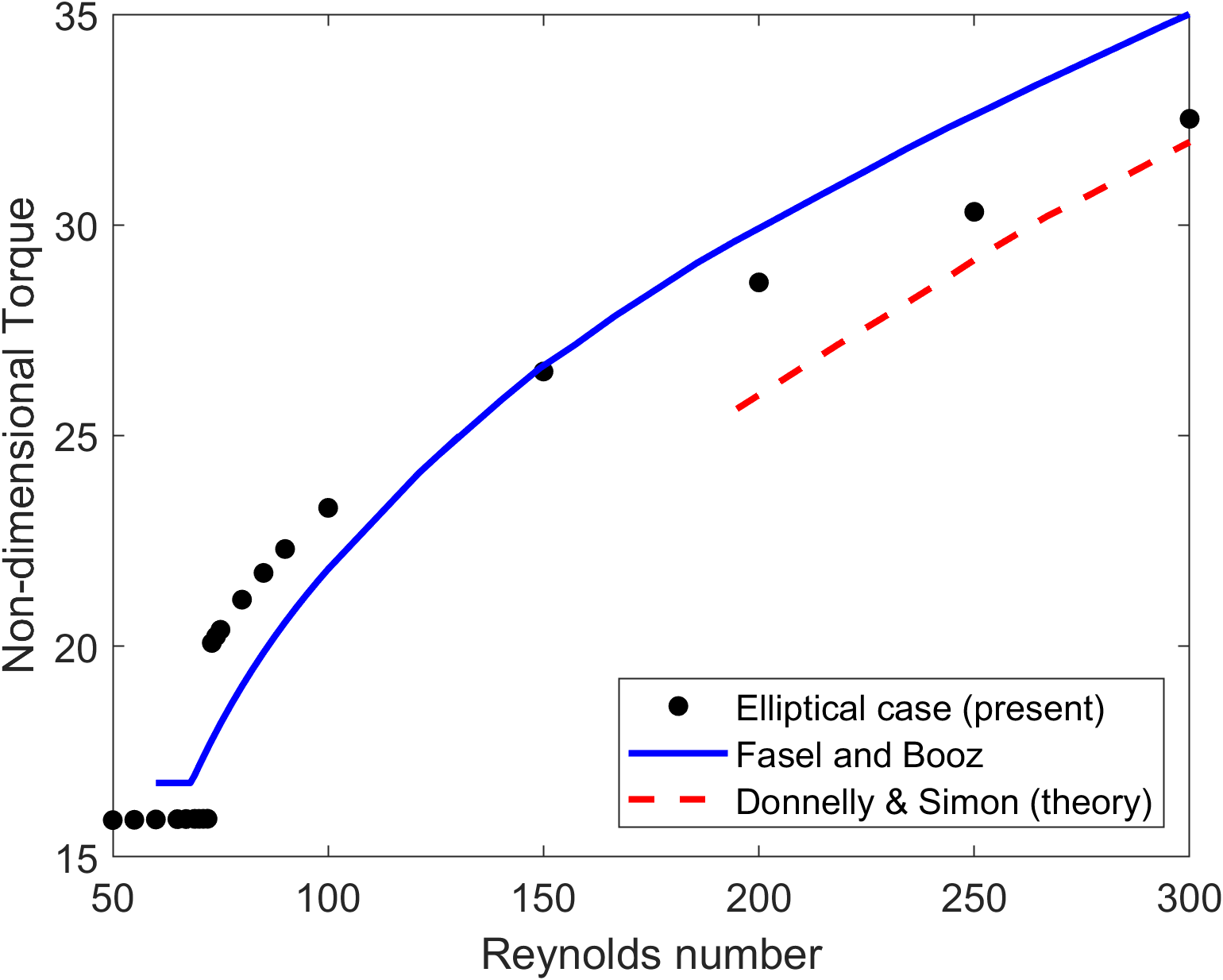}
         \caption{}
         \label{fig:TvsReplot}
    \end{subfigure}
    \hfill
    \begin{subfigure}{0.45\textwidth}
         \centering
         \includegraphics[width=\textwidth]{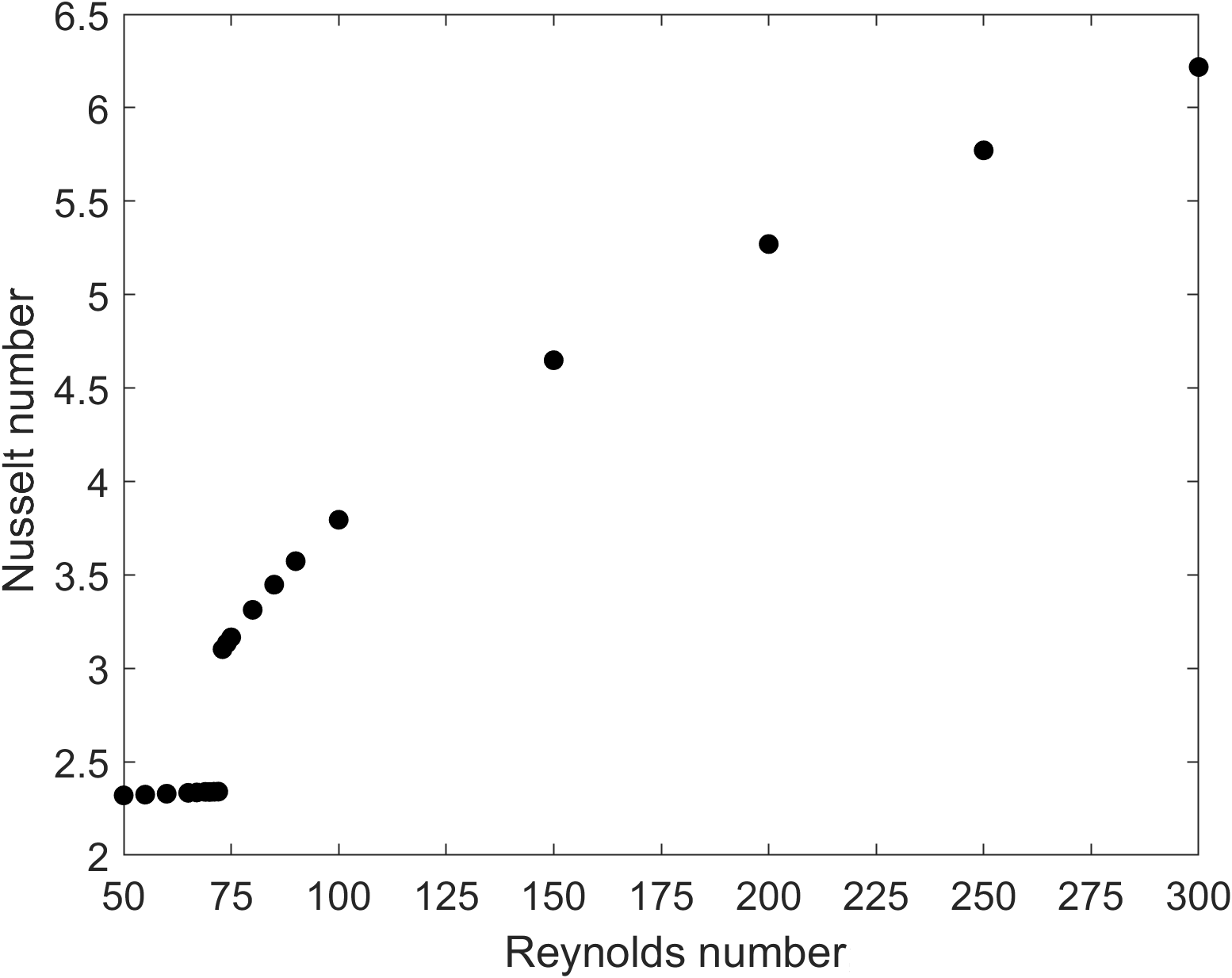}
         \caption{}
         \label{fig:NuvsReplot}
    \end{subfigure}
    \caption{a) Plot of non-dimensional torque vs Reynolds number for the present case with elliptical outer cylinder in comparison with simple Taylor Couette flow from \citet{donnelly1960empirical}, and \citet{fasel_booz_1984}; and b) Nusselt number vs Reynolds number for the present case}
    \label{fig:Non-T_and_NuvsRe}
\end{figure}

\begin{figure}[h]
    \centering
    \begin{subfigure}{0.45\textwidth}
        \centering
        \includegraphics[width=\textwidth]{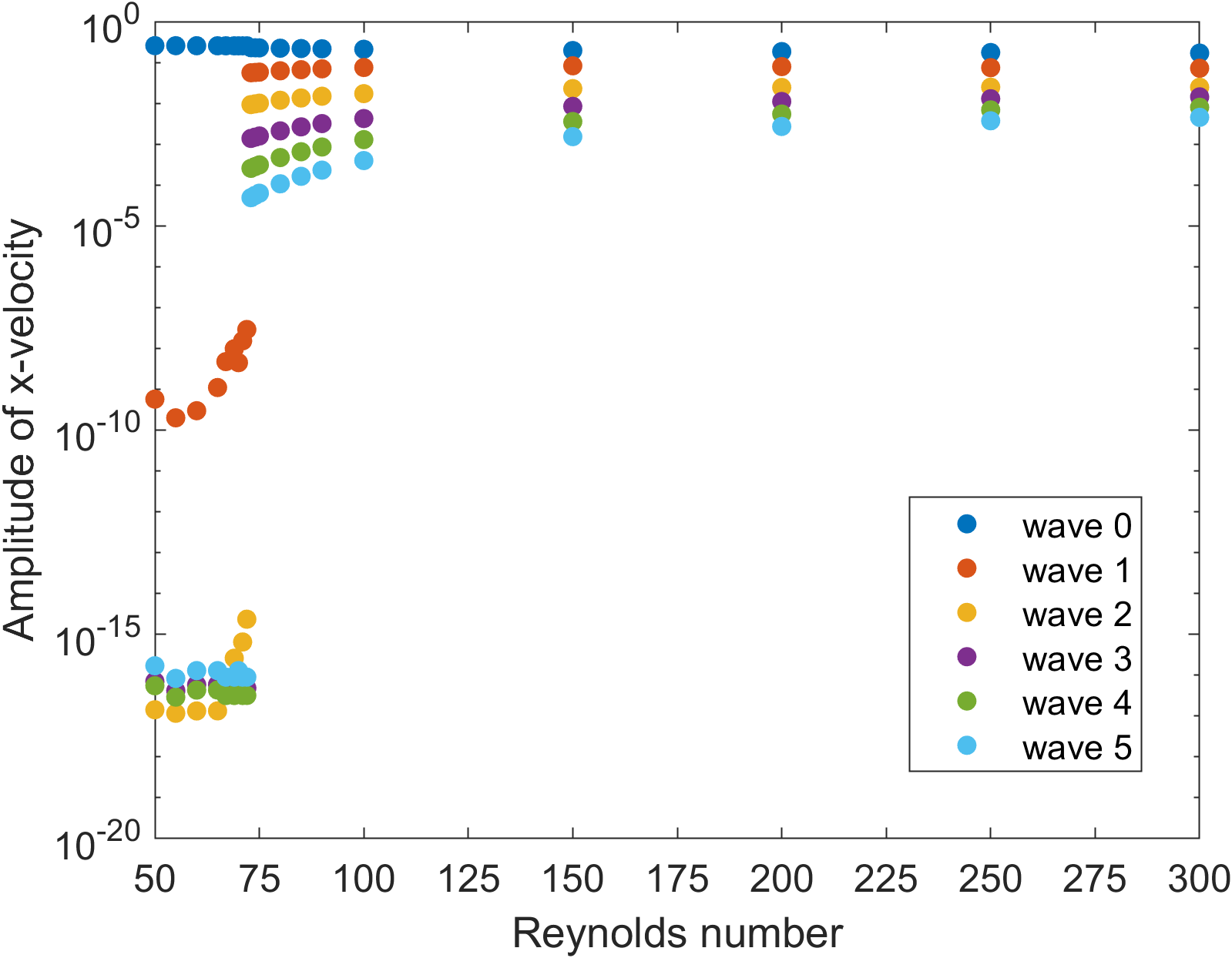}
    \caption{}
    \label{fig:wave_vs_Rex}
    \end{subfigure}
    \hfill
    \begin{subfigure}{0.45\textwidth}
        \centering
    \includegraphics[width=\textwidth]{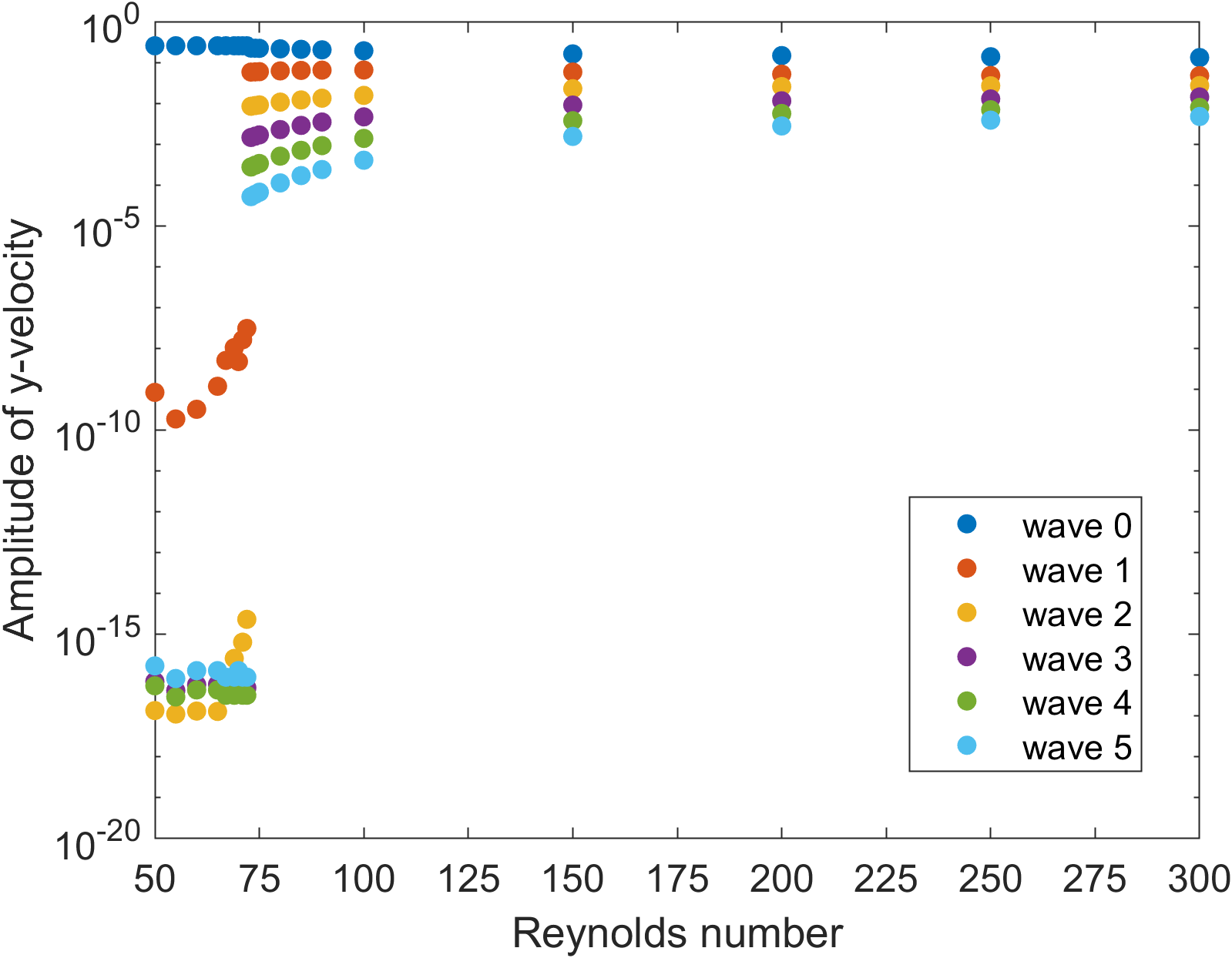}
    \caption{}
    \label{fig:wave_vs_Rey}
    \end{subfigure}
    \hfill
    \par \bigskip
    \begin{subfigure}{0.45\textwidth}
        \centering
    \includegraphics[width=\textwidth]{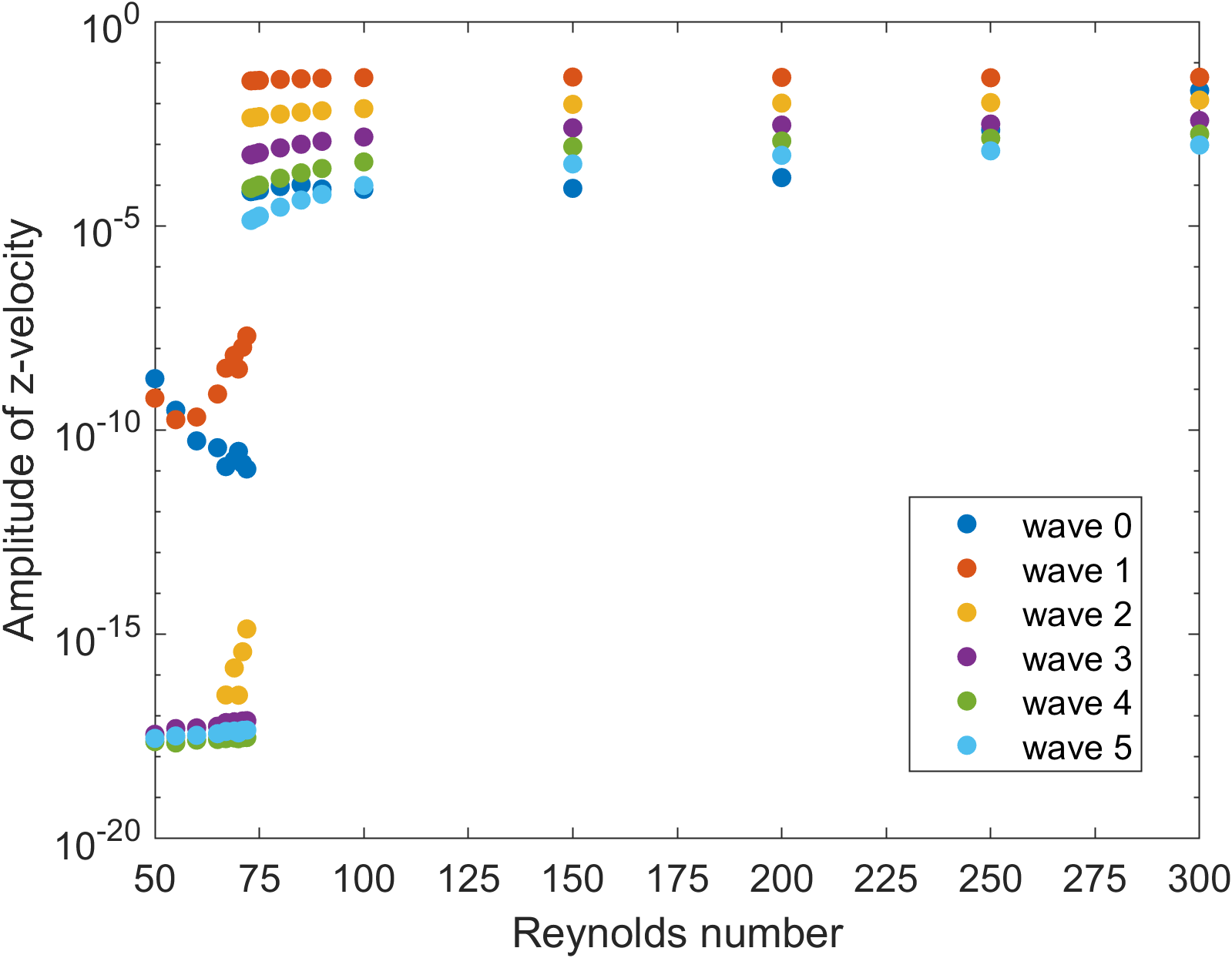}
    \caption{}
    \label{fig:wave_vs_Rez}
\end{subfigure}
\caption{Plots of amplitudes of first six Fourier modes averaged over the entire domain for the three velocity components against Reynolds number: a) averaged amplitude of the x-velocity vs Reynolds number; b) averaged amplitude of the y-velocity vs Reynolds number; c) averaged amplitude of the z-velocity vs Reynolds number}
        \label{fig:wave_vs_Re}
\end{figure}

\Cref{fig:TvsReplot,fig:NuvsReplot} show the non-dimensional torque and Nusselt number as a function of the Reynolds number. The non-dimensional torque and heat transfer are computed as given in \cref{eqn:non-dimensional_torque,eqn:Nusselt_number}.
From \cref{fig:TvsReplot} we observe that as expected the non-dimensional torque is constant for the subcritical Reynolds numbers and suddenly increases nonlinearly with Reynolds number. In a similar manner, the Nusselt number jumps when transition occurs to a Taylor-Couette flow from a base Couette flow. The base two-dimensional flow was previously presented in \cite{unnikrishnan2022shear} where several flow patterns are documented for varying flow and geometric parameters. Both the Nusselt number and the non-dimensional torque have a monotonic increasing trend with Reynolds number once transition to a supercritical state occurs. The transition is seen to occur at a Reynolds number between 70 and 75, although in this study we did not precisely determine its value. \Cref{fig:wave_vs_Re} plots the amplitudes of the first five oscillatory modes and the zeroth mode for the Reynolds numbers computed. It can be seen that the amplitudes of the oscillatory modes suddenly jump to moderate values at Reynolds number between 70 and 75, consistent with observations of torque and Nusselt number. The values of amplitude in the \cref{fig:wave_vs_Re} are not scaled. However to observe the bifurcation, the actual amplitudes are not required. We observe again that the first bifurcation occurs at a Reynolds number between 70 and 75, transitioning from a Couette flow to a Taylor-Couette flow. 

\subsection{Plots of streamlines}
\begin{figure}[H]
     \centering
     \begin{subfigure}[b]{0.55\textwidth}
         \centering
         \includegraphics[height=60mm]{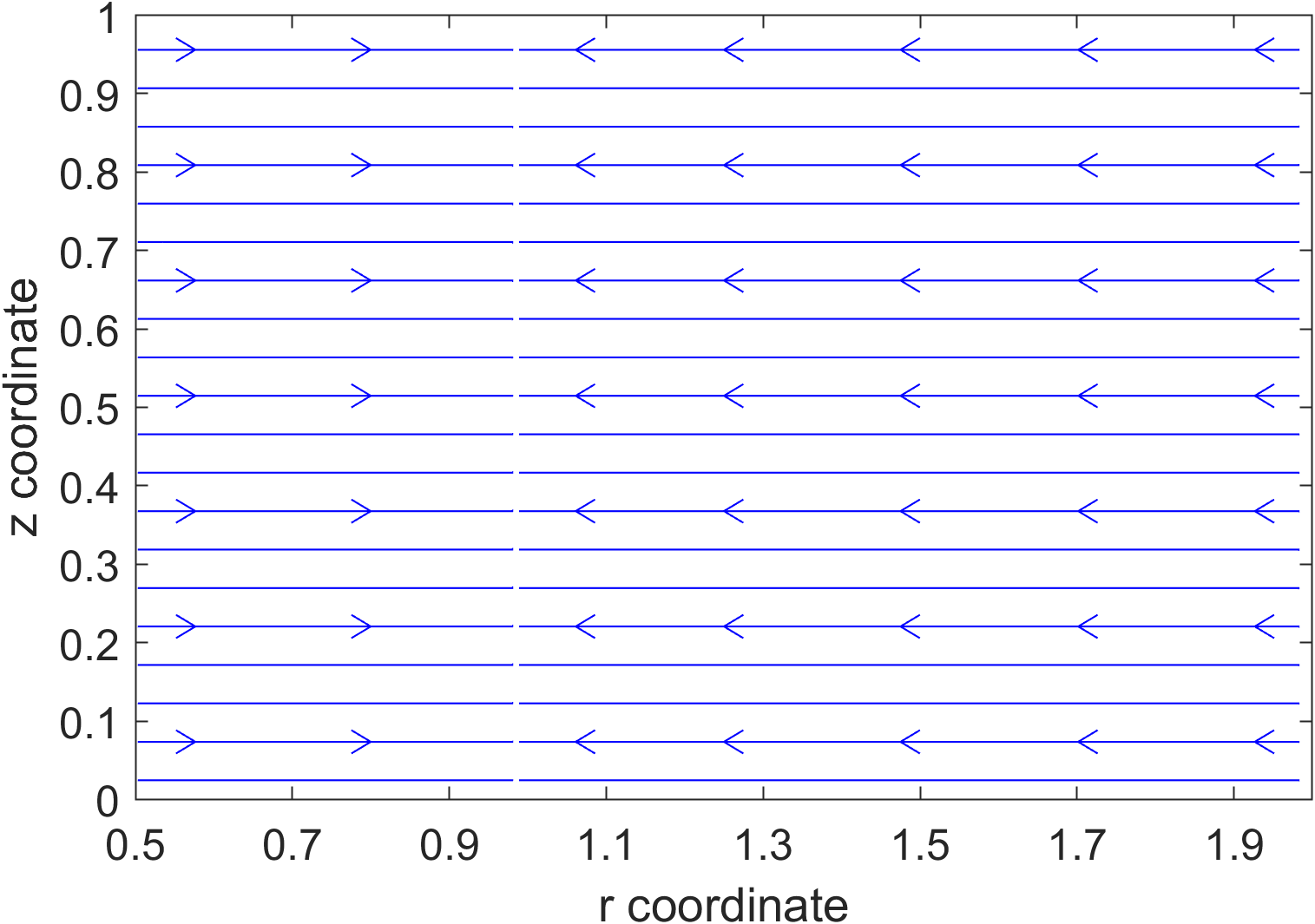}
         \caption{}
         \label{}
     \end{subfigure}
     \hfill
     \begin{subfigure}[b]{0.35\textwidth}
         \centering
         \includegraphics[height=60mm]{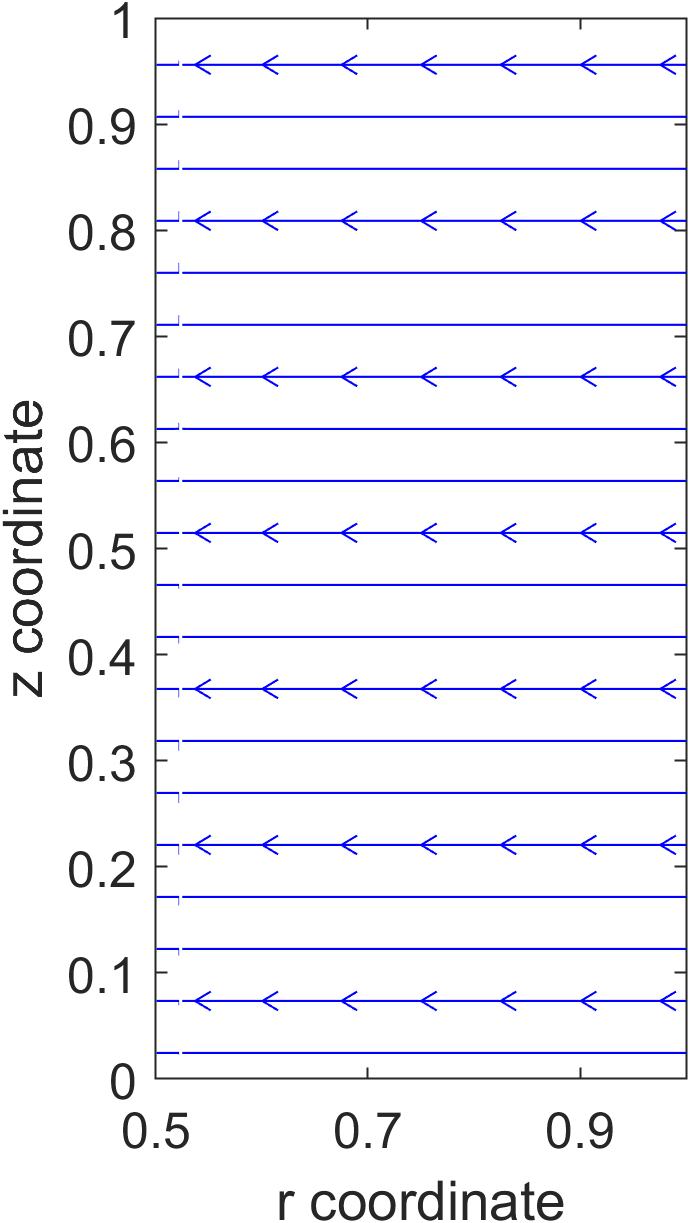}
         \caption{}
         \label{}
     \end{subfigure}
     \hfill
     \caption{Streamlines in a section along: a) the semi-major axis; b) the semi-minor axis; for Re = 70}
        \label{fig:streamlines_1}
\end{figure}

\begin{figure}[H]
     \centering
      \begin{subfigure}[b]{0.55\textwidth}
         \centering
         \includegraphics[height=60mm]{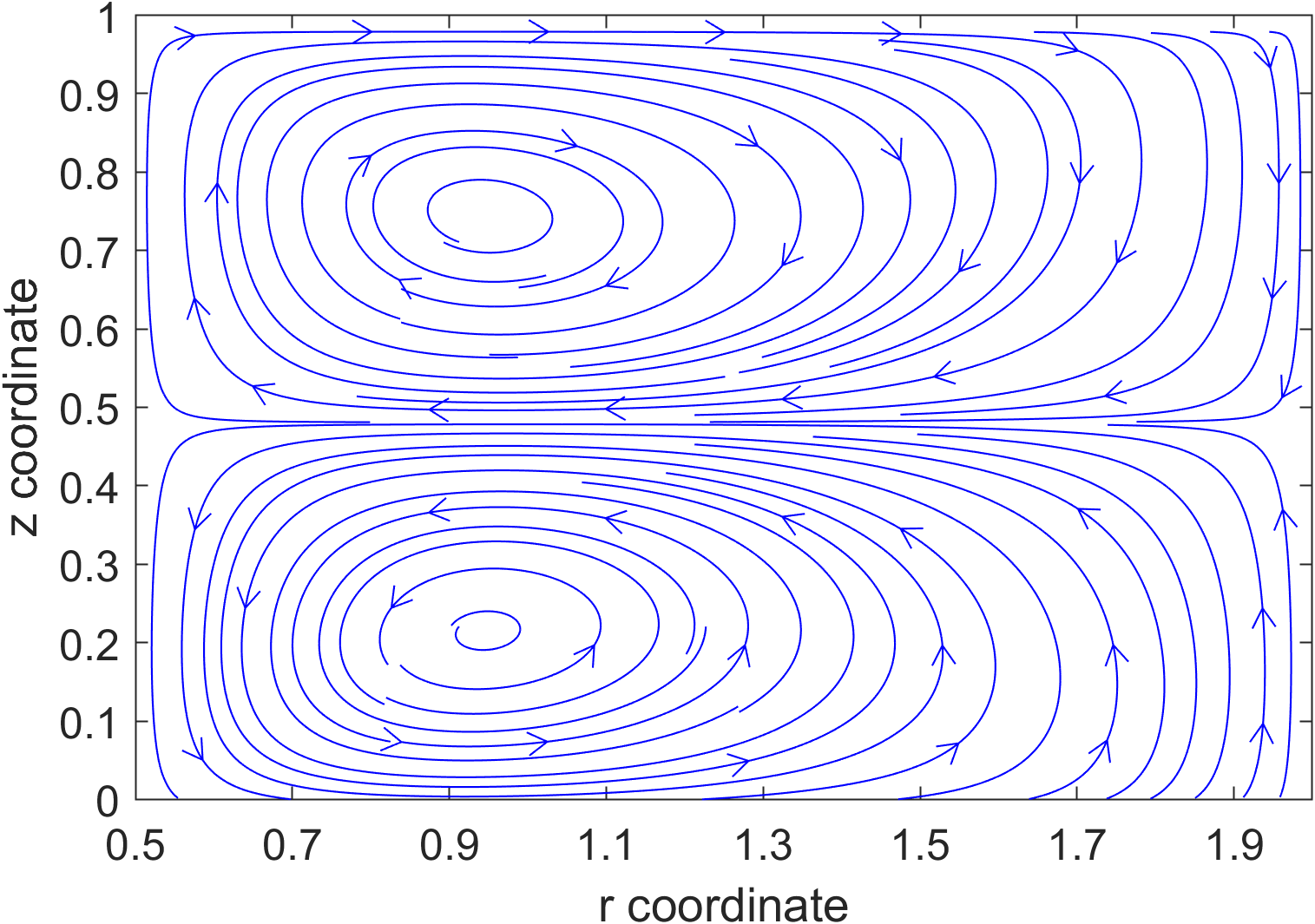}
         \caption{}
         \label{}
     \end{subfigure}
     \hfill
     \begin{subfigure}[b]{0.35\textwidth}
         \centering
         \includegraphics[height=60mm]{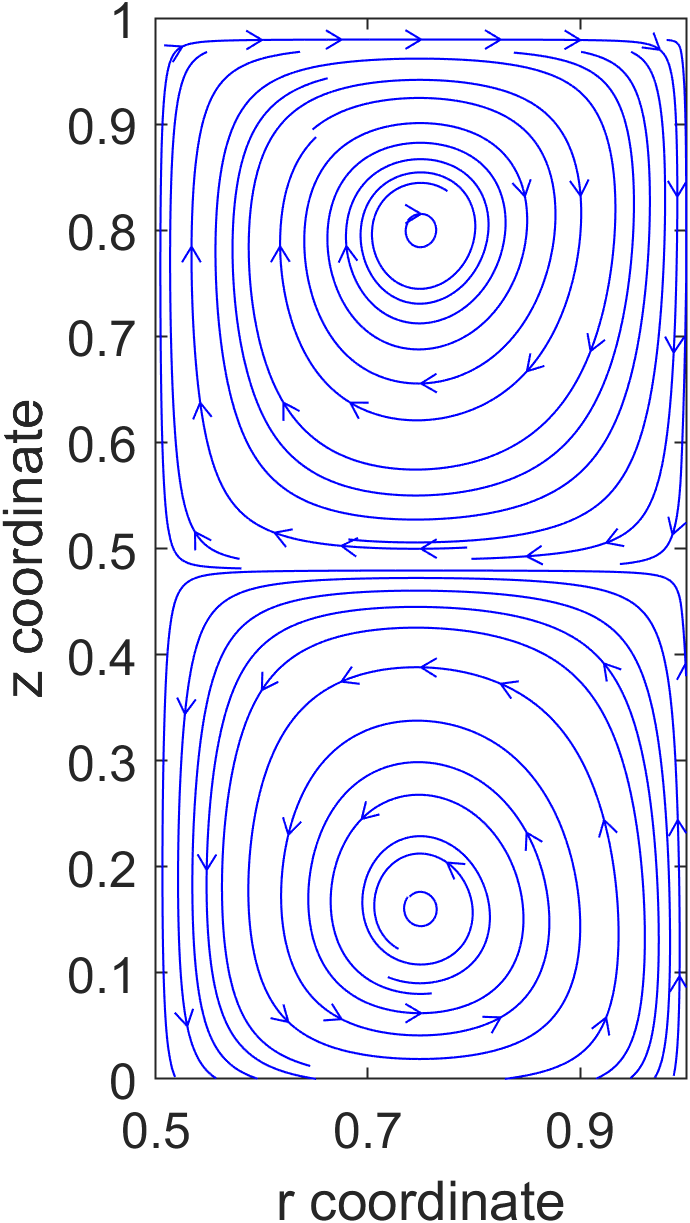}
         \caption{}
         \label{}
     \end{subfigure}
     \hfill
     \caption{Streamlines in a section along: a) the semi-major axis; b) the semi-minor axis; for Re = 75}
        \label{fig:streamlines_2}
\end{figure}

\begin{figure}[H]
     \centering
     \begin{subfigure}[b]{0.55\textwidth}
         \centering
         \includegraphics[height=60mm]{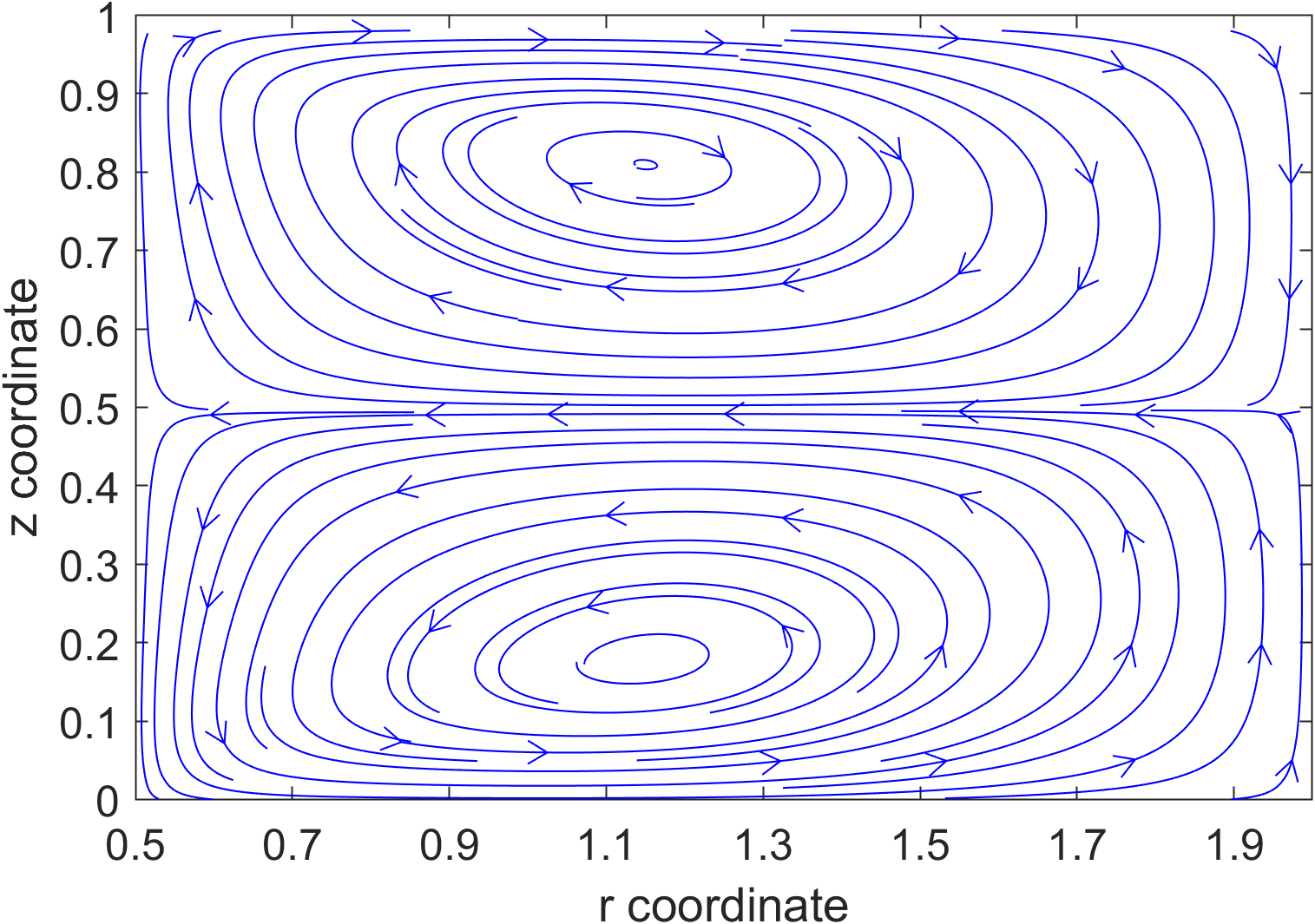}
         \caption{}
         \label{}
     \end{subfigure}
     \hfill
     \begin{subfigure}[b]{0.35\textwidth}
         \centering
         \includegraphics[height=60mm]{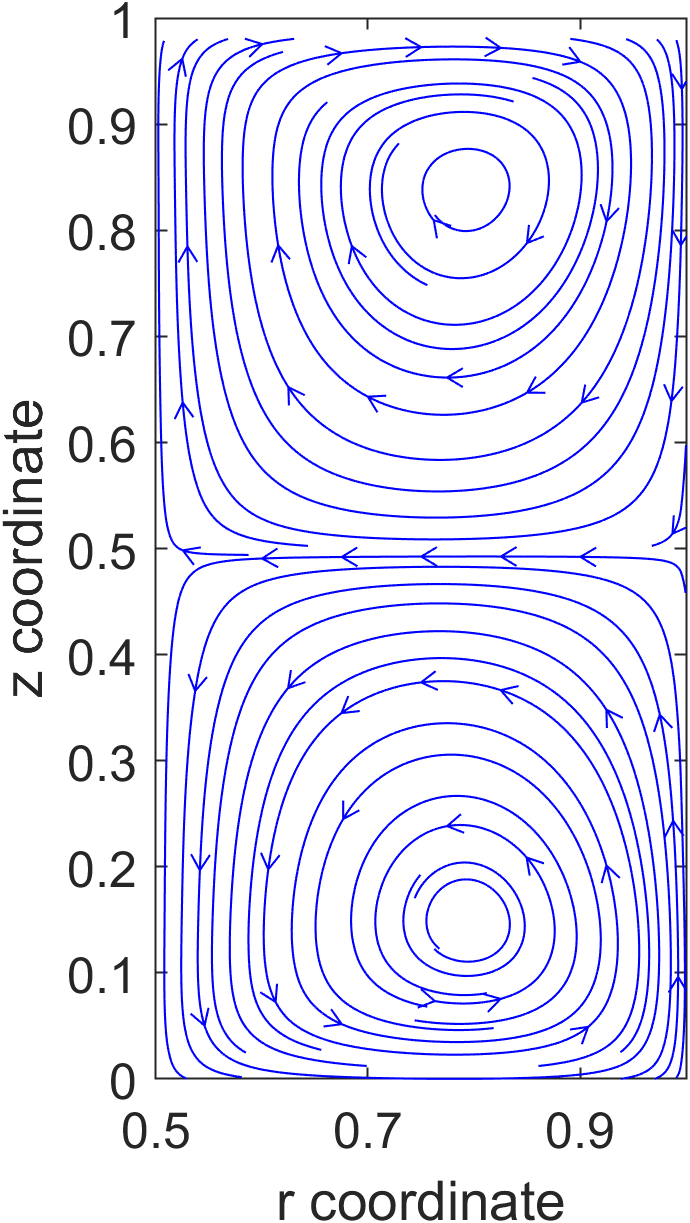}
         \caption{}
         \label{}
     \end{subfigure}
     \hfill
     \caption{Streamlines in a section along: a) the semi-major axis; b) the semi-minor axis; for Re = 200}
        \label{fig:streamlines_4}
\end{figure}

\begin{figure}[H]
     \centering
     \begin{subfigure}[b]{0.55\textwidth}
         \centering
         \includegraphics[height=60mm]{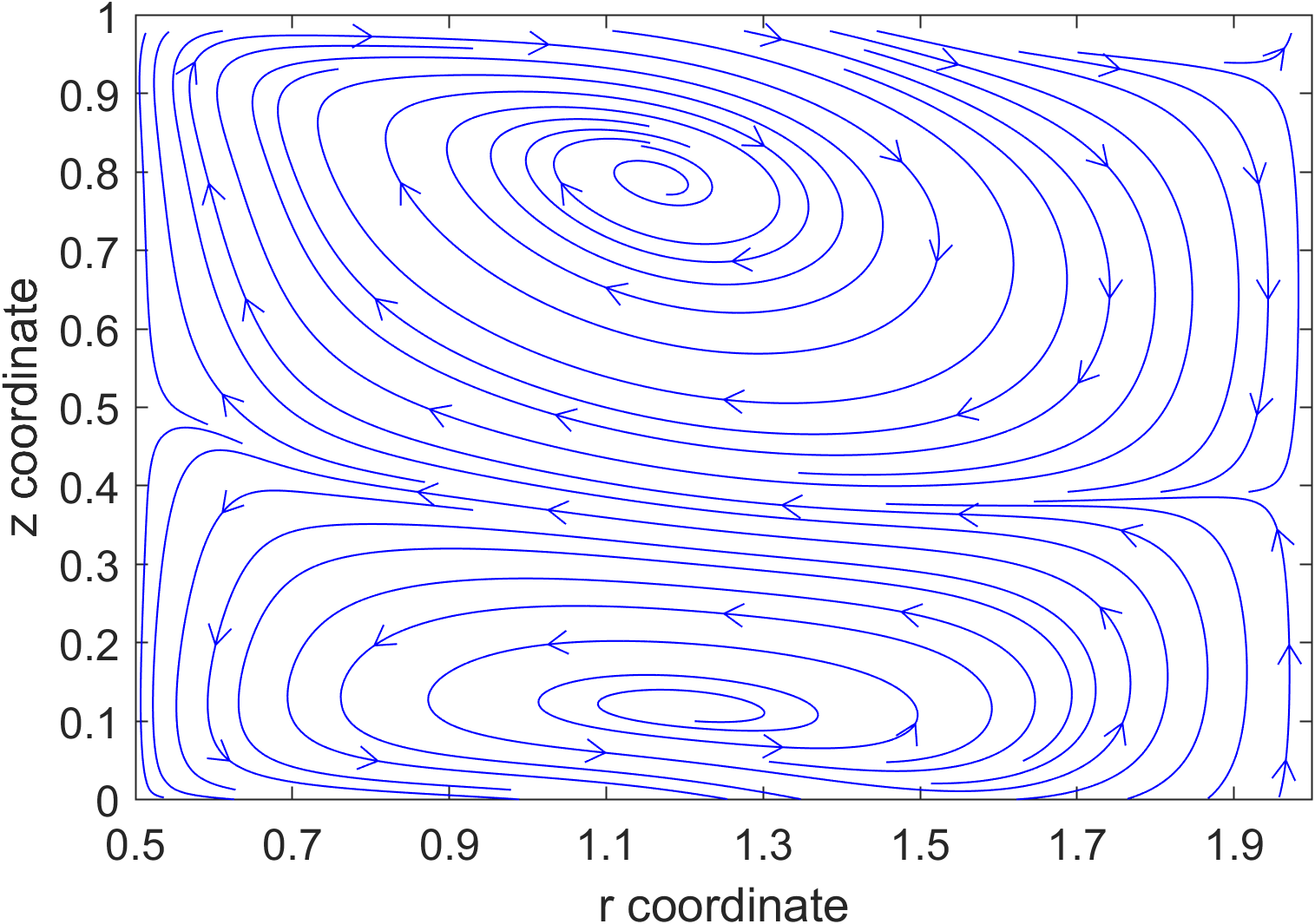}
         \caption{}
         \label{}
     \end{subfigure}
     \hfill
     \begin{subfigure}[b]{0.35\textwidth}
         \centering
         \includegraphics[height=60mm]{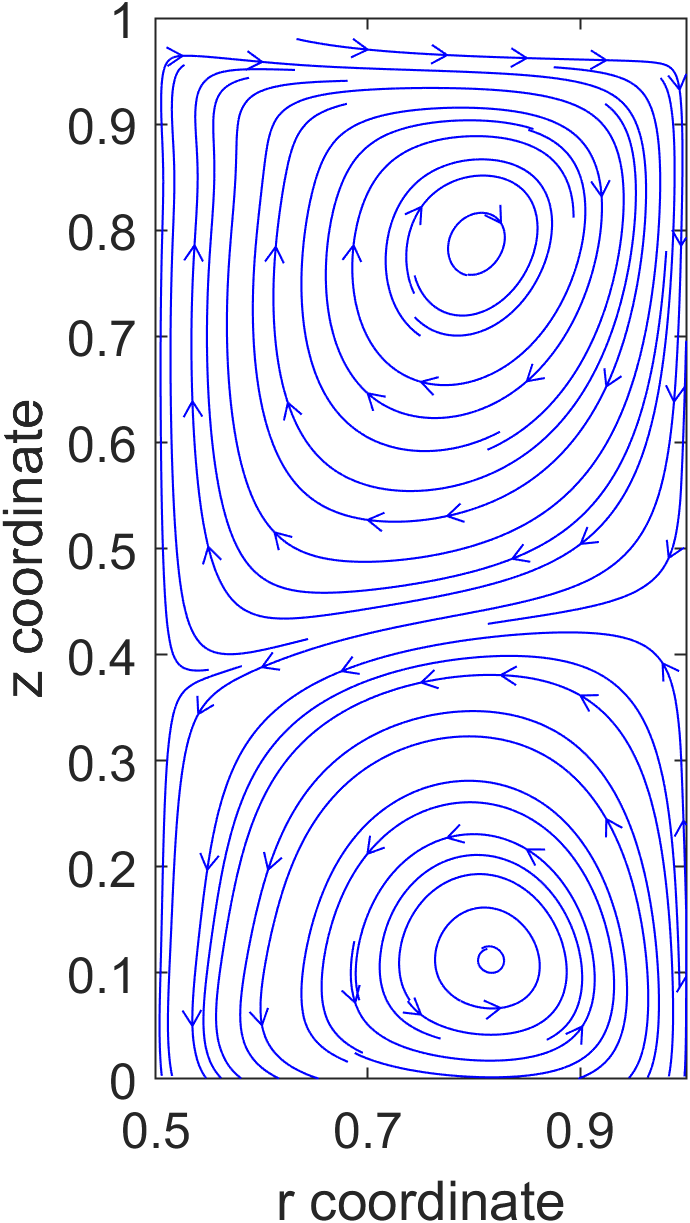}
         \caption{}
         \label{}
     \end{subfigure}
     \caption{Streamlines in a section along: a) the semi-major axis; b) the semi-minor axis; for Re = 300}
        \label{fig:streamlines_5}
\end{figure}

In \cref{fig:streamlines_1,fig:streamlines_2,fig:streamlines_4,fig:streamlines_5}, we present the streamlines for progressively increasing Reynolds numbers from 70 to 300 (the flow is seen to be steady at all these Reynolds numbers). Below the critical Reynolds number, the streamlines show an outward flow along the semi-major axis and an inward flow along the semi-minor axis. This is due to the expansion and contraction of the flow as it squeezes into the smaller gap and then expands to the larger cross-sectional area. The inward streamlines reflect a secondary vortex formed by the shear of the primary rotating flow caused by the rotating cylinder. We observe the same patterns for Re = 50, 60 and 70. The first Reynolds number where the Taylor vortices appear is 75, although they may have developed at a slightly smaller Reynolds number.  
At Re = 75 and beyond, the streamlines in the ($r-z$) plane show the formation of Taylor cells. As in the case of concentric circular cylinders, a periodic pair of counter-rotating vortices is formed with a height equal to the width of the smaller gap (at the minor axis). However, the vortices are no longer axisymmetric (as in the case of circular cylinders), but expand and contract between the major and minor axes of the ellipse. It can be seen that on the semi-major axis, the vortex occupies the entire gap with a vortex height of $0.5$ and a width of $1.0$ units. In the direction of the minor axis however, the cell is more of a square shape with width and height of $0.5$. The expansion of the Taylor cells into the larger gap is evident from these ($r-z$) plane plots.



\subsection{Axial velocity distributions}
\subsubsection{Contours of axial velocity in r-z planes}

\begin{figure}[H]
     \centering
    \hfill
     \begin{subfigure}[b]{0.49\textwidth}
         \centering
         \includegraphics[height=50mm]{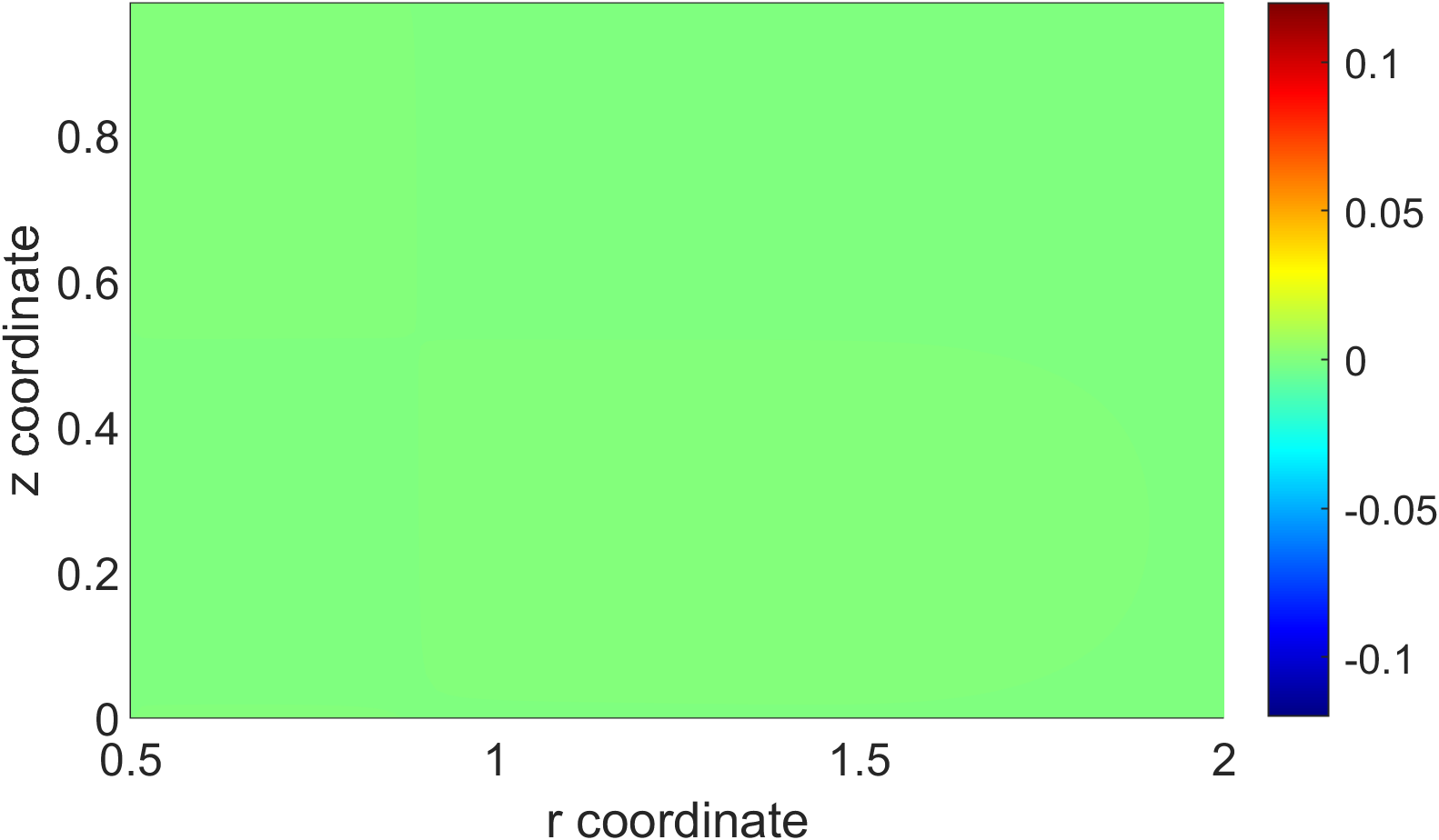}
         \caption{Re = 70}
         \label{}
     \end{subfigure}
     \hfill
     \begin{subfigure}[b]{0.49\textwidth}
         \centering
         \includegraphics[height=50mm]{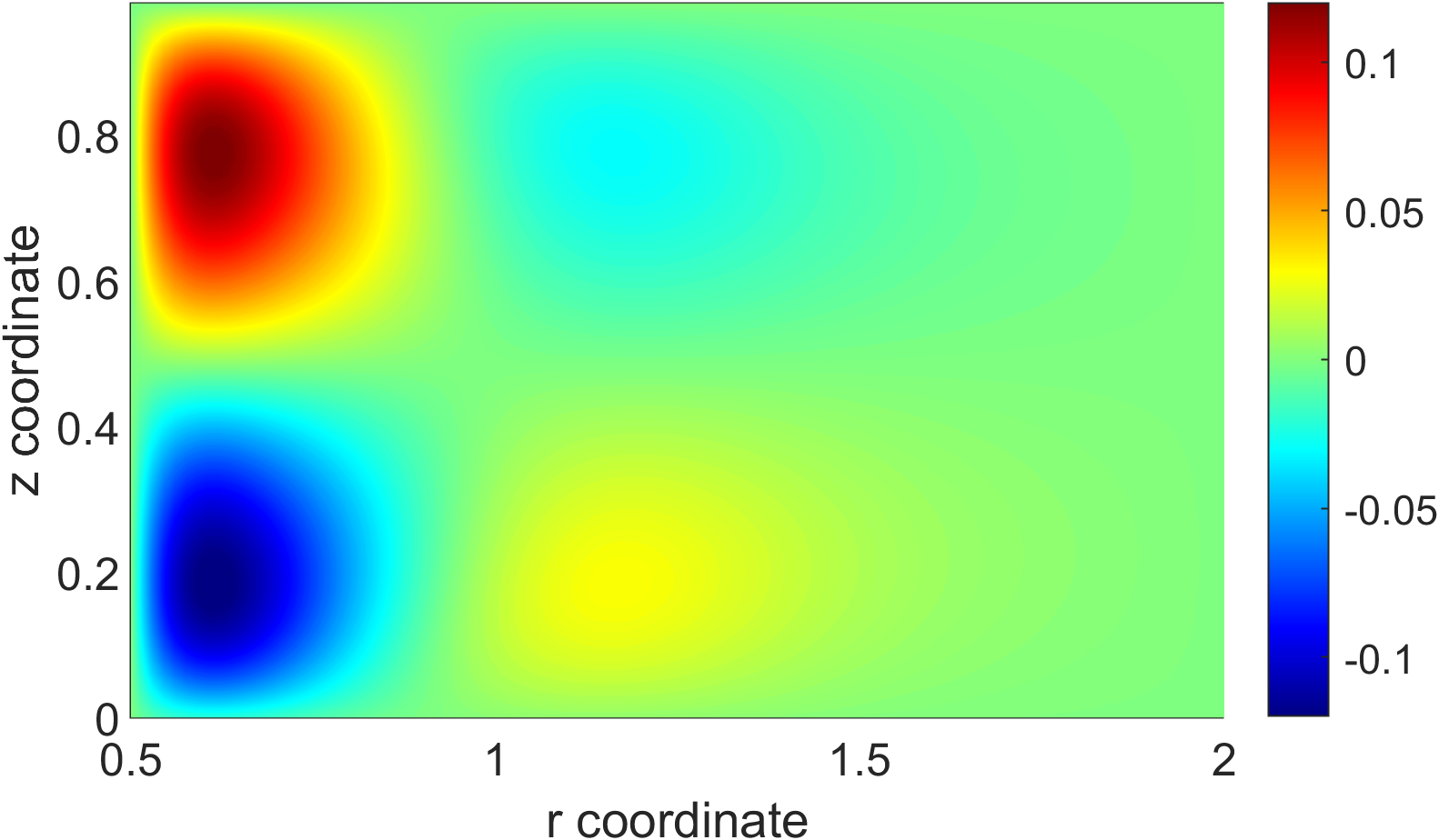}
         \caption{Re = 75}
         \label{}
     \end{subfigure}
     \par\bigskip
    \vspace{4mm}
     \begin{subfigure}[b]{0.49\textwidth}
         \centering
         \includegraphics[height=50mm]{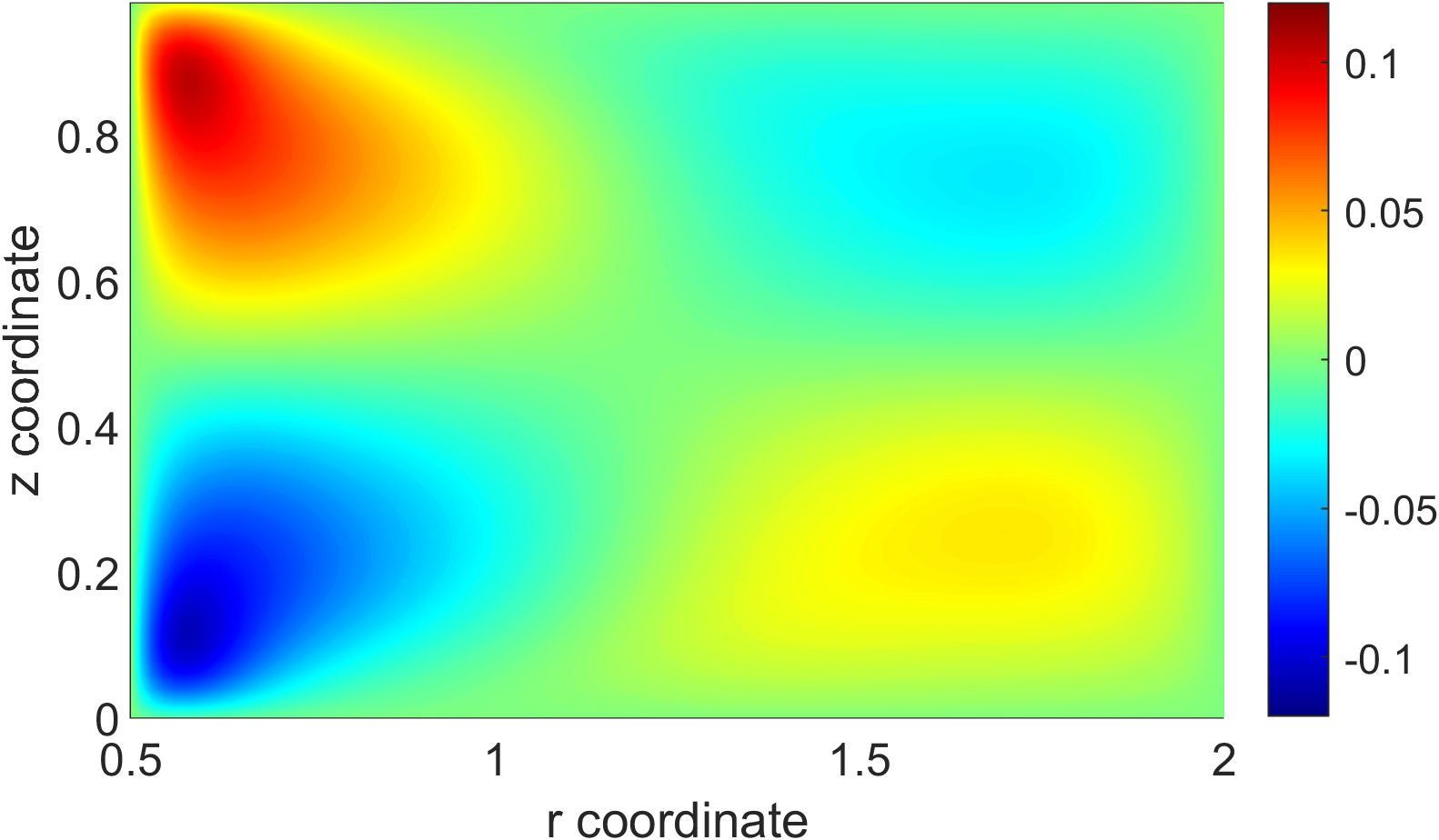}
         \caption{Re = 200}
         \label{}
     \end{subfigure}
     \begin{subfigure}[b]{0.49\textwidth}
         \centering
         \includegraphics[height=50mm]{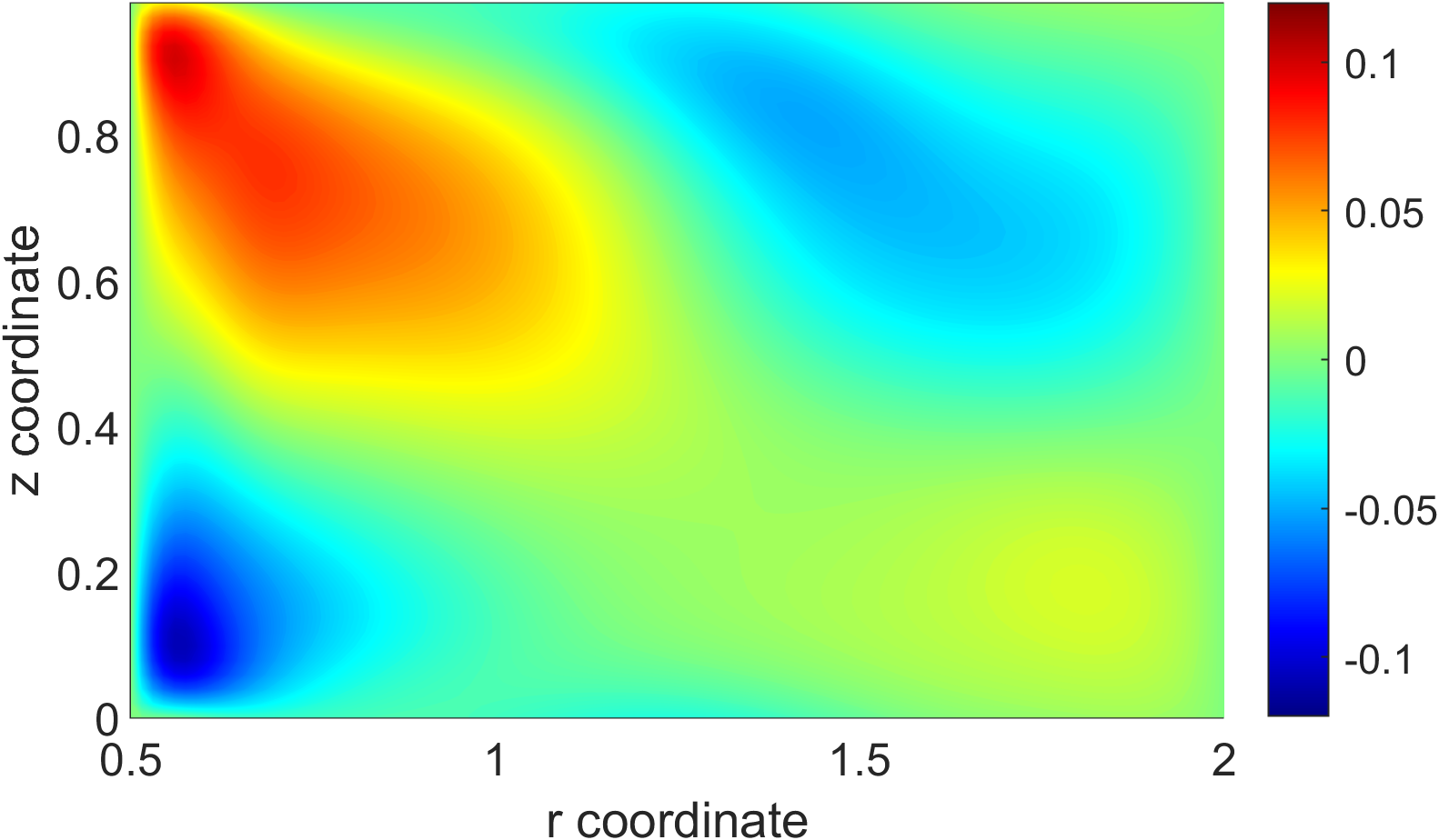}
         \caption{Re = 300}
         \label{}
     \end{subfigure}
     \caption{Contours of axial velocity in the plane along semi major axis of ellipse}
        \label{fig:z_velocity_major_axis}
\end{figure}

\begin{figure}[H]
     \centering
     \begin{subfigure}[b]{0.24\textwidth}
         \centering
         \includegraphics[width = 0.9\textwidth]{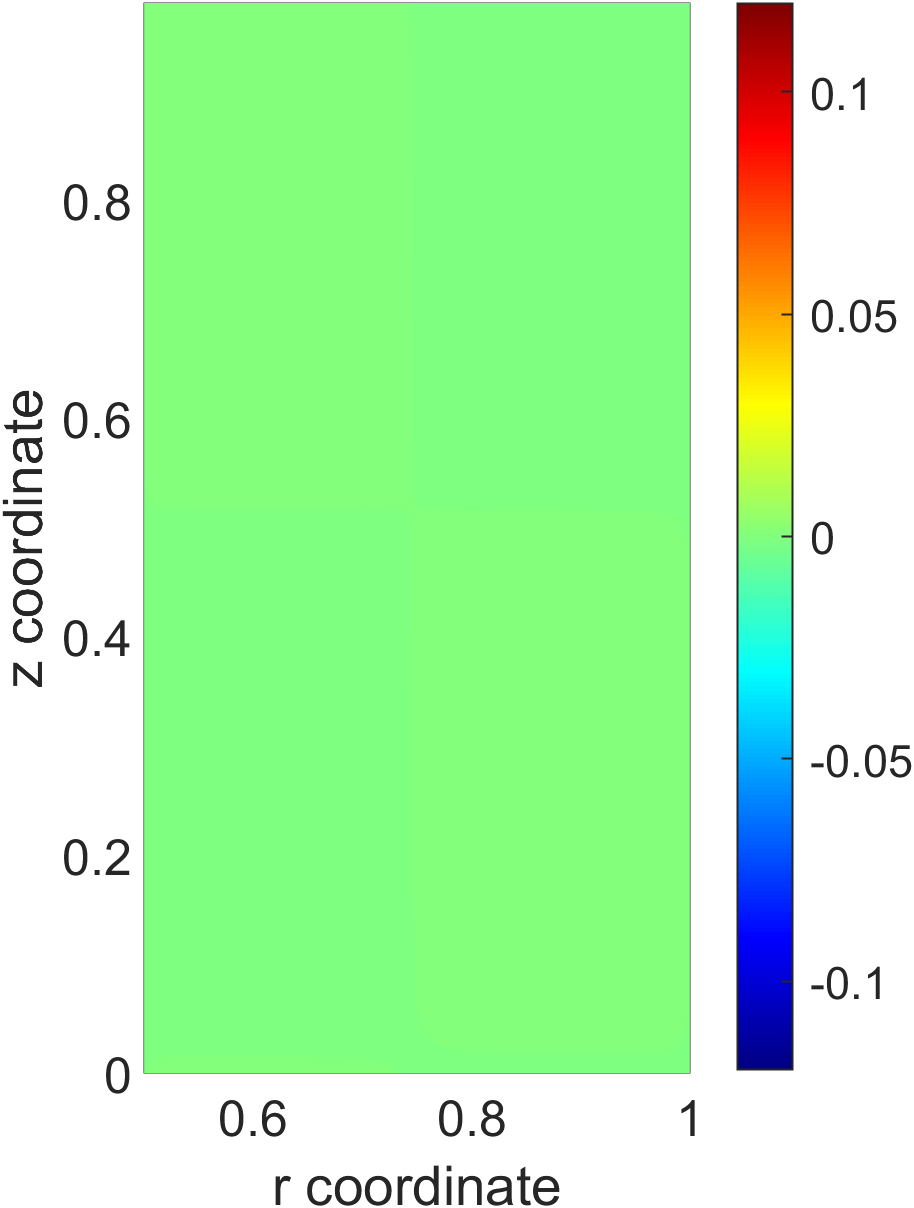}
         \caption{Re = 70}
         \label{}
     \end{subfigure}
     \hfill
     \begin{subfigure}[b]{0.24\textwidth}
         \centering
         \includegraphics[width = 0.9\textwidth]{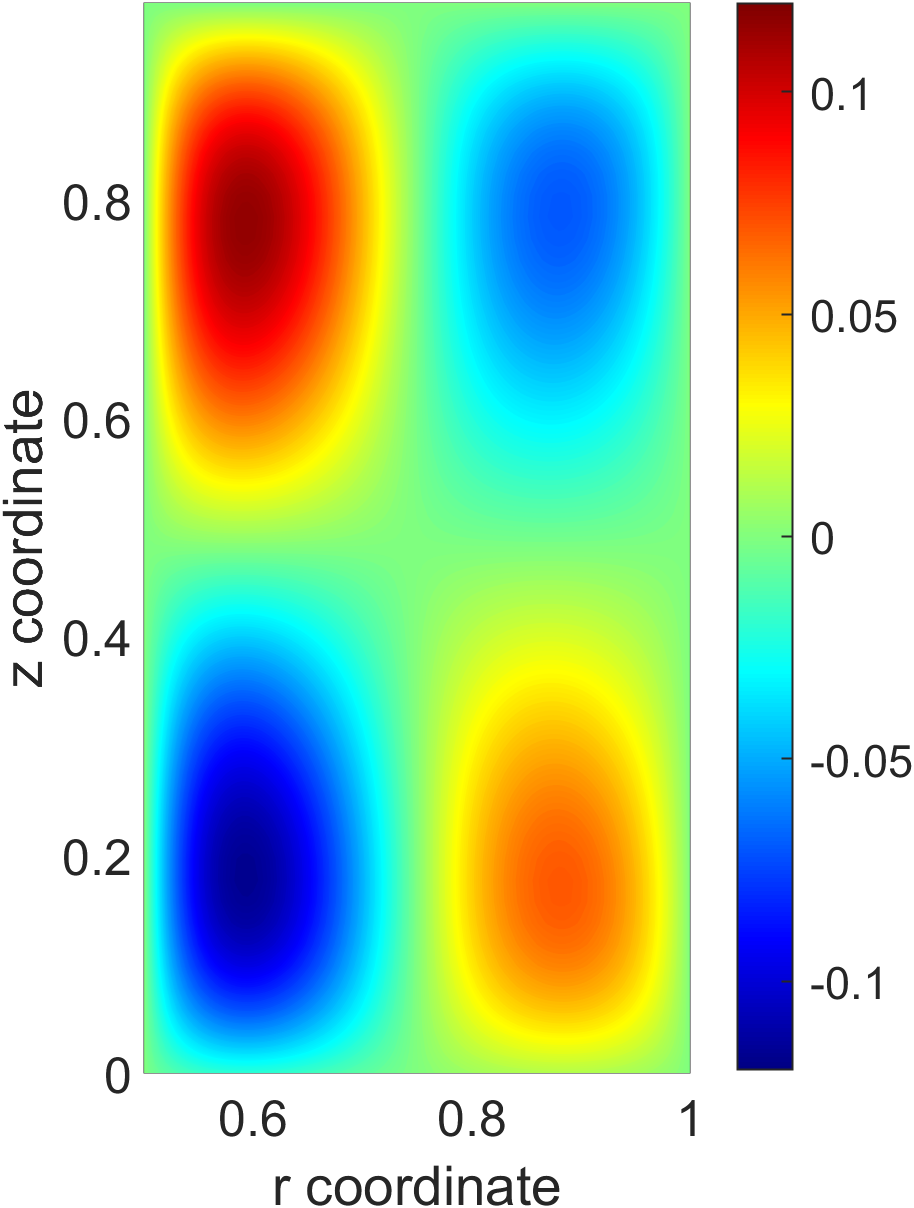}
         \caption{Re = 75}
         \label{}
     \end{subfigure}
     \hfill
     \begin{subfigure}[b]{0.24\textwidth}
         \centering
         \includegraphics[width = 0.9\textwidth]{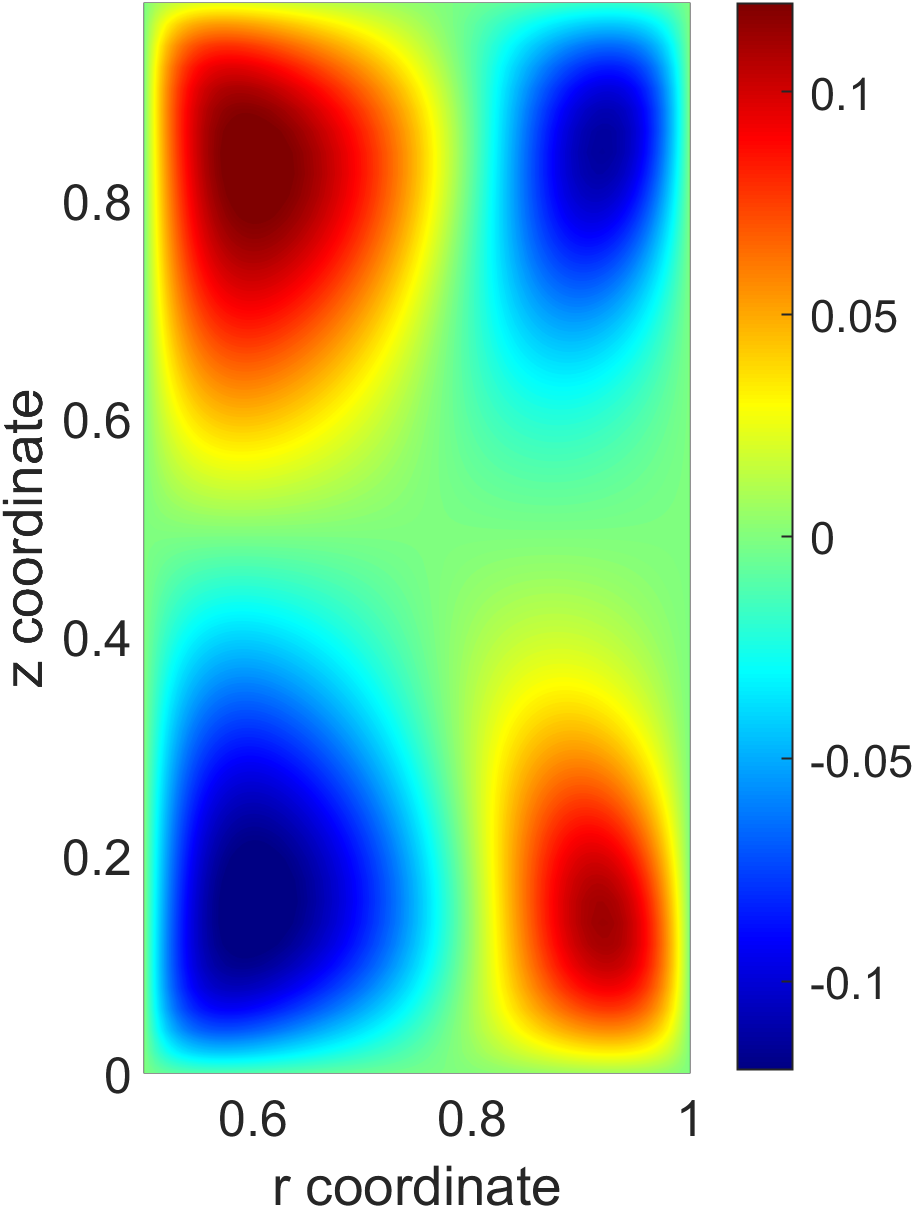}
         \caption{Re = 200}
         \label{}
     \end{subfigure}
     \hfill
     \begin{subfigure}[b]{0.24\textwidth}
         \centering
         \includegraphics[width = 0.9\textwidth]{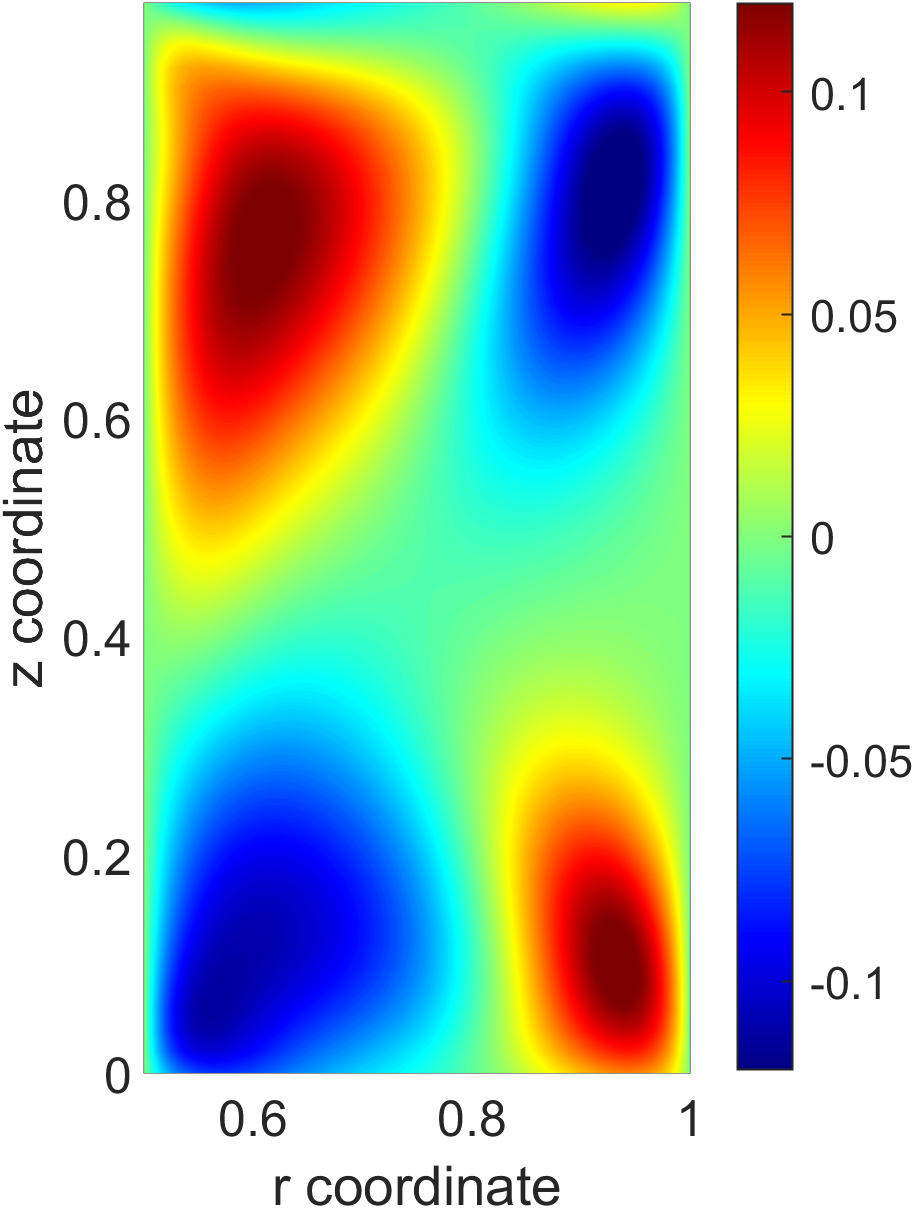}
         \caption{Re = 300}
         \label{}
     \end{subfigure}
     \caption{Contours of axial velocity in the plane along semi minor axis of ellipse}
        \label{fig:z_velocity_minor_axis}
\end{figure}

\Cref{fig:z_velocity_major_axis,fig:z_velocity_minor_axis} show the contours of axial velocity in ($r-z$) planes cutting the major and minor axes for four of the several Reynolds numbers computed. The plots reveal the same structure of the Taylor cells shown by the plots of streamlines but now indicate the magnitudes of the axial velocities. The pair of vortices show the upward and downward motion of the flow. As seen earlier, there is no axial flow at Reynolds number below 75. At Re = 75, we first notice the appearance of the Taylor cells. This value of Reynolds number is close to the value also observed for concentric circular cylinders. We notice that along the major axis, the center of the Taylor cells is located approximately at r = 0.95. The downward axial flow traverses radially and reaches the outer wall and occupies the entire gap. We see only one single Taylor cell which is elongated in the radial direction along the major axis. At the outer wall, the radial flow turns upwards and flows to the inner wall. This pattern is the opposite for the clockwise rotating cell.  As the flow occupies the outer wall region, the axial velocities become smaller because of the larger area for the flow. As the Reynolds number is increased, the Taylor cells get somewhat distorted, and the centers of the cells move closer to the outer wall ($r \approx 1.1$). The cells continue to distort with increase in Reynolds number and at $Re = 300$, the flow transitions to a complex structure. We observe that the flow is however still steady at this Reynolds number. \Cref{fig:z_velocity_minor_axis} shows the results for the plane cutting the minor axis of the ellipse. Here, the gap is $0.5$ only and the cell structure resembles the circular cylinder case. Even at $Re = 300$, the cells appear well-structured. Note that the Reynolds number is defined by the gap at the minor axis, hence the Reynolds number defined by the gap along the major axis will be three times larger. Hence the flow in the major axis gap may have transitioned to some form of spiral vortices.  

\subsubsection{Contours of axial velocity in z-$\theta$ planes: $r = 0.8$}

\begin{figure}[H]
    \centering
    \begin{subfigure}{0.45\textwidth}
        \centering
        \includegraphics[height=50mm]{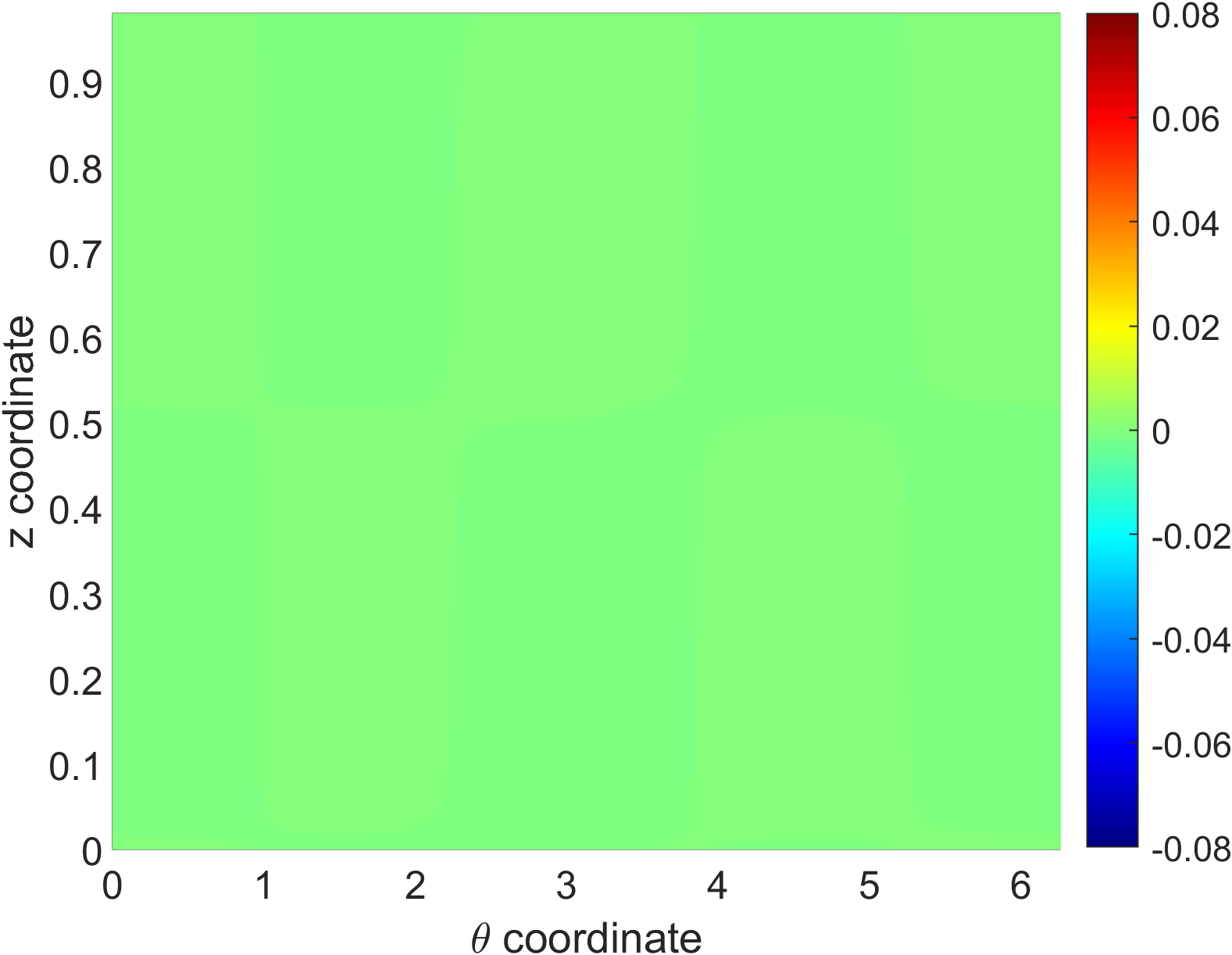}
        \caption{$Re = 70$}
        \label{}
    \end{subfigure}
    \hfill
    \begin{subfigure}{0.45\textwidth}
        \centering
        \includegraphics[height=50mm]{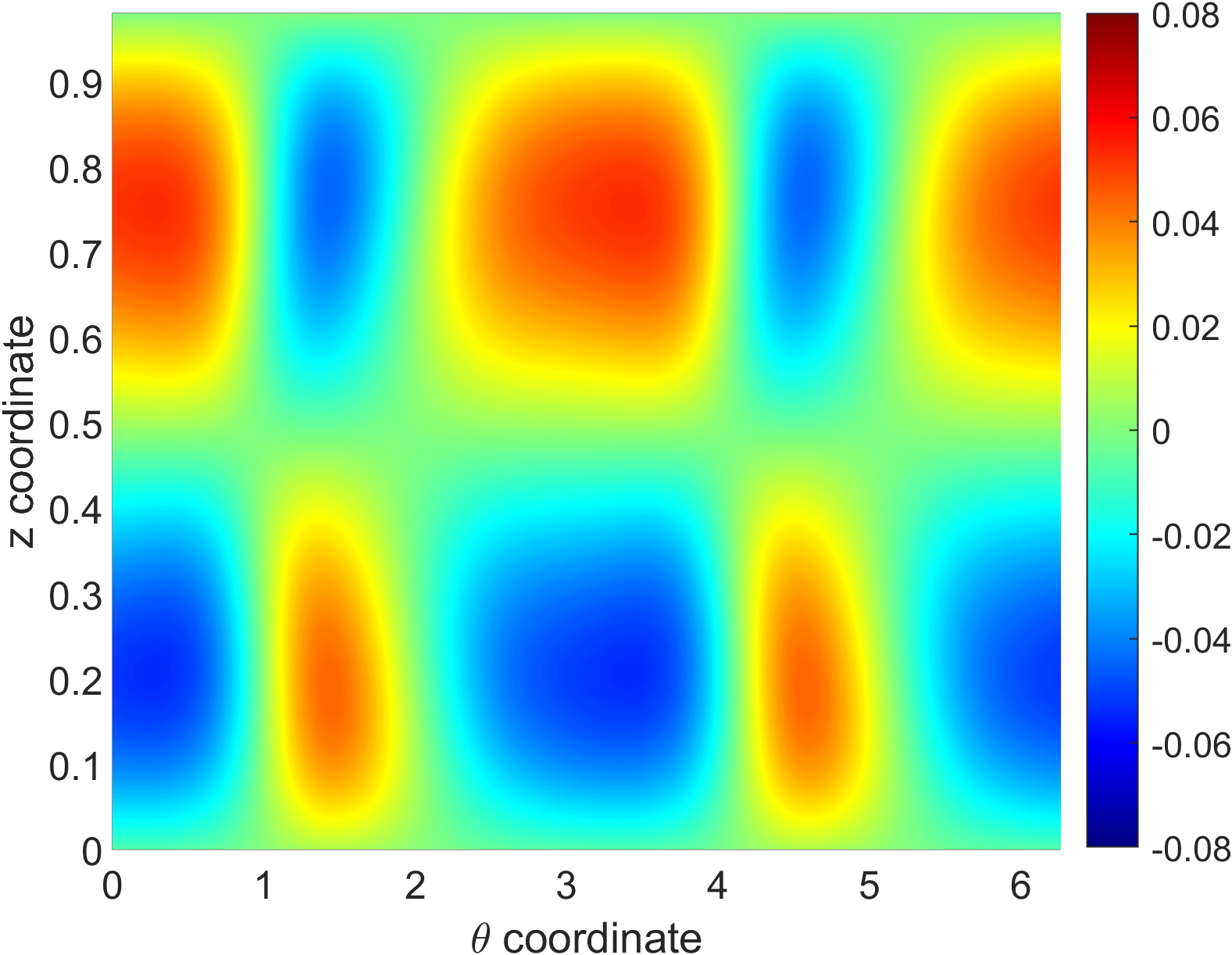}
        \caption{$Re = 75$}
        \label{}
    \end{subfigure}
    \par\bigskip
    \begin{subfigure}{0.45\textwidth}
        \centering
        \includegraphics[height=50mm]{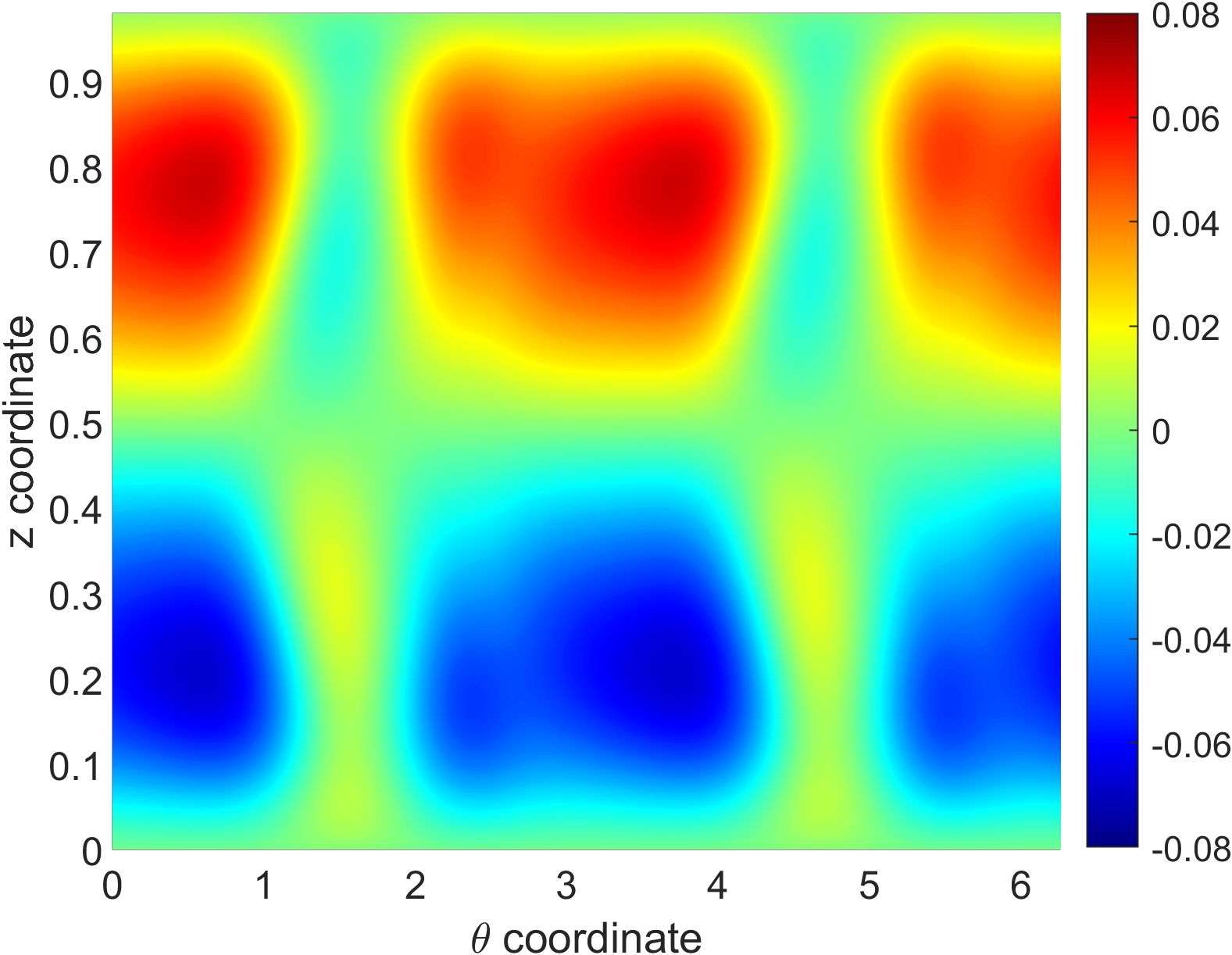}
        \caption{$Re = 200$}
        \label{}
    \end{subfigure}
    \hfill
    \begin{subfigure}{0.45\textwidth}
        \centering
        \includegraphics[height=50mm]{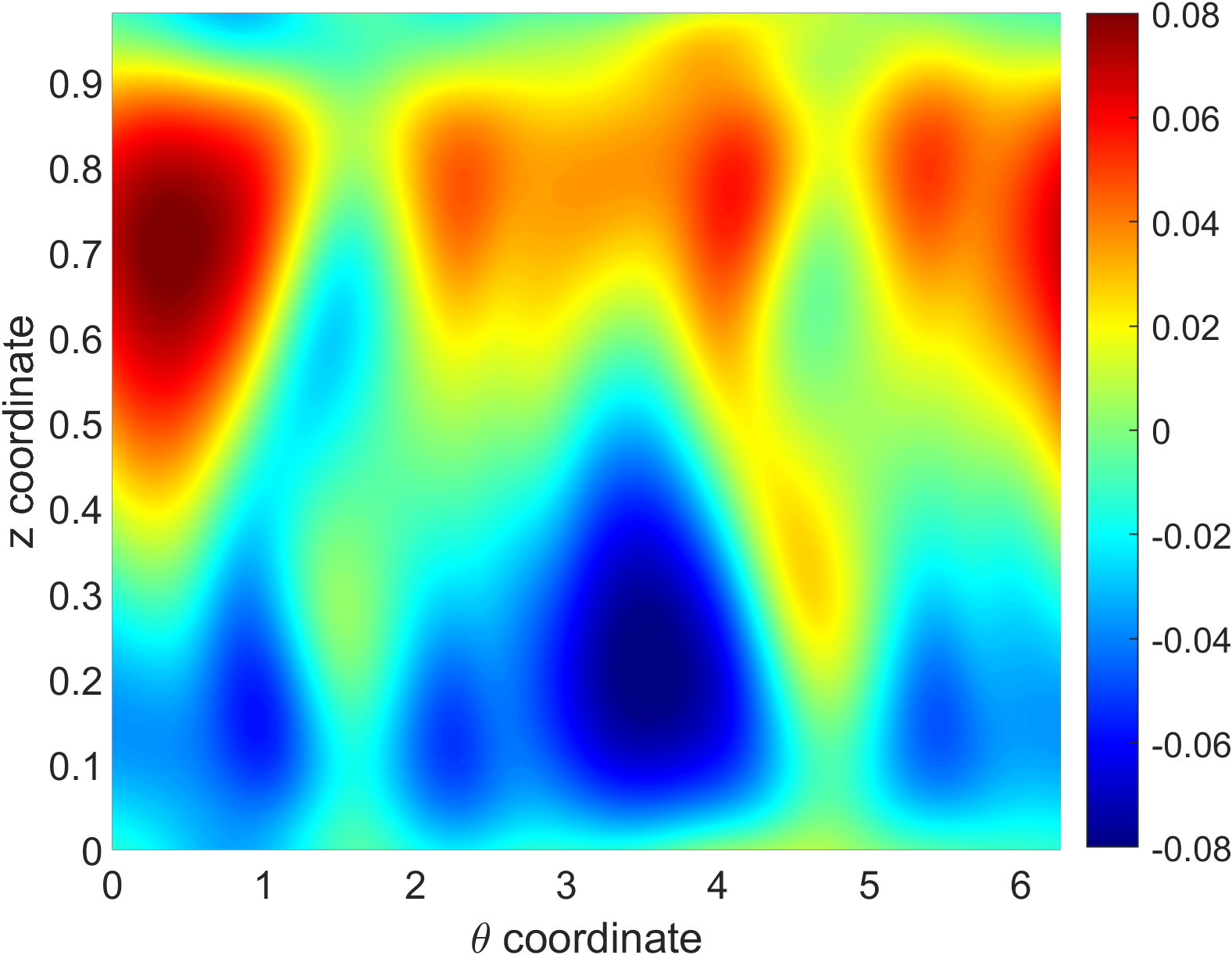}
        \caption{$Re = 300$}
        \label{}
    \end{subfigure}
    \caption{Contours of z-velocity for different Reynolds numbers in a constant radius plane where radius, $r = 0.8$}
    \label{fig:zvelocity_ztheta_section}
\end{figure}

\Cref{fig:zvelocity_ztheta_section} show the contours of axial velocity in the ($z-\theta$) planes. We have arbitrarily selected a radius of $0.8$ ($r_i=0.5$, $b = 1.0$, and $a=2.0$) and plot the contours of the axial velocity. For $Re = 75$, the first Re considered in the super-critical region, we see alternate regions of positive and negative velocities. This contrasts with horizontal bands throughout the $\theta$ space seen for the concentric circular cylinder case. The alternate regions are due to the shift in the vortex centers as the Taylor cells move from the major to the minor gaps. We have plotted the contours for $r = 0.8$, which corresponds to the left of the vortex center in the major gap with upward (positive) velocity above $z=0.5$ and negative axial velocity below around $z=0.5$. However, in the minor gap, the center of the vortex moves closer to $r=0.75$ (center of the gap) and therefore at $r=0.8$, a downward flow in the upper region and a upward flow in the region below is seen. It is still one single toroidal vortex with the toroid converging and diverging between the major and minor gaps. Because we have considered a constant absolute radius, we observe these alternating patterns. The same patterns are seen for $Re = 100$ (not shown here) and for $Re = 200$.  For $Re = 300$, the above pattern gets distorted with break-up of the cell structure. As mentioned earlier, this may be a prelude to the onset of unsteadiness in the flow, although the flow at $Re = 300$ converged to a steady state.

\subsubsection{Contours of axial velocity in r-$\theta$ planes}

\Cref{fig:zvel_clock_Mid-section_a,fig:zvel_section_inbtw_a,fig:zvel_cr_section_a} show contours of axial velocities in three constant z planes ($r - \theta$) for $Re = 75$. As expected, the Taylor cells are mirror symmetric about the vertical plane, but are antisymmetric about the horizontal plane. The plots for $Re = 200$ (\cref{fig:zvel_clock_Mid-section_b,fig:zvel_section_inbtw_b,fig:zvel_cr_section_b}) show similar features as for $Re = 75$. However, we observe small amounts of distortion of the inner concentric flow, accompanied by longer ‘tongues’ in the flow along the outer walls. With increase of Reynolds number to $300$ (\cref{fig:zvel_clock_Mid-section_c,fig:zvel_section_inbtw_c,fig:zvel_cr_section_c}), the inner rotating flow gets further distorted with also changes in the outer tongues. These distortions again point to the forthcoming unsteadiness in the flow beyond $Re = 300$. The flow has also lost symmetry (mirror symmetry) about the vertical (horizontal) axes.

\begin{figure}[H]
    \centering
    \includegraphics[width=0.45\textwidth]{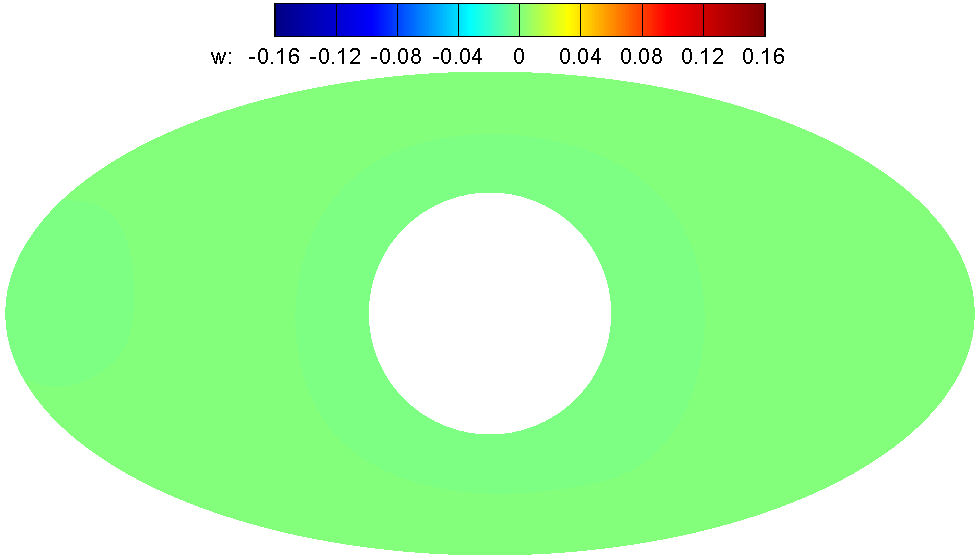}
    \caption{Contour of axial velocity for a subcritical Reynolds number of 70}
    \label{fig:zvel_rtheta}
\end{figure}

\begin{figure}[H]
    \centering
    \begin{subfigure}{0.45\textwidth}
        \centering
        \includegraphics[width=\textwidth]{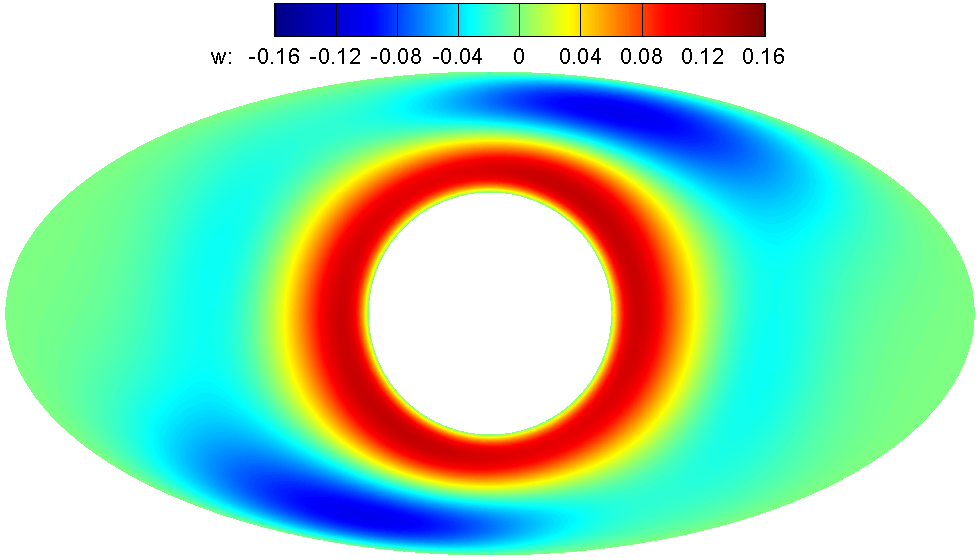}
        \caption{Re = 75}
        \label{fig:zvel_clock_Mid-section_a}
    \end{subfigure}
    \hfill
    \begin{subfigure}{0.45\textwidth}
        \centering
        \includegraphics[width=\textwidth]{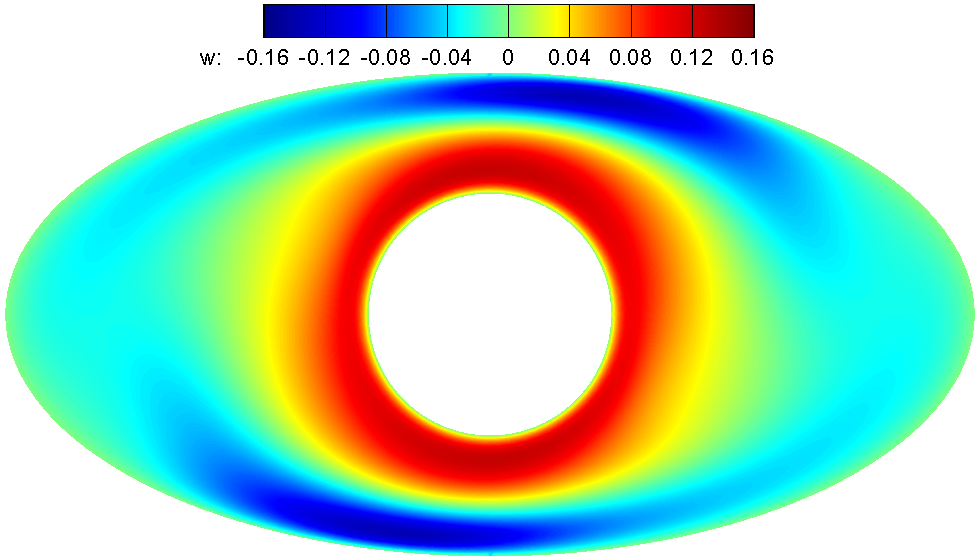}
        \caption{Re = 200}
        \label{fig:zvel_clock_Mid-section_b}
    \end{subfigure}
    \hfill
    \par\bigskip
    \begin{subfigure}{0.45\textwidth}
        \centering
        \includegraphics[width=\textwidth]{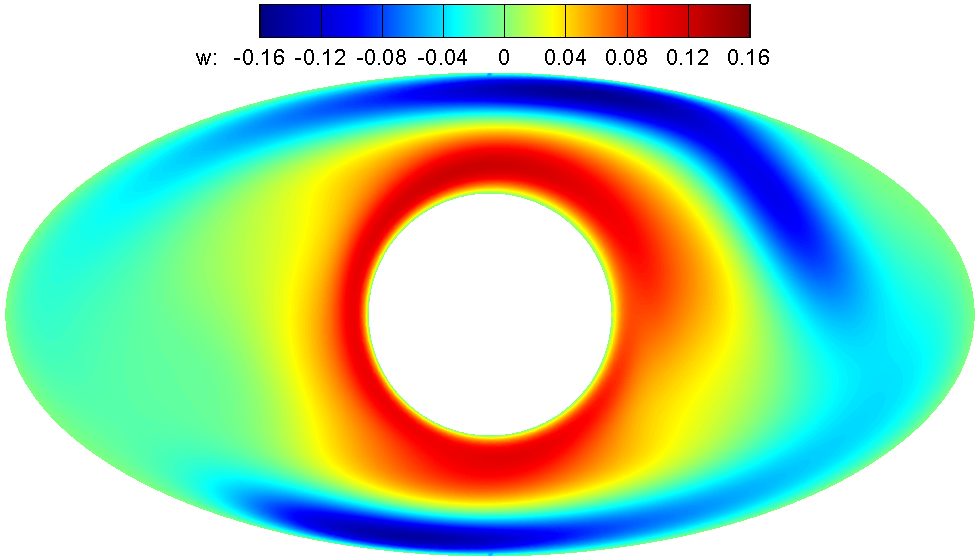}
        \caption{Re = 300}
        \label{fig:zvel_clock_Mid-section_c}
    \end{subfigure}
    \caption{Mid-section cutting through the center of Taylor cell rotating clockwise for supercritical stage}
    \label{fig:zvel_clock_Mid-section}
\end{figure}

\begin{figure}[H]
    \centering
    \begin{subfigure}{0.45\textwidth}
        \centering
        \includegraphics[width=\textwidth]{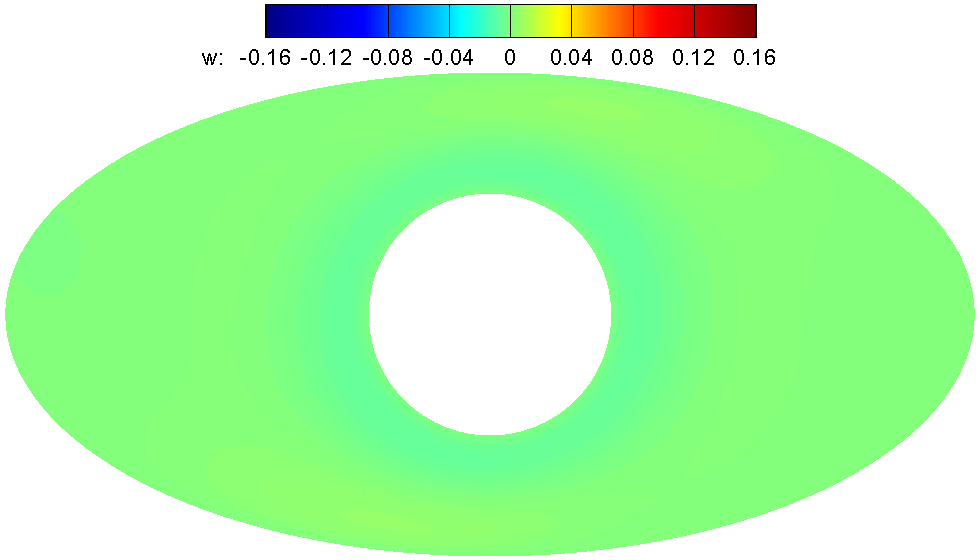}
        \caption{Re = 75}
        \label{fig:zvel_section_inbtw_a}
    \end{subfigure}
    \hfill
    \begin{subfigure}{0.45\textwidth}
        \centering
        \includegraphics[width=\textwidth]{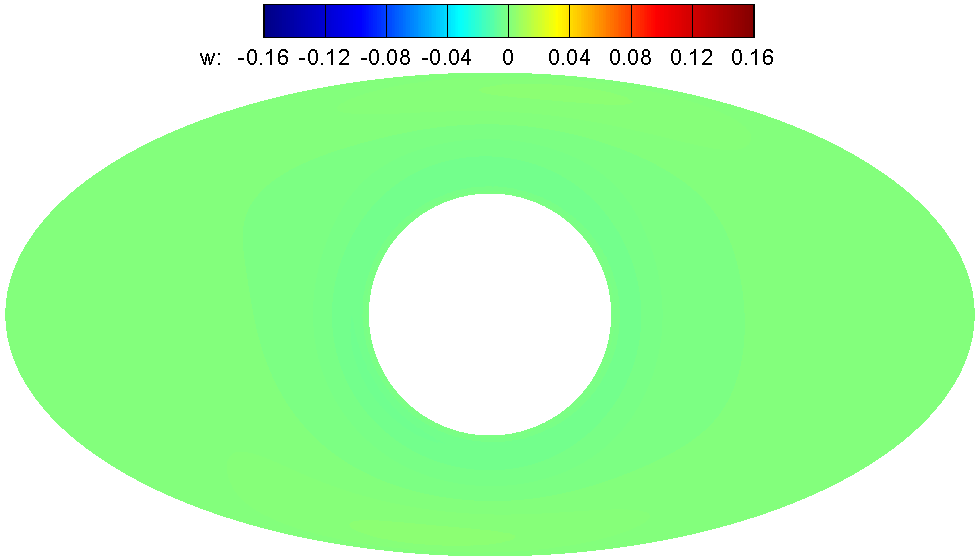}
        \caption{Re = 200}
        \label{fig:zvel_section_inbtw_b}
    \end{subfigure}
    \hfill
    \par\bigskip
    \begin{subfigure}{0.45\textwidth}
        \centering
        \includegraphics[width=\textwidth]{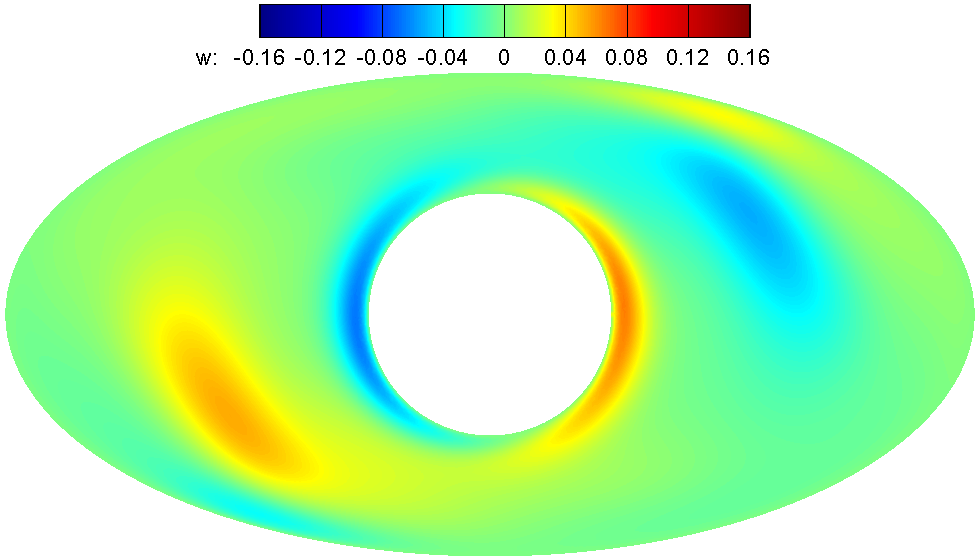}
        \caption{Re = 300}
        \label{fig:zvel_section_inbtw_c}
    \end{subfigure}
    \caption{Section between Taylor cells for supercritical stage}
    \label{fig:zvel_section_inbtw}
\end{figure}

\begin{figure}[H]
    \centering
    \begin{subfigure}{0.45\textwidth}
        \centering
        \includegraphics[width=\textwidth]{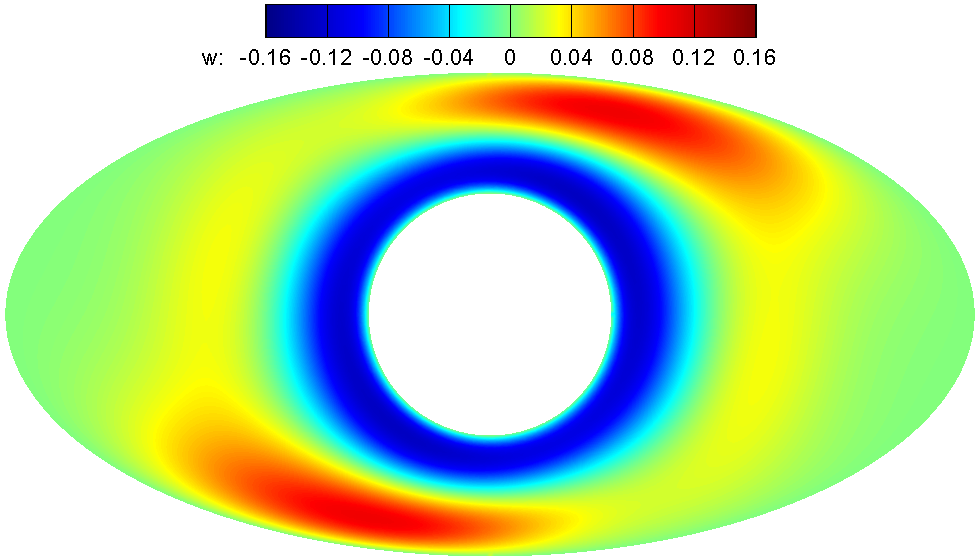}
        \caption{Re = 75}
        \label{fig:zvel_cr_section_a}
    \end{subfigure}
    \hfill
    \begin{subfigure}{0.45\textwidth}
        \centering
        \includegraphics[width=\textwidth]{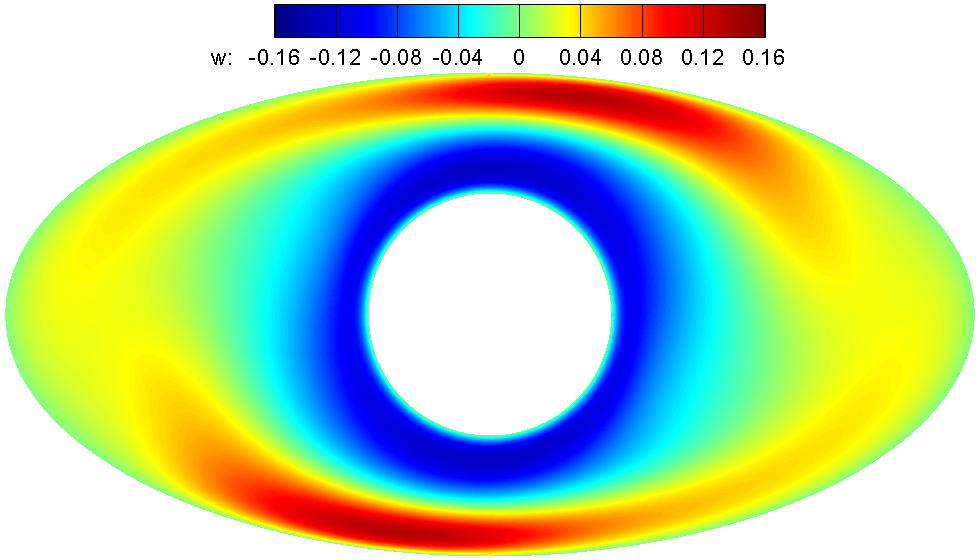}
        \caption{Re = 200}
        \label{fig:zvel_cr_section_b}
    \end{subfigure}
    \hfill
    \par\bigskip
    \begin{subfigure}{0.45\textwidth}
        \centering
        \includegraphics[width=\textwidth]{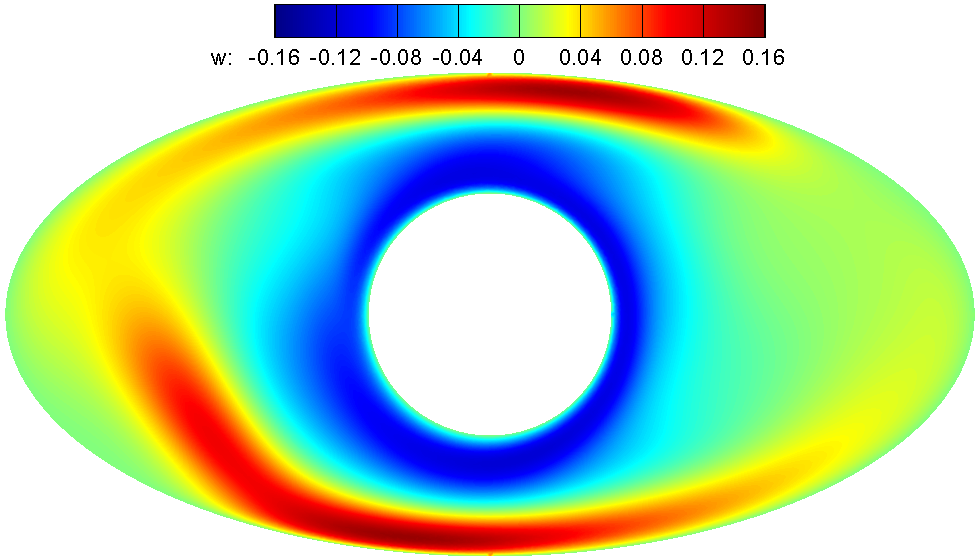}
        \caption{Re = 300}
        \label{fig:zvel_cr_section_c}
    \end{subfigure}
    \caption{Mid-section cutting through the center of Taylor cell rotating counter clockwise for supercritical stage}
    \label{fig:zvel_cr_section}
\end{figure}

\subsubsection{Isosurfaces of axial velocity}

\begin{figure}[H]
    \centering
    \begin{subfigure}{0.24\textwidth}
        \centering
        \includegraphics[height=50mm]{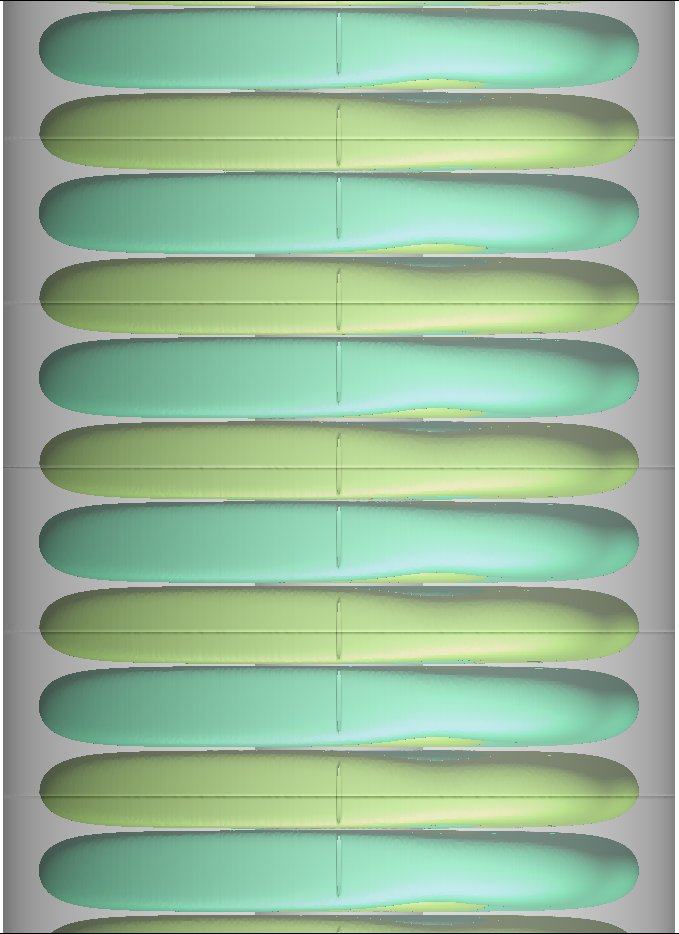}
        \caption{Re = 75}
        \label{}
    \end{subfigure}
    \hfill
    \begin{subfigure}{0.24\textwidth}
        \centering
        \includegraphics[height=50mm]{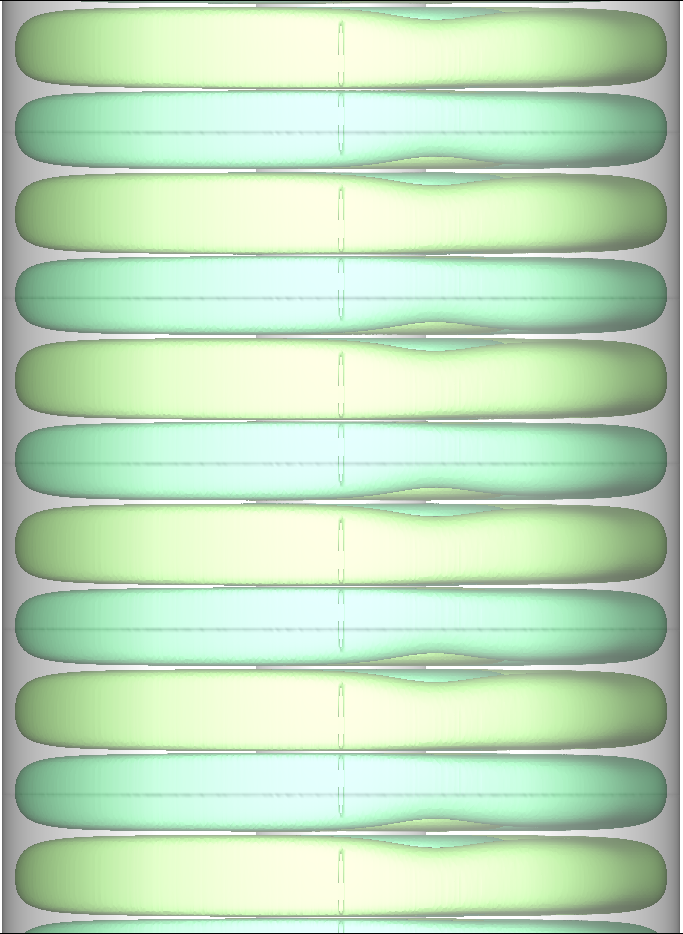}
        \caption{Re = 100}
        \label{}
    \end{subfigure}
    \hfill
    \begin{subfigure}{0.24\textwidth}
        \centering
        \includegraphics[height=50mm]{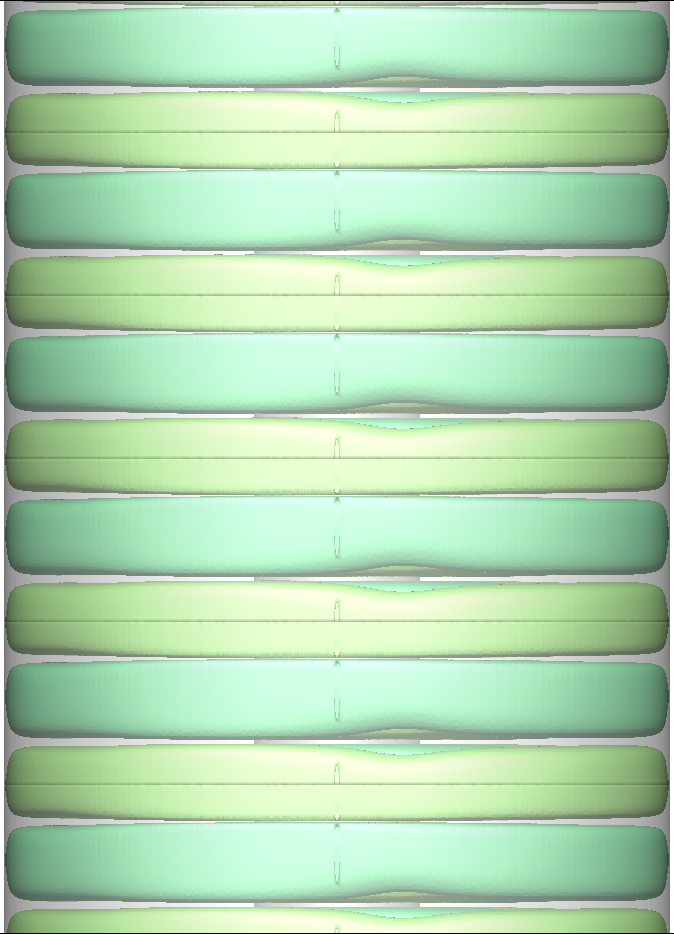}
        \caption{Re = 200}
        \label{}
    \end{subfigure}
    \hfill
    \begin{subfigure}{0.24\textwidth}
        \centering
        \includegraphics[height=50mm]{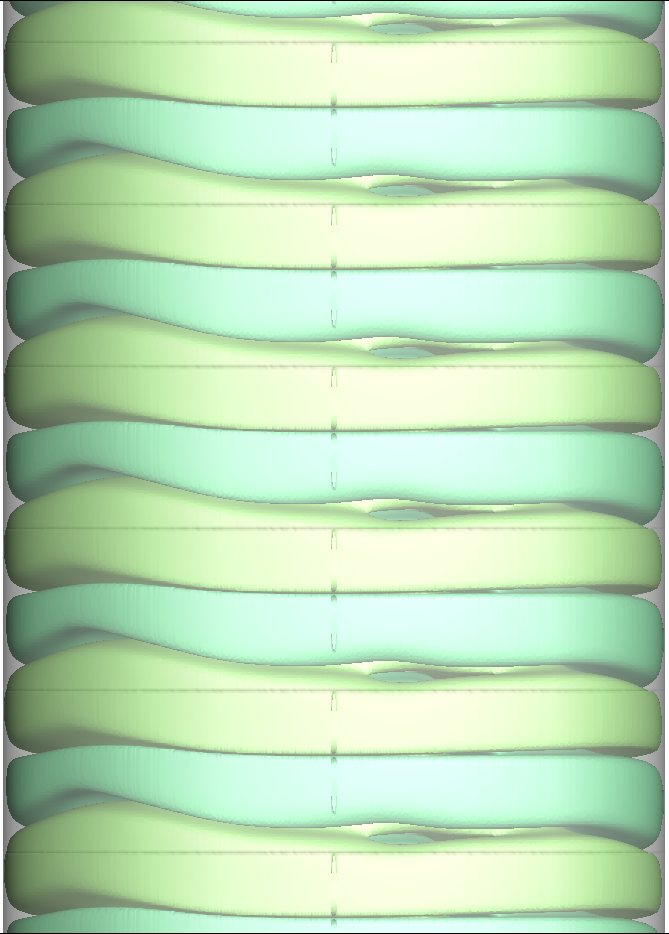}
        \caption{Re = 300}
        \label{}
    \end{subfigure}
    \caption{Isosurfaces of axial velocity for $w = \pm 0.005$ viewed from $x-z$ plane (view on the major axis section of ellipse)}
    \label{fig:iso_zvel_major_axis}
\end{figure}

\begin{figure}[H]
    \centering
    \begin{subfigure}{0.2\textwidth}
        \centering
        \includegraphics[height=50mm]{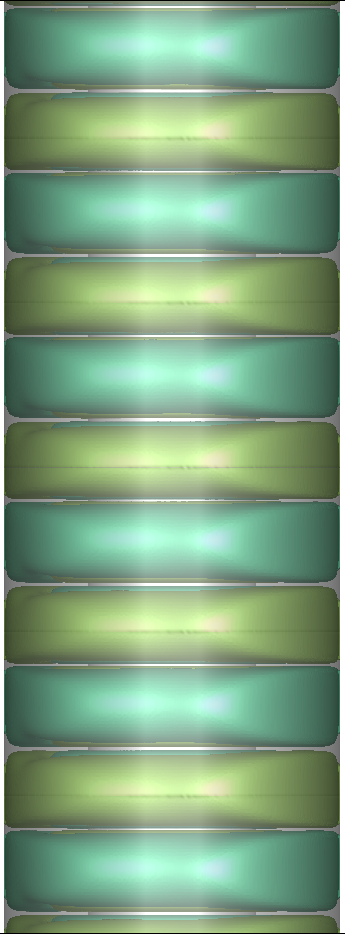}
        \caption{Re = 75}
        \label{}
    \end{subfigure}
    \begin{subfigure}{0.2\textwidth}
        \centering
        \includegraphics[height=50mm]{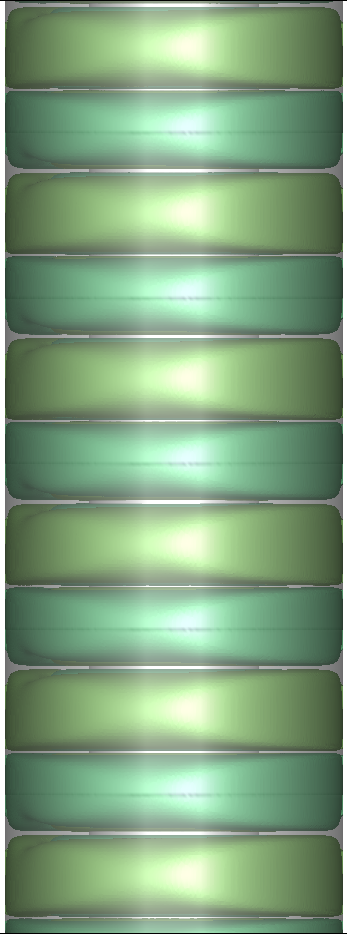}
        \caption{Re = 100}
        \label{}
    \end{subfigure}
    \begin{subfigure}{0.2\textwidth}
        \centering
        \includegraphics[height=50mm]{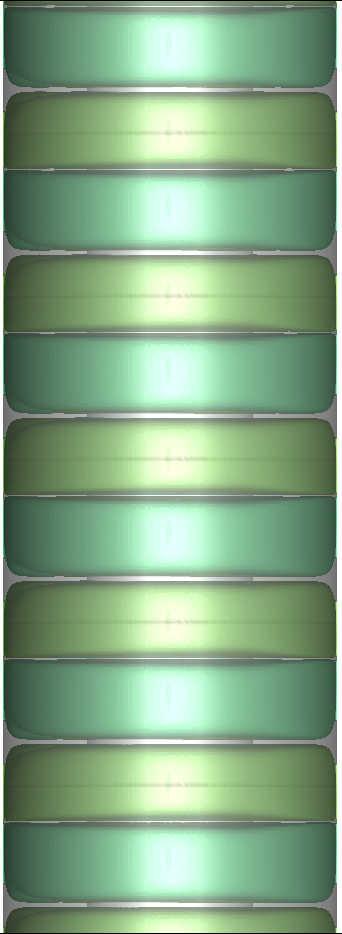}
        \caption{Re = 200}
        \label{}
    \end{subfigure}
    \begin{subfigure}{0.2\textwidth}
        \centering
        \includegraphics[height=50mm]{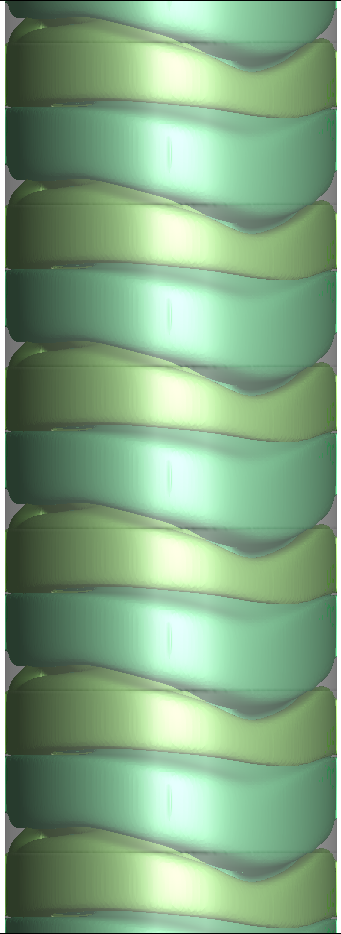}
        \caption{Re = 300}
        \label{fig:iso_zvel_minor_axis_d}
    \end{subfigure}
    \caption{Isosurfaces of axial velocity for $w = \pm 0.005$ viewed from $y-z$ plane (view on the minor axis section of ellipse)}
    \label{fig:iso_zvel_minor_axis}
\end{figure}

\Cref{fig:iso_zvel_major_axis,fig:iso_zvel_minor_axis} present the isosurfaces of axial velocity for different supercritical Reynolds number. The tubular structures are isosurfaces of axial velocity for two different velocities alternately being $+0.005$ and $-0.005$ respectively. Two of them together constitutes on Taylor cell. The Taylor cells appear to be shifting to a wavy structure as Reynolds number changes from 200 to 300. This is more pronounced in the view on minor axis section as in \cref{fig:iso_zvel_minor_axis}. There is a small pinching of the tubular structures seen in \cref{fig:iso_zvel_major_axis}, as they cross the mid section. This is the location where the fluid exits the small gap area where the pressure is higher. \citet{fenstermacher_swinney_gollub_1979} has observed higher order instabilities as the oscillations and twisting of Taylor vortex cells in simple Taylor Couette flow before they transition to turbulence. The first transition from Taylor vortex is to the wavy vortex flow, characterised by waviness in azimuthal direction. We observe twist and oscillations, characteristics of wavy vortex flow, in the elliptical case as can be observed in \cref{fig:iso_zvel_minor_axis_d} for a Reynolds number of 300. From here the flow becomes unsteady as the Reynolds number reaches 350. Turbulent cases are not simulated in this study. 

\subsection{Pressure distributions}
\subsubsection{Contours of pressure in r-z planes}

\begin{figure}[H]
     \centering
     \begin{subfigure}[b]{0.49\textwidth}
         \centering
         \includegraphics[height=50mm]{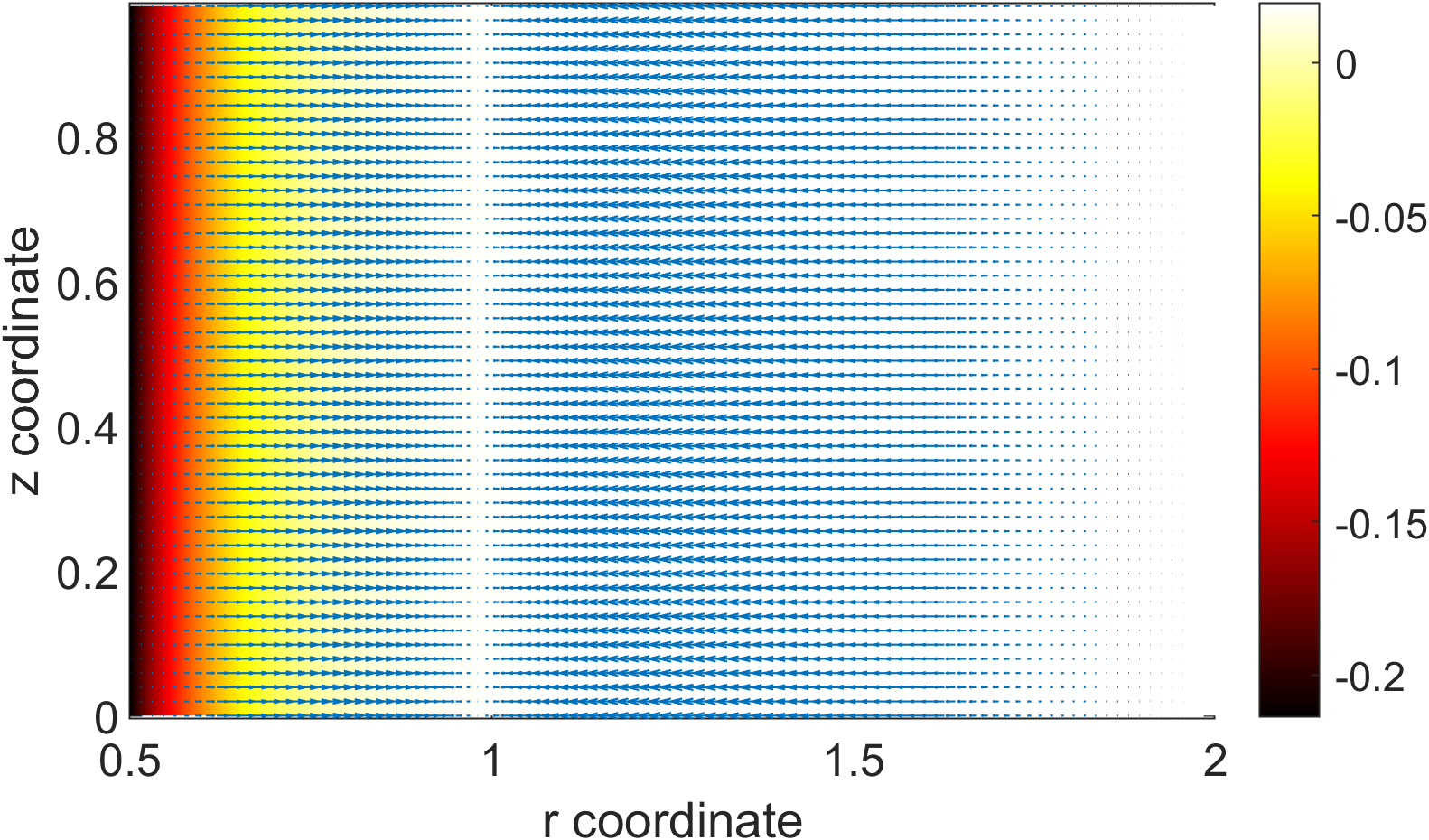}
         \caption{Re = 70}
         \label{}
     \end{subfigure}
     \hfill
     \begin{subfigure}[b]{0.49\textwidth}
         \centering
         \includegraphics[height=50mm]{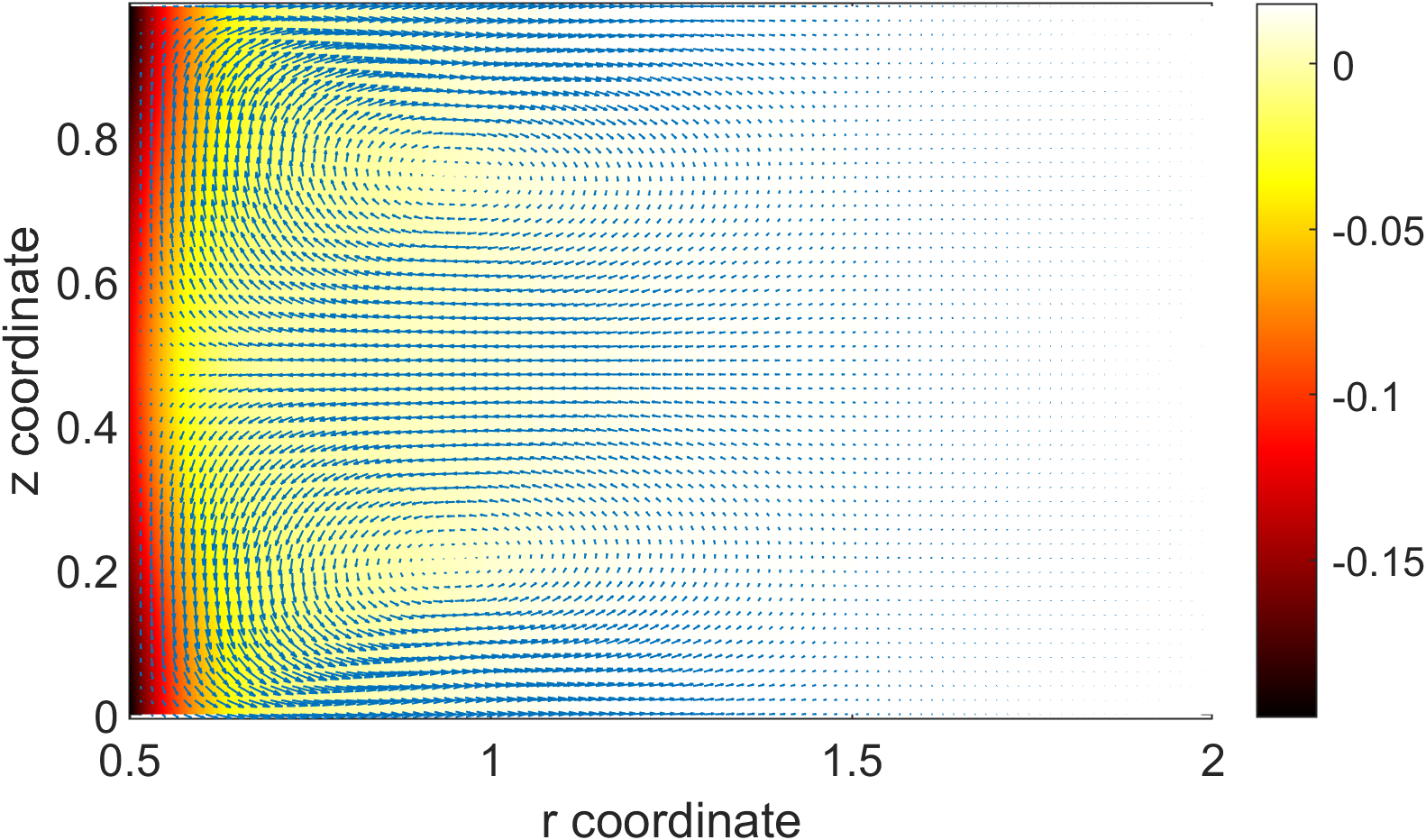}
         \caption{Re = 75}
         \label{}
     \end{subfigure}
     \par\bigskip
     \begin{subfigure}[b]{0.49\textwidth}
         \centering
         \includegraphics[height=50mm]{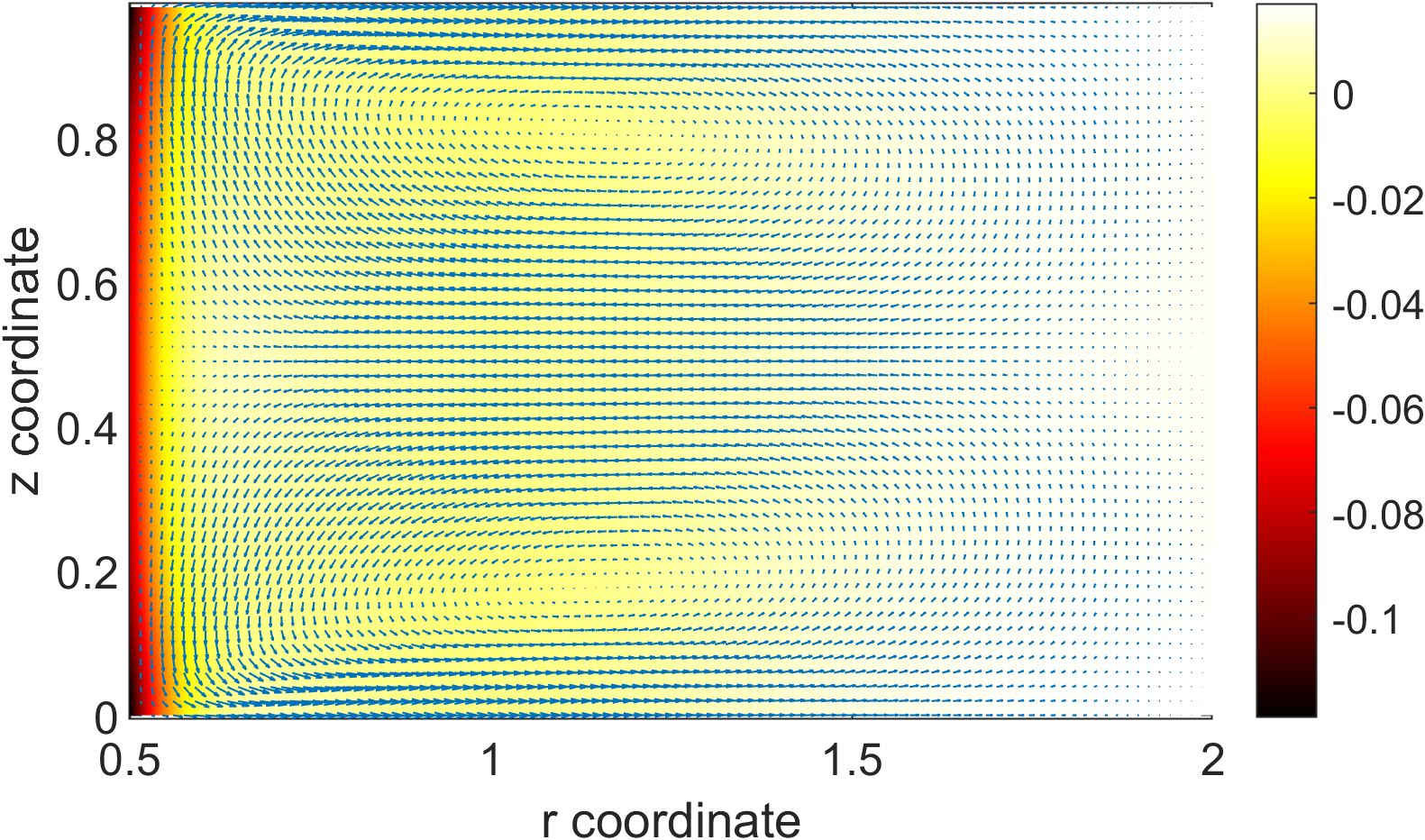}
         \caption{Re = 200}
         \label{}
     \end{subfigure}
     \hfill
     \begin{subfigure}[b]{0.49\textwidth}
         \centering
         \includegraphics[height=50mm]{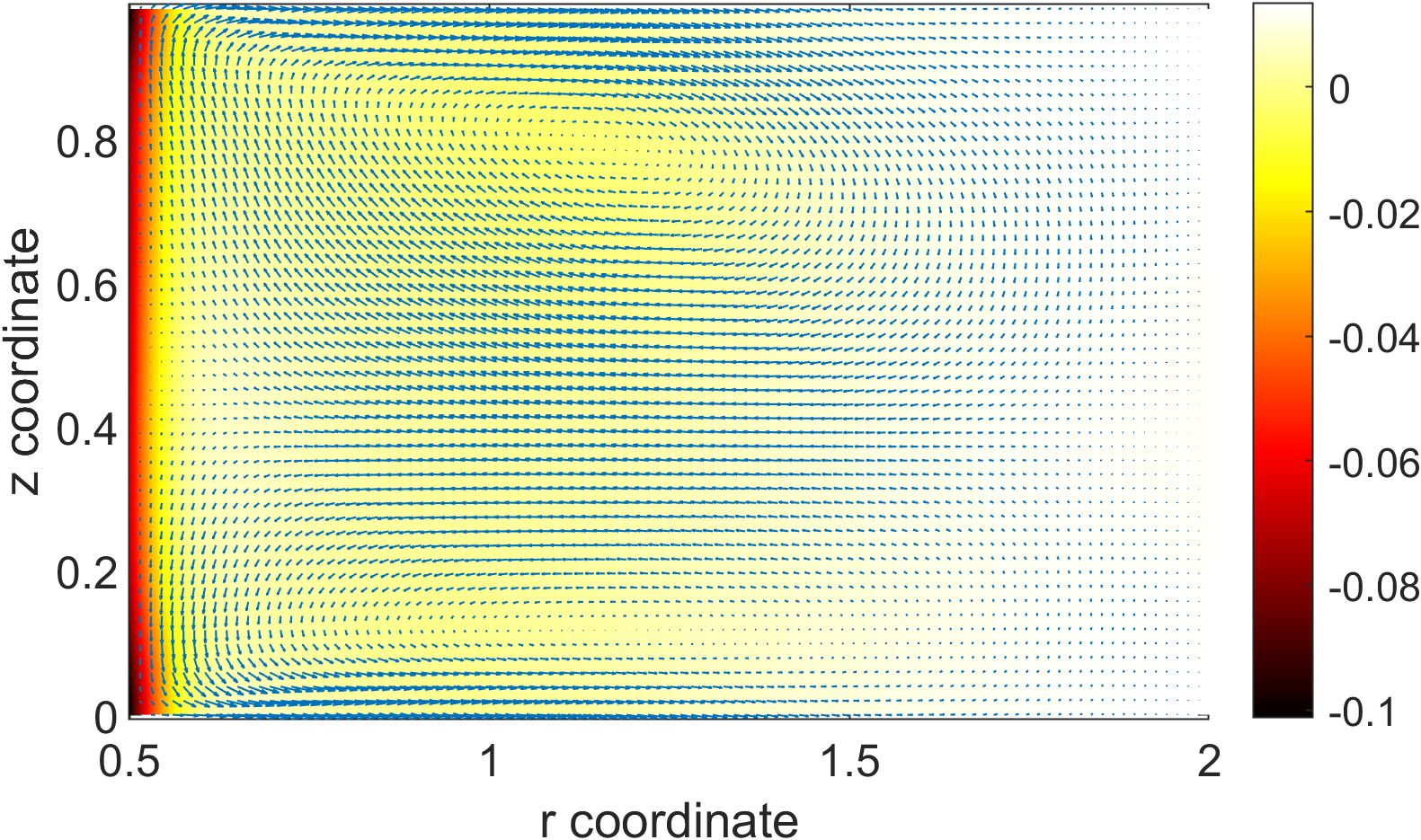}
         \caption{Re = 300}
         \label{}
     \end{subfigure}
     \caption{Pressure plots in the plane along semi major axis of ellipse}
        \label{fig:pressure_plots_0deg}
\end{figure}

\begin{figure}[H]
     \centering
     \begin{subfigure}[b]{0.24\textwidth}
         \centering
         \includegraphics[width = 0.9\textwidth]{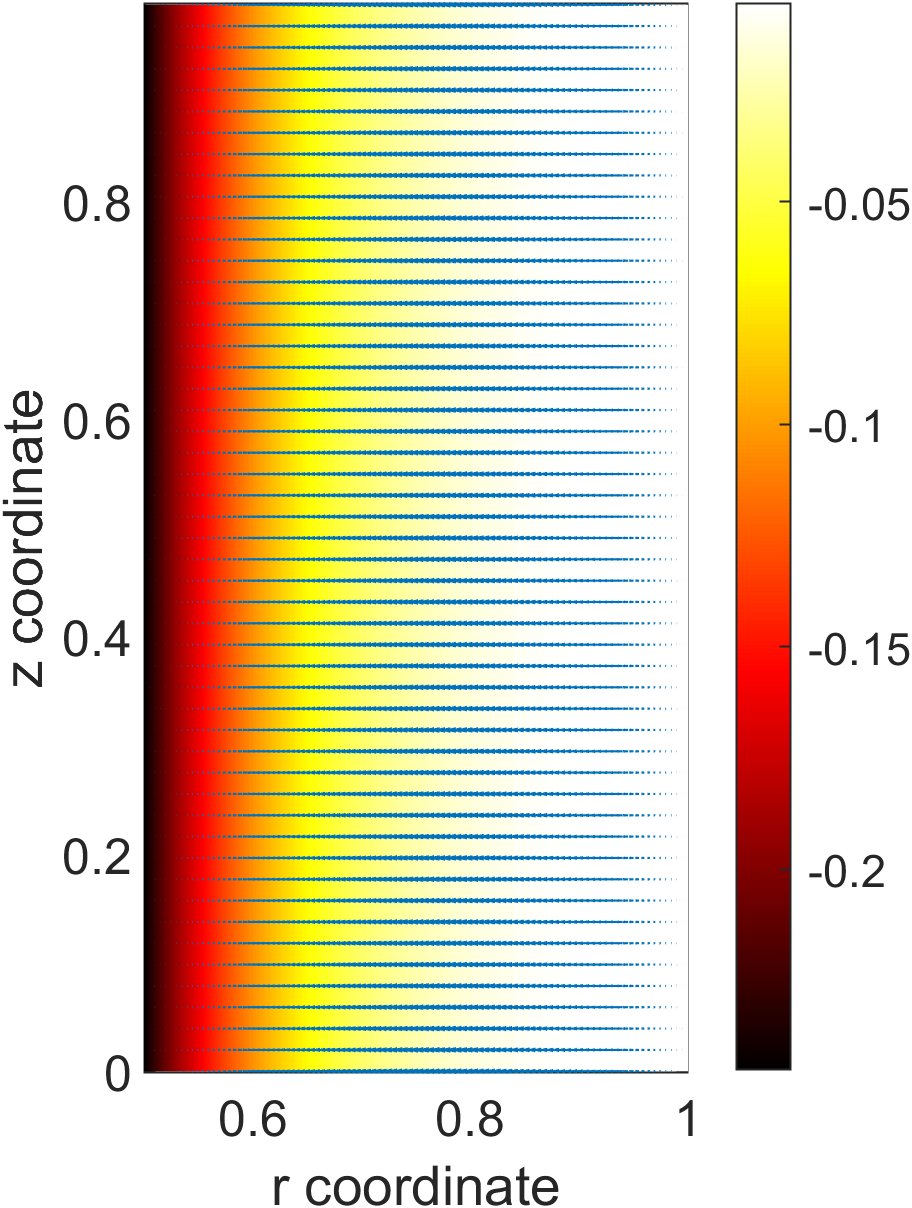}
         \caption{Re = 70}
         \label{}
     \end{subfigure}
     \hfill
     \begin{subfigure}[b]{0.24\textwidth}
         \centering
         \includegraphics[width =0.9\textwidth]{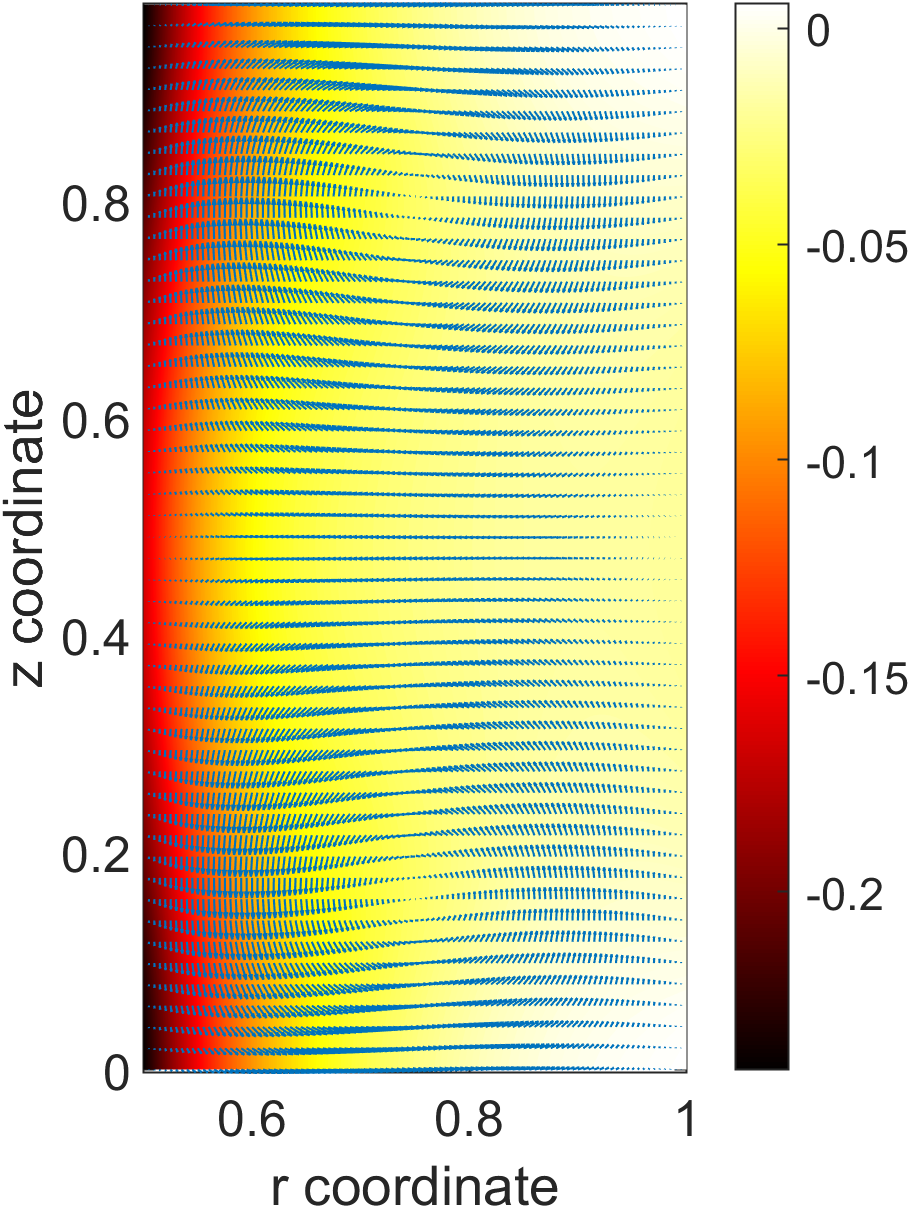}
         \caption{Re = 75}
         \label{}
     \end{subfigure}
     \hfill
     \begin{subfigure}[b]{0.24\textwidth}
         \centering
         \includegraphics[width = 0.9\textwidth]{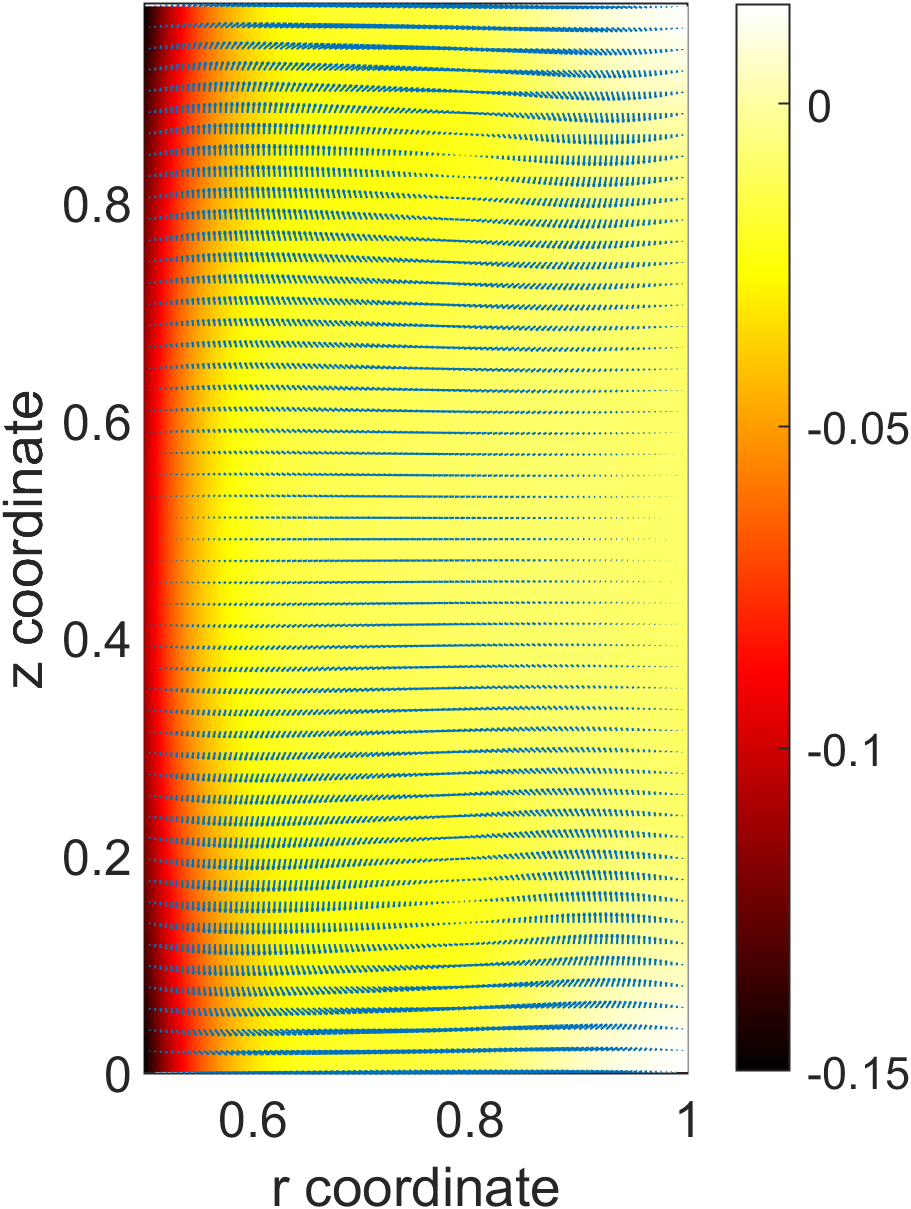}
         \caption{Re = 200}
         \label{}
     \end{subfigure}
     \hfill
     \begin{subfigure}[b]{0.24\textwidth}
         \centering
         \includegraphics[width = 0.9\textwidth]{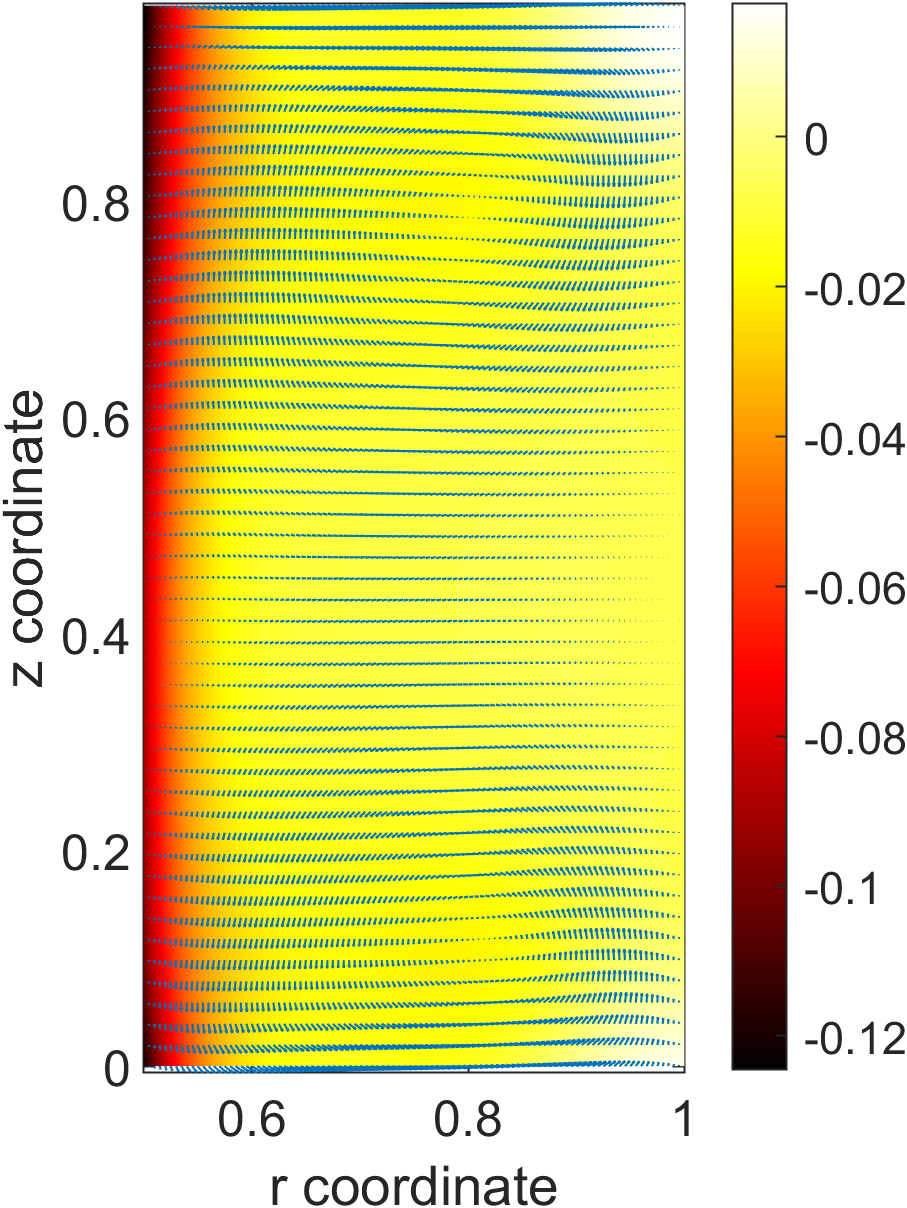}
         \caption{Re = 300}
         \label{}
     \end{subfigure}
     \caption{Pressure plots in the plane along semi minor axis of ellipse}
        \label{fig:pressure_plots_90deg}
\end{figure}

At subcritical Reynolds numbers, the pressure distribution is primarily due to the rotational flow. For the concentric Couette flow at subcritical Reynolds number, the distribution is axisymmetric and one-dimensional. However, for the case of the elliptical outer cylinder, the pressure distribution is non-axisymmetric and two-dimensional.  As the Reynolds number increases to the supercritical value, a three-dimensional pressure distribution consistent with the velocity distribution and the streamlines evolves. \Cref{fig:pressure_plots_0deg,fig:pressure_plots_90deg} show the pressure distributions in the ($r-z$) plane along the semi-major and semi-minor axes, along with the velocity vectors. At subcritical $Re = 70$, the pressure varies only with radius as there are no Taylor cells. When the Taylor cells form at $Re=75$ and beyond, the pressure on the inner cylinder distorts to increase at the dividing point between the lower and upper vortices. The higher pressure moves the downward and upward moving flows radially towards the inner cylinder. As shown for the axial-velocity contours, the pressure contours (and velocity vectors) become distorted at $Re = 300$. Until such a $Re$, the flow in the semi-major ($r-z$) plane is very much well-structured. The pressure distributions shown in the semi-minor ($r-z$) planes are comparatively well structured even at $Re = 300$. The higher pressure occurs at $z$ around $0.5$ which is the junction of the upward and downward flow.

\subsubsection{Isosurfaces of pressure}

\begin{figure}[H]
    \centering
    \begin{subfigure}{0.24\textwidth}
        \centering
        \includegraphics[height=50mm]{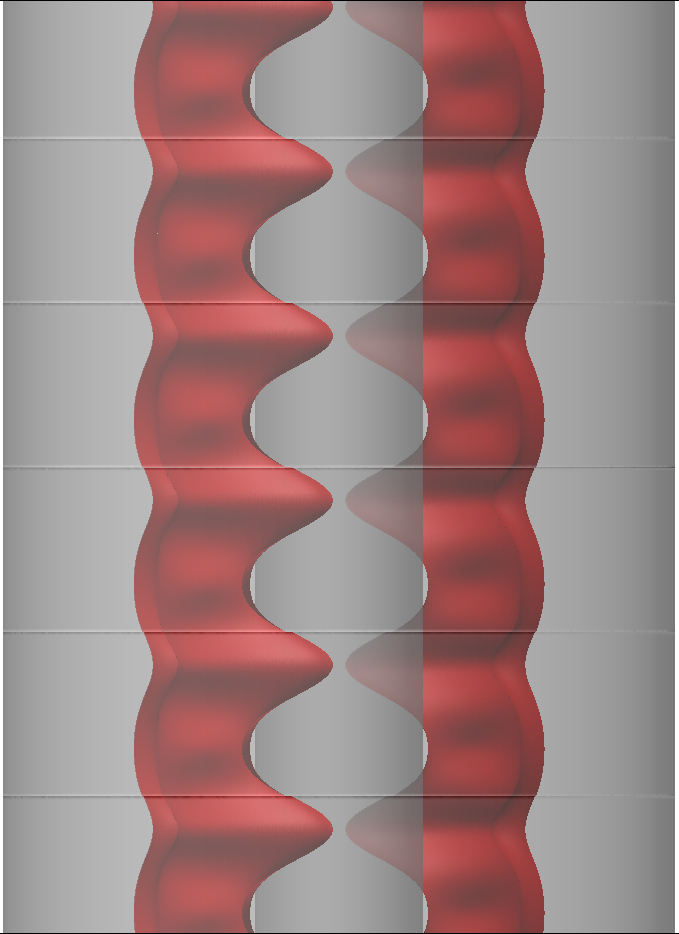}
        \caption{Re = 75}
        \label{}
    \end{subfigure}
    \hfill
    \begin{subfigure}{0.24\textwidth}
        \centering
        \includegraphics[height=50mm]{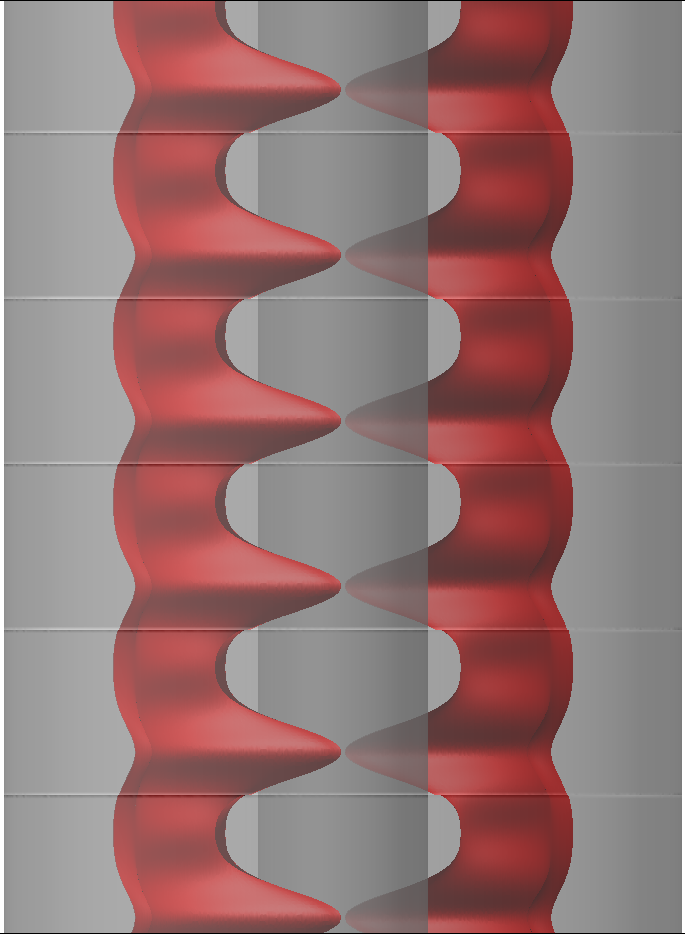}
        \caption{Re = 100}
        \label{}
    \end{subfigure}
    \hfill
    \begin{subfigure}{0.24\textwidth}
        \centering
        \includegraphics[height=50mm]{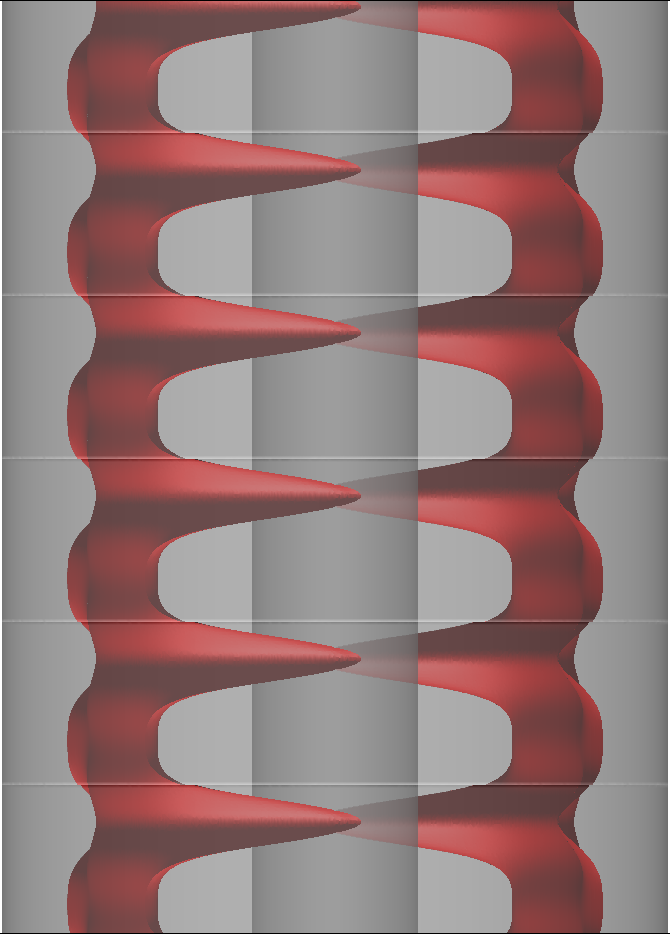}
        \caption{Re = 200}
        \label{}
    \end{subfigure}
    \hfill
    \begin{subfigure}{0.24\textwidth}
        \centering
        \includegraphics[height=50mm]{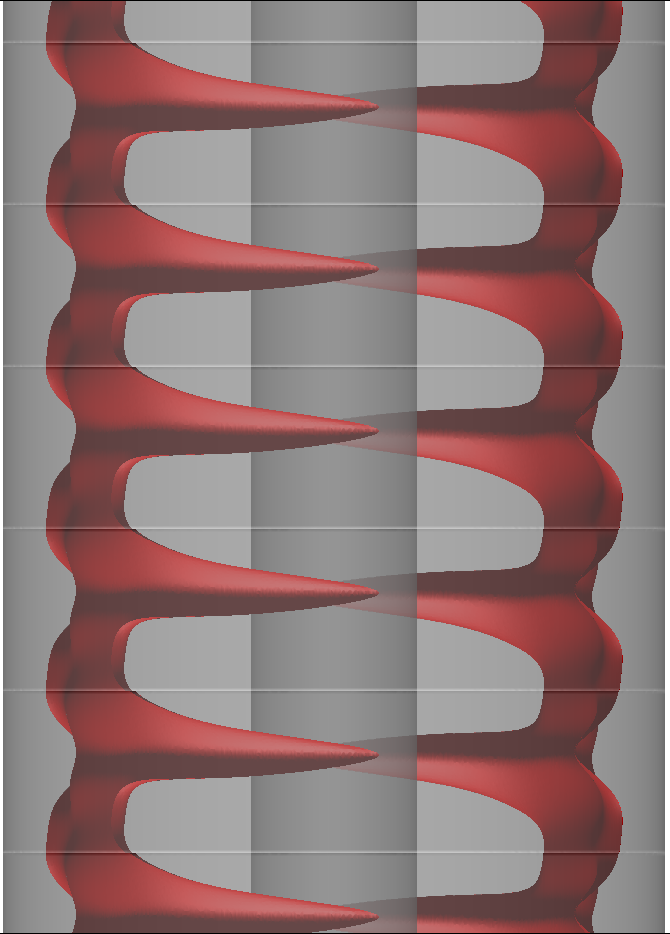}
        \caption{Re = 300}
        \label{}
    \end{subfigure}
    \caption{Isosurfaces of pressure for $p = 0.01$, viewed from $x-z$ plane (view on the major axis section of ellipse)}
    \label{fig:iso_p_major_axis}
\end{figure}

\begin{figure}[H]
    \centering
    \begin{subfigure}{0.2\textwidth}
        \centering
        \includegraphics[height=50mm]{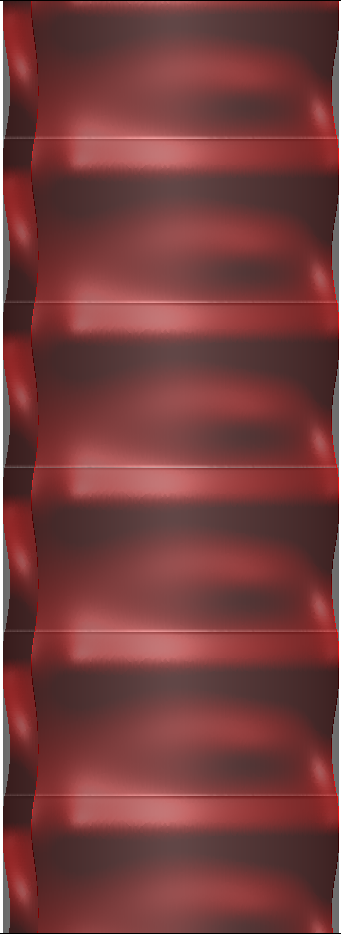}
        \caption{Re = 75}
        \label{}
    \end{subfigure}
    \begin{subfigure}{0.2\textwidth}
        \centering
        \includegraphics[height=50mm]{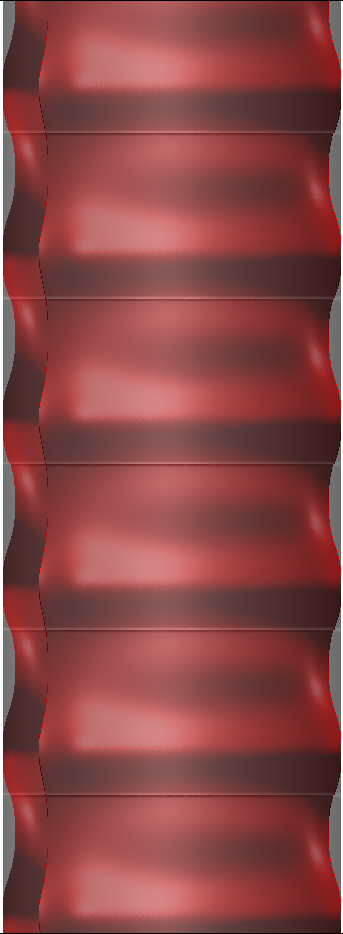}
        \caption{Re = 100}
        \label{}
    \end{subfigure}
    \begin{subfigure}{0.2\textwidth}
        \centering
        \includegraphics[height=50mm]{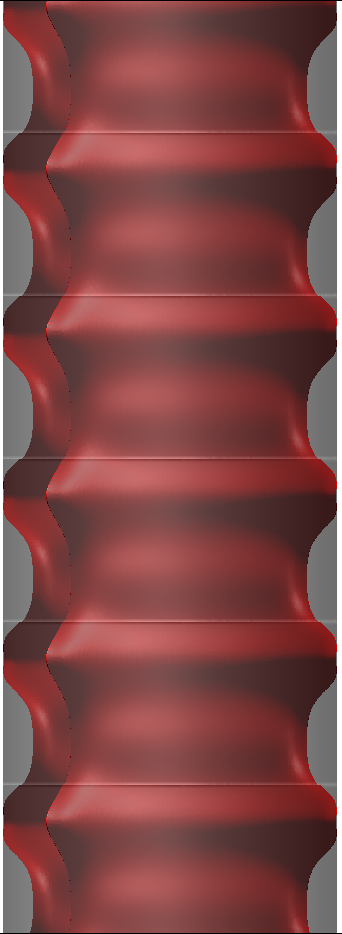}
        \caption{Re = 200}
        \label{}
    \end{subfigure}
    \begin{subfigure}{0.2\textwidth}
        \centering
        \includegraphics[height=50mm]{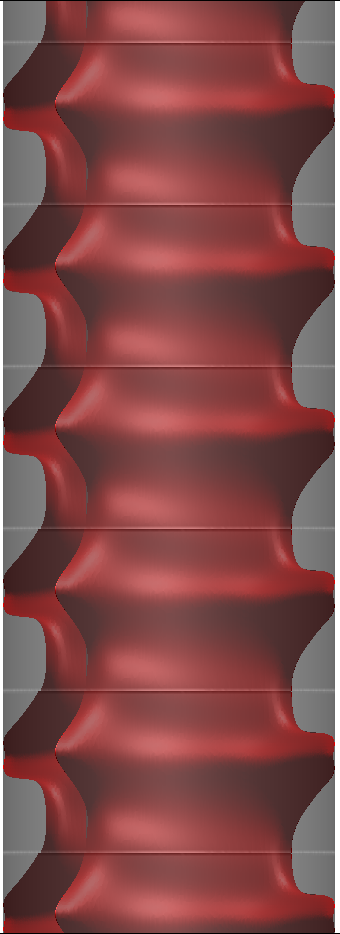}
        \caption{Re = 300}
        \label{}
    \end{subfigure}
    \caption{Isosurfaces of pressure for $p = 0.01$, viewed from $y-z$ plane (view on the minor axis section of ellipse)}
    \label{fig:iso_p_minor_axis}
\end{figure}

\Cref{fig:iso_p_major_axis,fig:iso_p_minor_axis} show the isosurfaces of the pressure in two views for pressure magnitude of $0.01$. The surfaces of the pressure show the distortions due to the Taylor cells. The bellow like shape of the isosurface indicate that there exists an alternating high pressure and low pressure regions along the axis. High pressure region are observed at the sections in between the Taylor cells and low pressure at the midsection of the Taylor cells. The formation of finger-like structures in the isosurfaces of pressure in \cref{fig:iso_p_major_axis} can be seen to form as the flow approaches the minor axis of ellipse. This region being the shortest gap acts as a venturi to increase the speed of flow in between and as a result in the decrease of pressure. The tips of these finger-like structures being at an axial section between the two counter rotating Taylor cells, the pressure is observed to reduce below a value of 0.01.
At $Re =300$ a slight deformation in the structure is observed, where the surface seems to be slightly inclined to the axis indicating formation of a spiral flow. 

\subsection{Vorticity distributions}
Vorticity, {\boldmath$\eta$}, is defined as the \textit{curl} of velocity vector ($\nabla \times \textbf{u}$) with components in $x, y,$ and $z$ directions. It measures the amount of rotation of a fluid element along the local axis. The magnitude of vorticity vector measures the strength of the local rotation of the fluid. In the following subsections, contours of vorticity magnitude in different planes and the isosurfaces of vorticity magnitude are presented.

\subsubsection{Contours of vorticity magnitude in r-z planes}

\begin{figure}[H]
     \centering
     \begin{subfigure}[b]{0.45\textwidth}
         \centering
         \includegraphics[height=50mm]{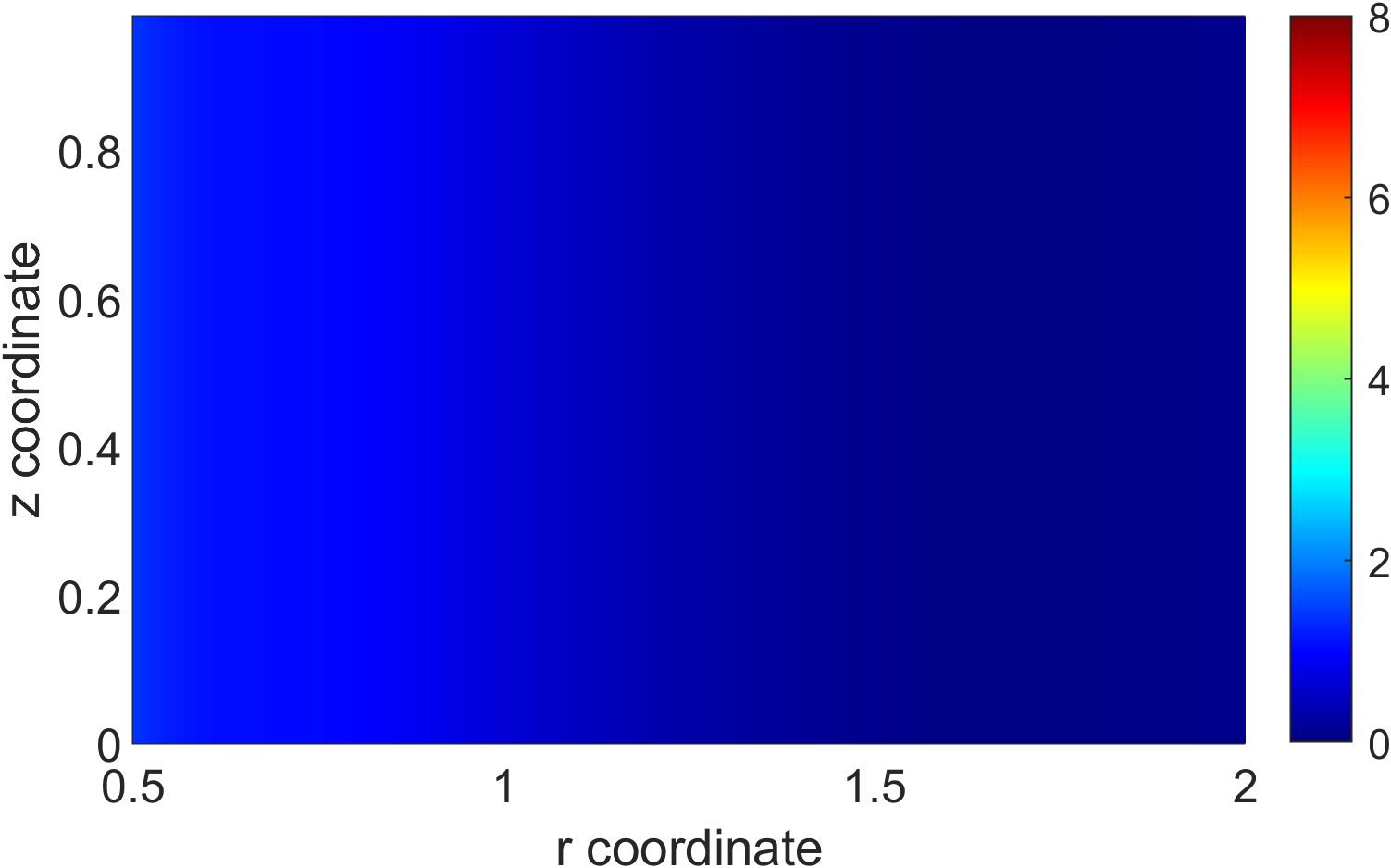}
         \caption{Re = 70}
         \label{}
     \end{subfigure}
     \hfill
     \begin{subfigure}[b]{0.45\textwidth}
         \centering
         \includegraphics[height=50mm]{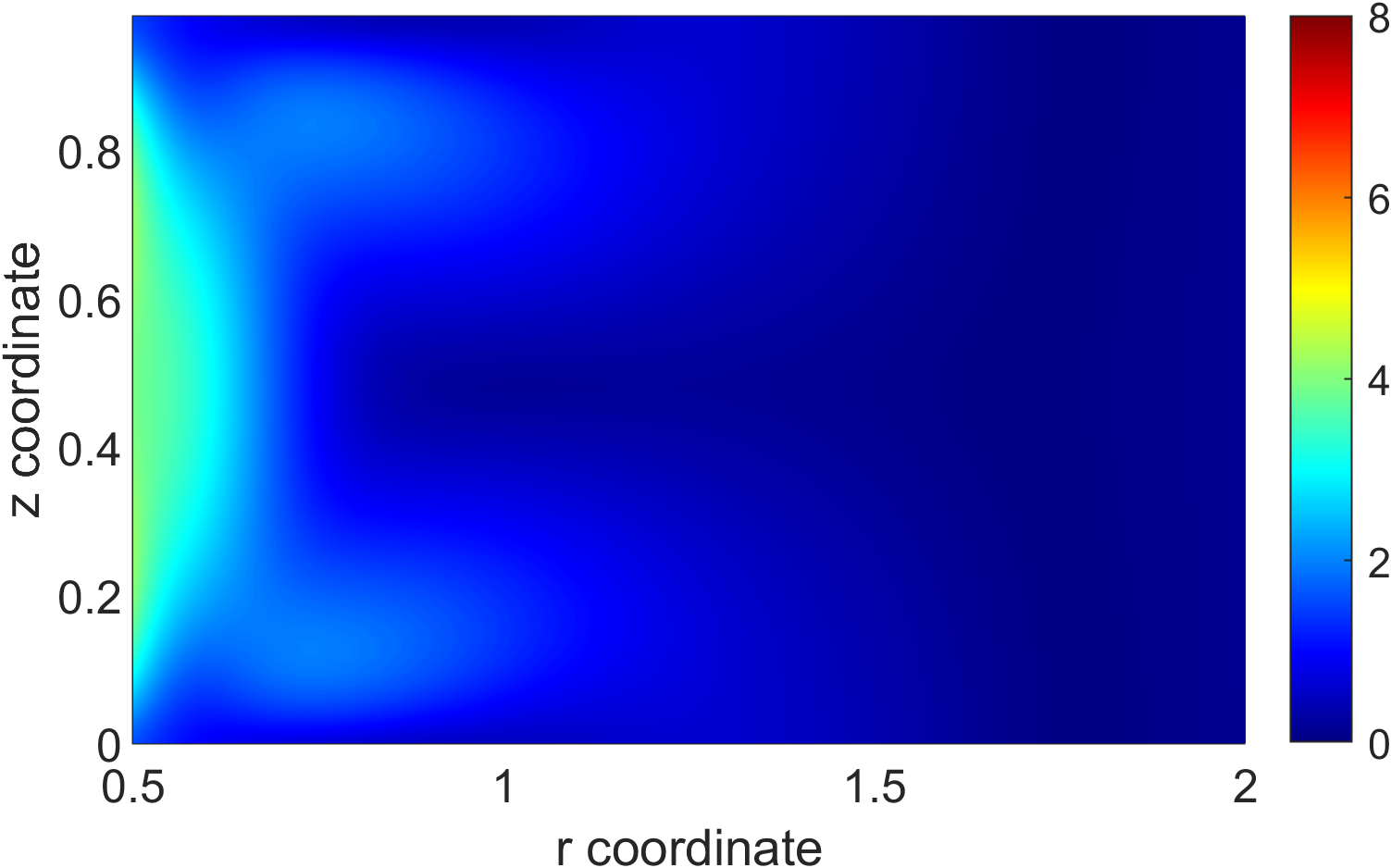}
         \caption{Re = 75}
         \label{}
     \end{subfigure}
     \hfill
     \par \bigskip
     \begin{subfigure}[b]{0.45\textwidth}
         \centering
         \includegraphics[height=50mm]{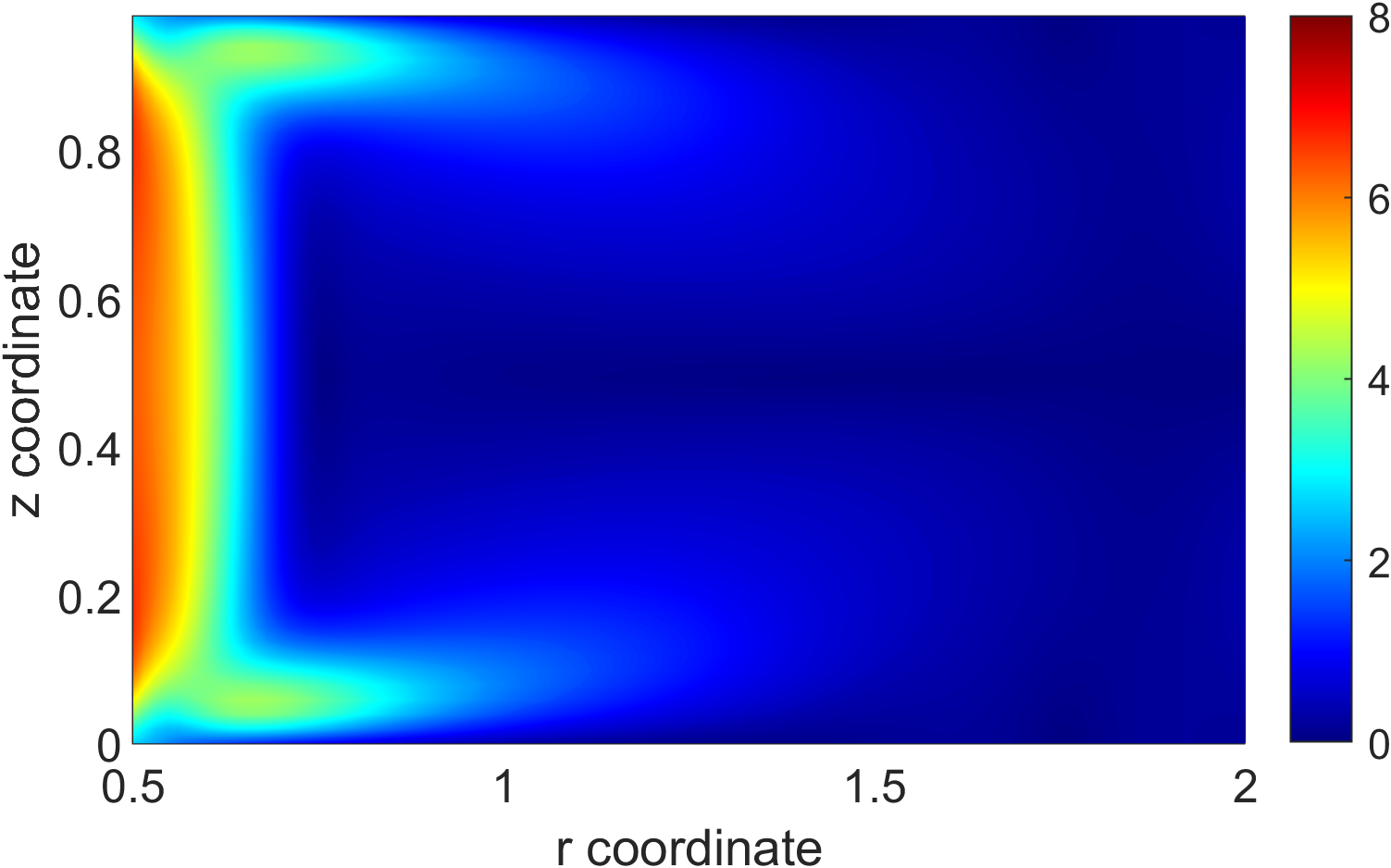}
         \caption{Re = 200}
         \label{}
     \end{subfigure}
     \hfill
     \begin{subfigure}[b]{0.45\textwidth}
         \centering
         \includegraphics[height=50mm]{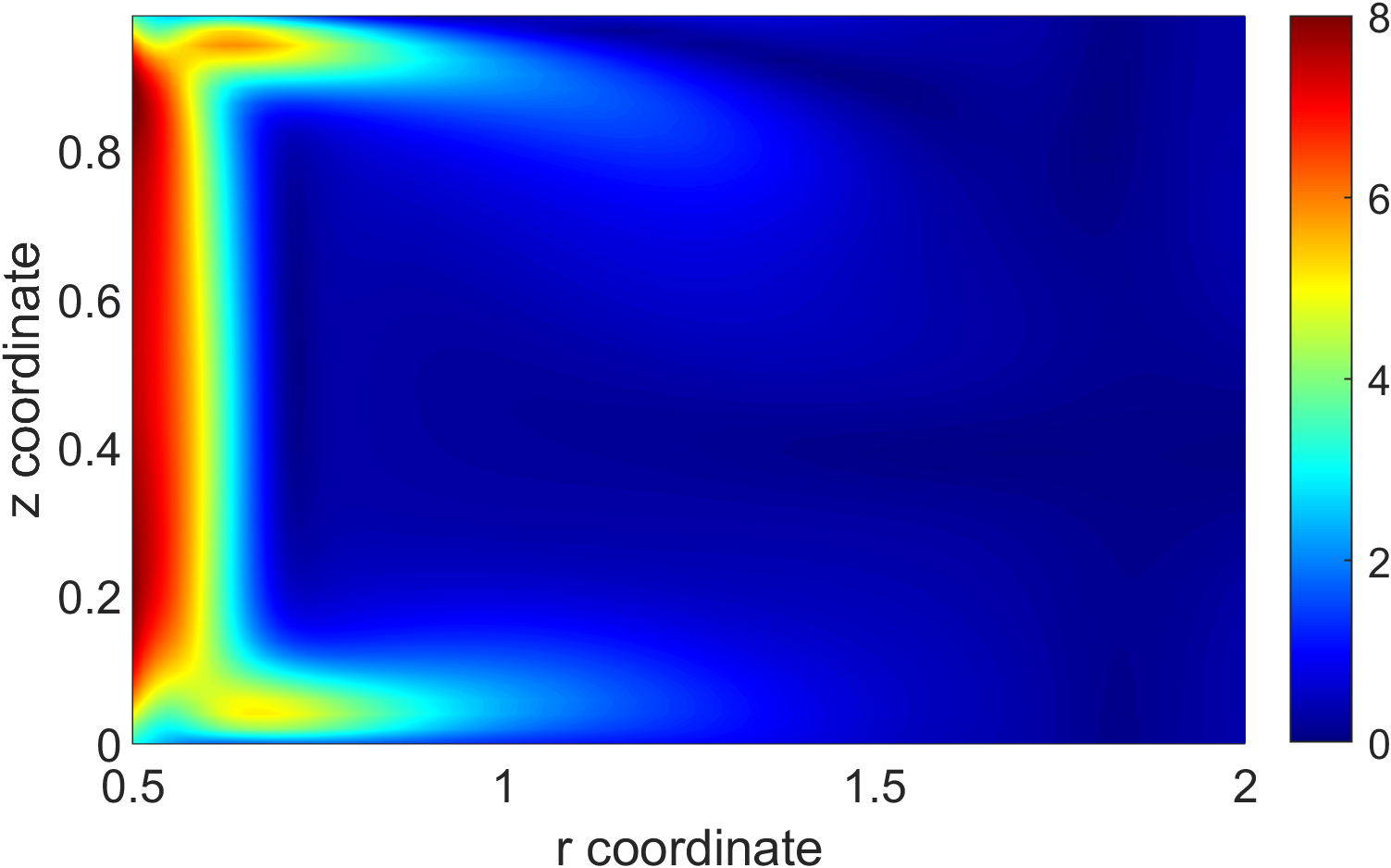}
         \caption{Re = 300}
         \label{}
     \end{subfigure}
     \caption{Contours of vorticity magnitude in the plane along semi major axis of ellipse}
        \label{fig:eta_rz_section_deg0}
\end{figure}

\begin{figure}[H]
     \centering
     \begin{subfigure}[b]{0.19\textwidth}
         \centering
         \includegraphics[height=50mm]{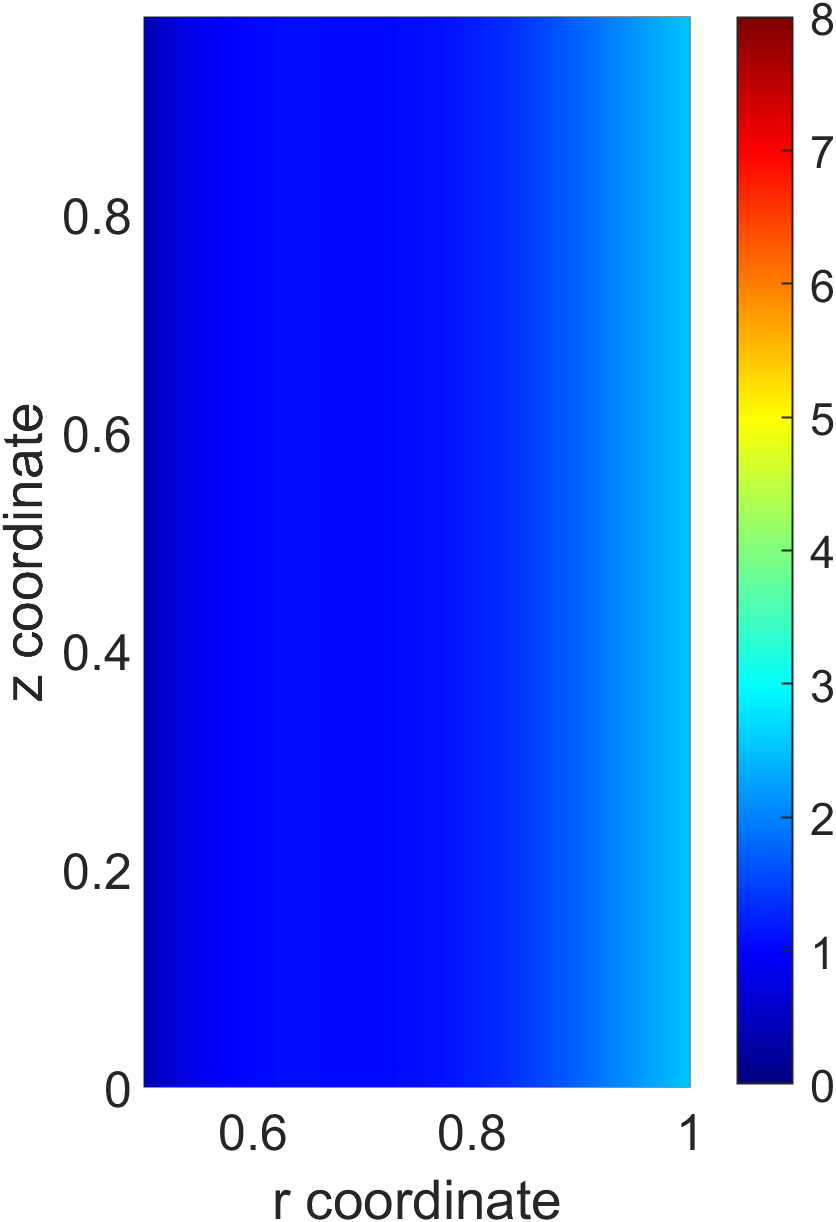}
         \caption{Re = 70}
         \label{}
     \end{subfigure}
     \hfill
     \begin{subfigure}[b]{0.19\textwidth}  
         \centering
         \includegraphics[height=50mm]{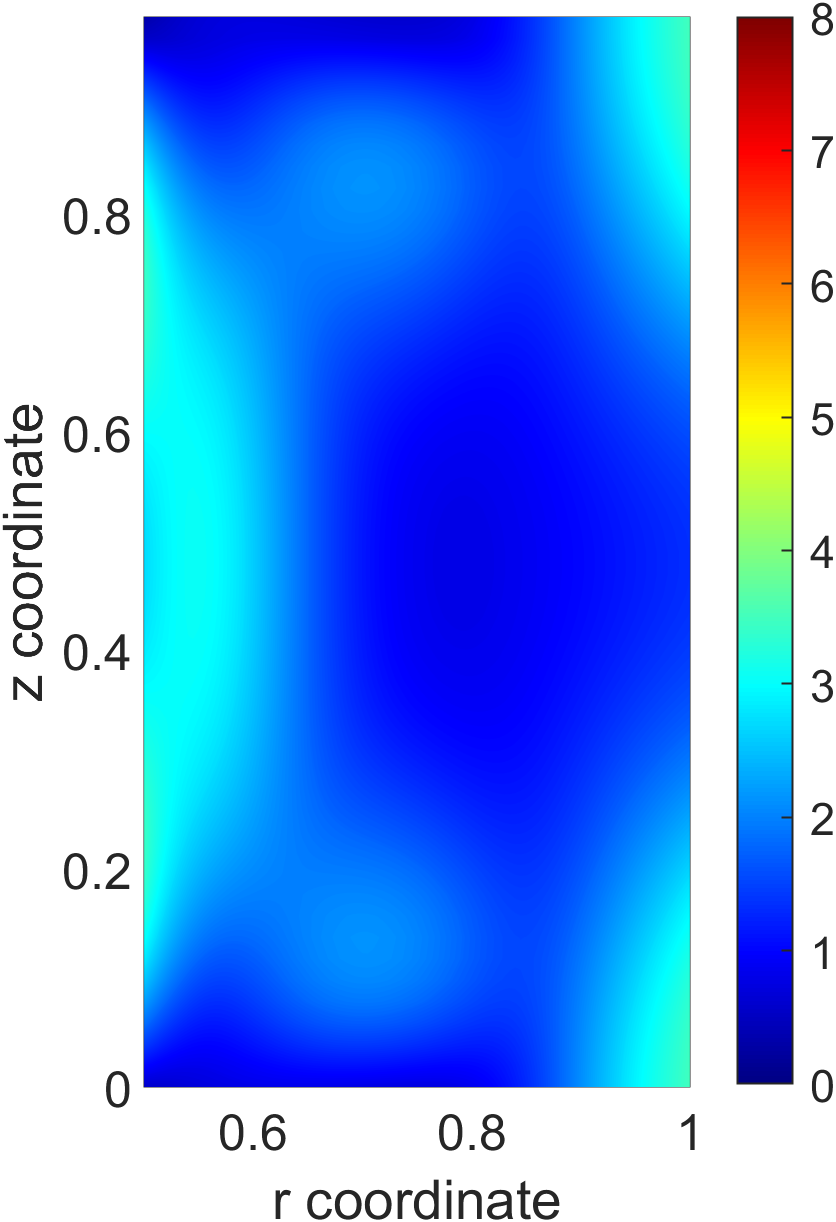}
         \caption{Re = 75}
         \label{}
     \end{subfigure}
     \hfill
     \begin{subfigure}[b]{0.19\textwidth}
         \centering
         \includegraphics[height=50mm]{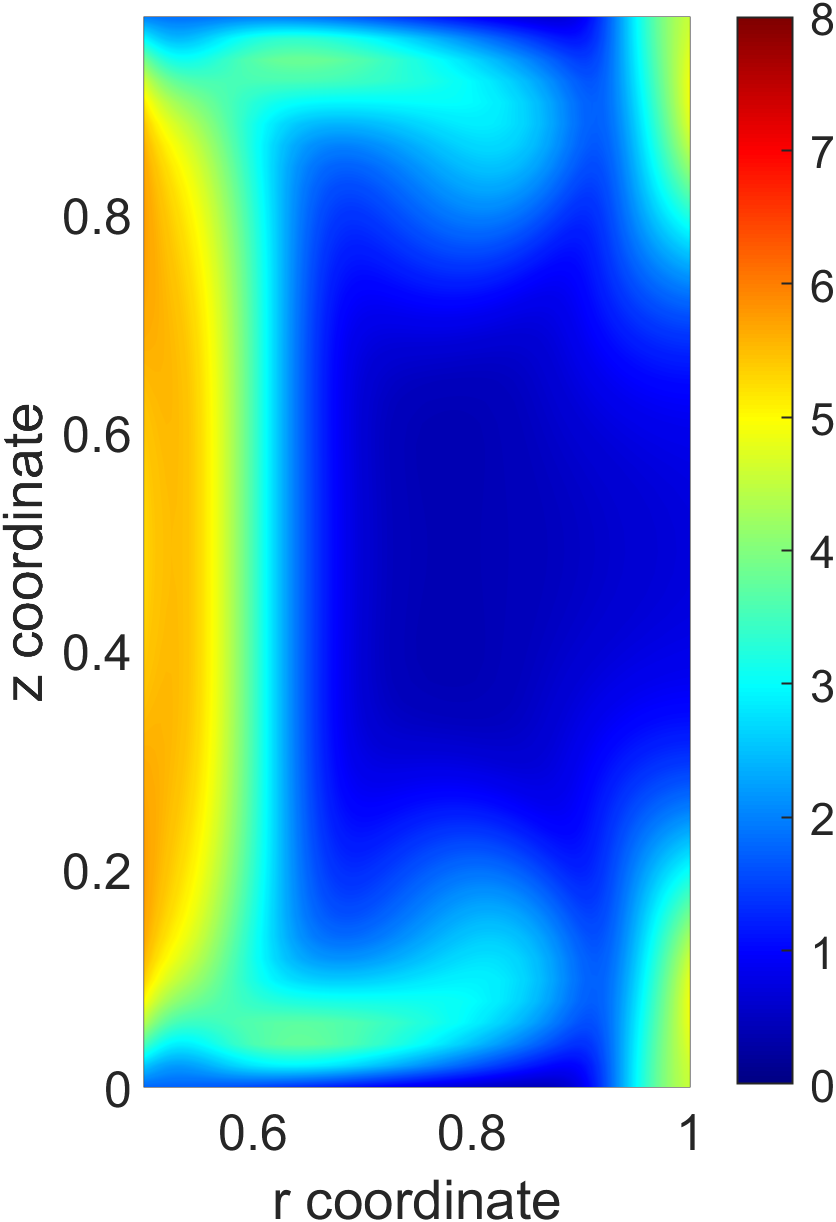}
         \caption{Re = 200}
         \label{}
     \end{subfigure}
     \hfill
     \begin{subfigure}[b]{0.19\textwidth}
         \centering
         \includegraphics[height=50mm]{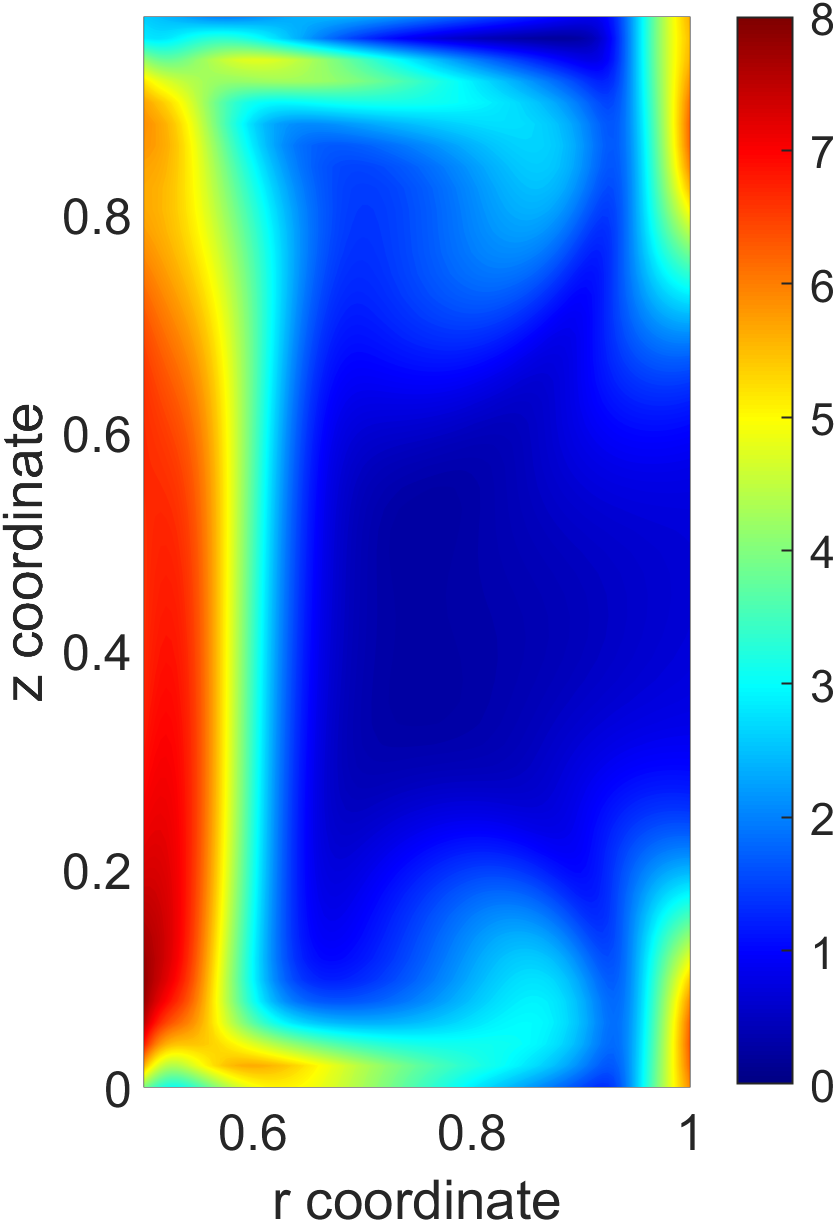}
         \caption{Re = 300}
         \label{}
     \end{subfigure}
     \caption{Contours of vorticity magnitude in the plane along semi minor axis of ellipse}
        \label{fig:eta_rz_section_deg90}
\end{figure}

At subcritical $Re$, the vorticity is only due to motion in the $x-y$ plane. Since the cross-section is an ellipse, we see differences in distribution along major and minor axes. In \cref{fig:eta_rz_section_deg0,fig:eta_rz_section_deg90}, we have plotted the contours of vorticity magnitude in two $r-z$ planes corresponding to major and minor axes. At subcritical $Re$ of $70$, the vorticity is quite small. However, at $Re=75$, due to the formation of the Taylor cells, the vorticity near the inner wall ($r=0.5$) increases due to the shear component $\frac{\partial w}{\partial r}$. This vorticity incresase as the cells strengthen with Reynolds number. Although the cells rotate in opposite direction, since we have plotted the magnitude, it is seen to be symmetric about $z=0.5$. At the top and bottom of the contours, we see "fingers" of vorticity, which are a result of the flow turning at the corners of the Taylor cells. These increase with Reynolds number and at $Re = 300$, we see some asymmetry probably due to the onset of spiral flow. We observe the same features in the minor gap, but the fingers of vorticity are stronger at the bottom for $Re = 300$. This may be because the spiral vortices in the major gap are realigning to be more axisymmetric in the minor gap. 

\subsubsection{Isosurfaces of vorticity magnitude}

More visual information can be obtained from surfaces of iso-vorticity in the major and minor gap regions. We show surfaces of vorticity magnitude of $2.0$ non-dimensional units. It can be seen that vorticity is concentrated near the inner rotating cylinder and in the return flows from the inner and outer cylinder. The slight tilt at $Re = 300$ is seen in the major gap with the vortices tilting upwards. The tilt of vorticity reinforce the same patterns seen for the pressure and axial velocity.

\begin{figure}[H]
    \centering
    \begin{subfigure}{0.24\textwidth}
        \centering
        \includegraphics[height=50mm]{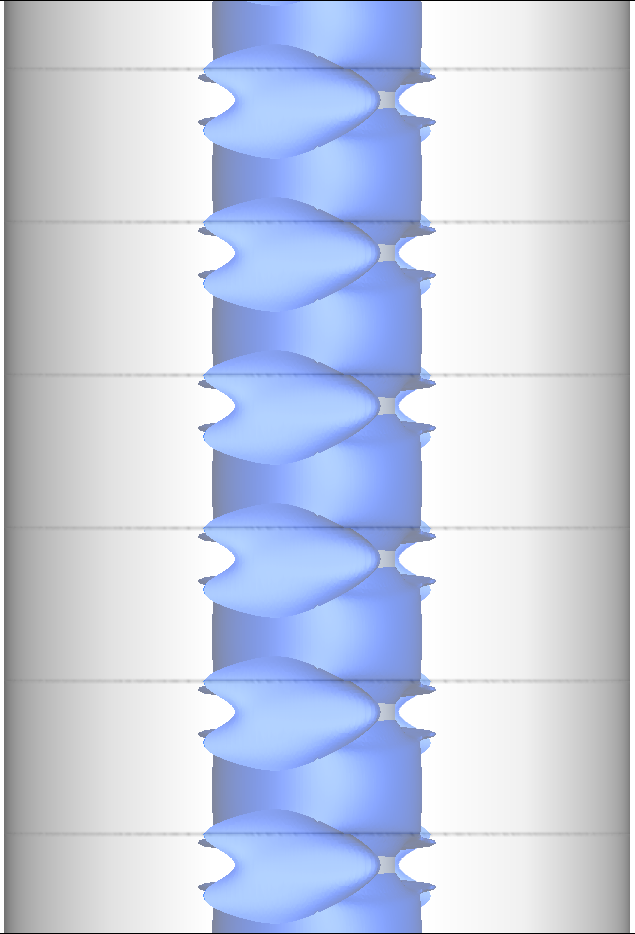}
        \caption{Re = 75}
        \label{}
    \end{subfigure}
    \hfill
    \begin{subfigure}{0.24\textwidth}
        \centering
        \includegraphics[height=50mm]{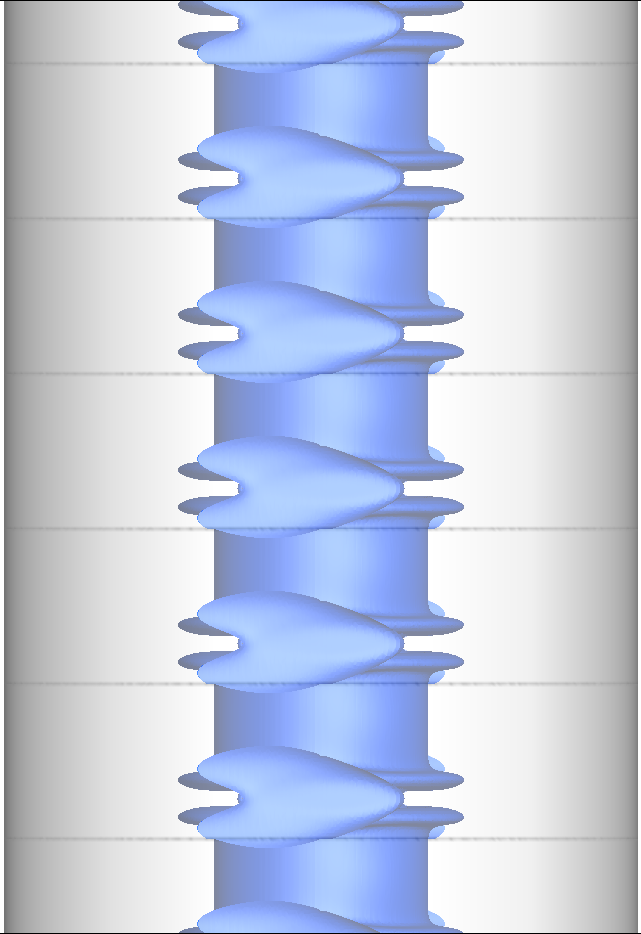}
        \caption{Re = 100}
        \label{}
    \end{subfigure}
    \hfill
    \begin{subfigure}{0.24\textwidth}
        \centering
        \includegraphics[height=50mm]{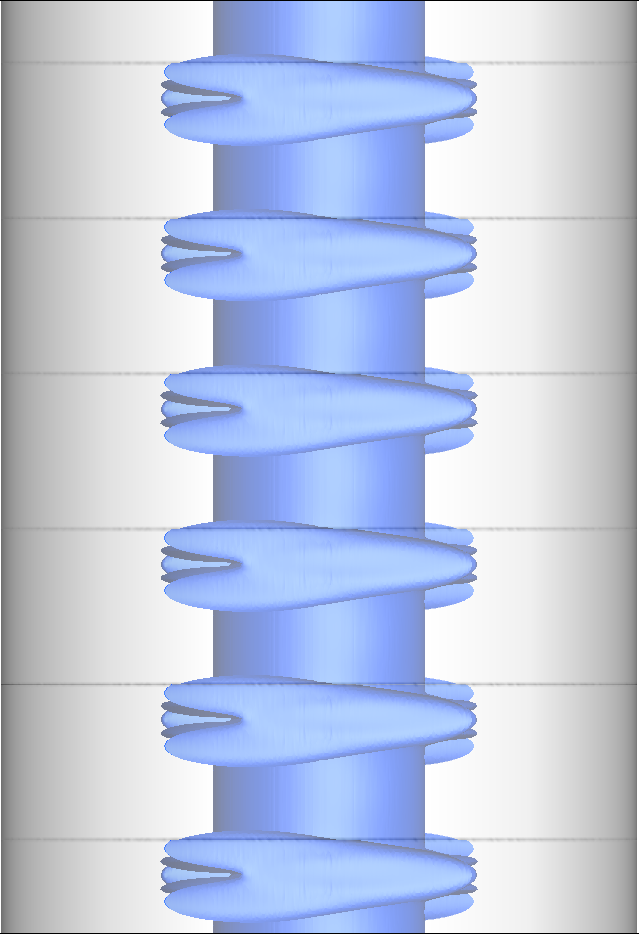}
        \caption{Re = 200}
        \label{}
    \end{subfigure}
    \hfill
    \begin{subfigure}{0.24\textwidth}
        \centering
        \includegraphics[height=50mm]{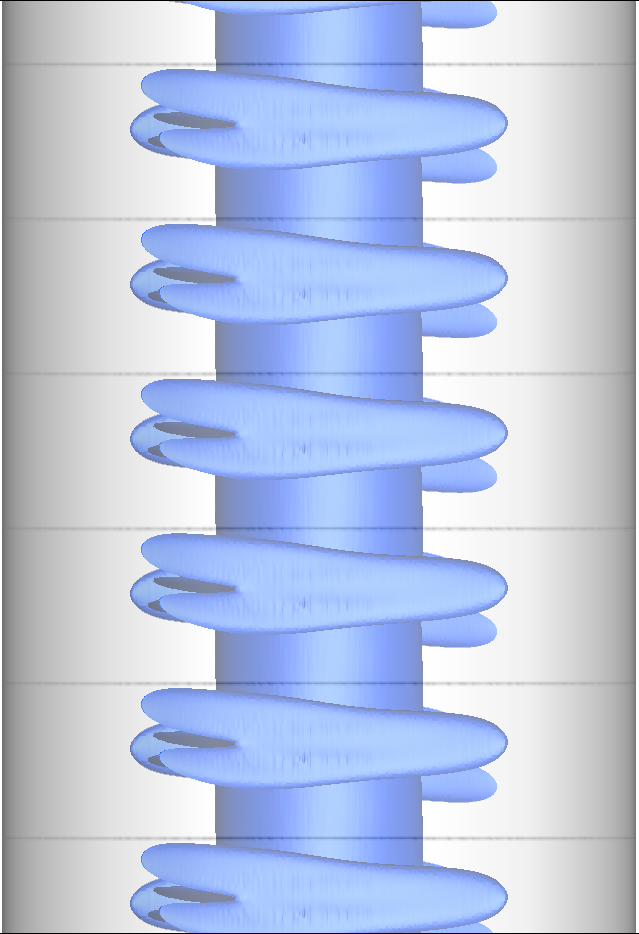}
        \caption{Re = 300}
        \label{}
    \end{subfigure}
    \caption{Isosurfaces of vorticity magnitude for $\eta = 2.0$, viewed from $x-z$ plane (view on the major axis section of ellipse)}
    \label{fig:iso_eta_major_axis}
\end{figure}

\begin{figure}[H]
    \centering
    \begin{subfigure}{0.2\textwidth}
        \centering
        \includegraphics[height=50mm]{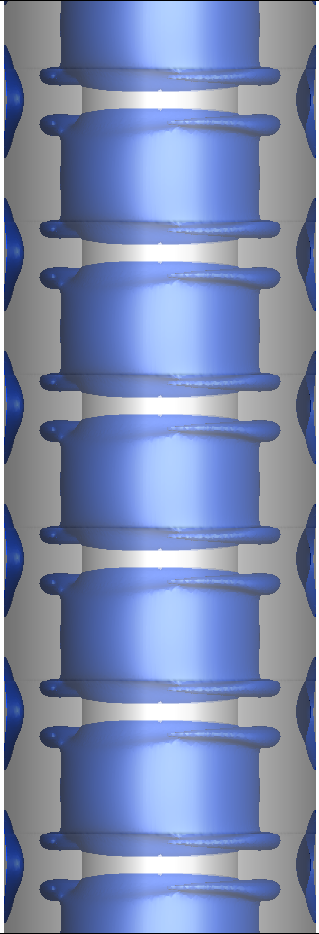}
        \caption{Re = 75}
        \label{}
    \end{subfigure}
    \begin{subfigure}{0.2\textwidth}
        \centering
        \includegraphics[height=50mm]{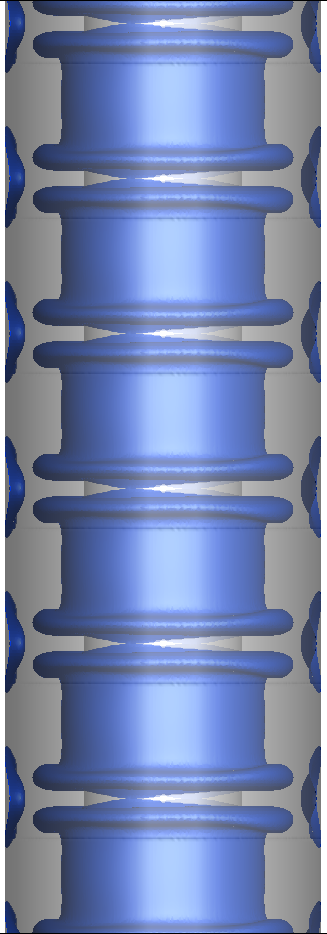}
        \caption{Re = 100}
        \label{}
    \end{subfigure}
    \begin{subfigure}{0.2\textwidth}
        \centering
        \includegraphics[height=50mm]{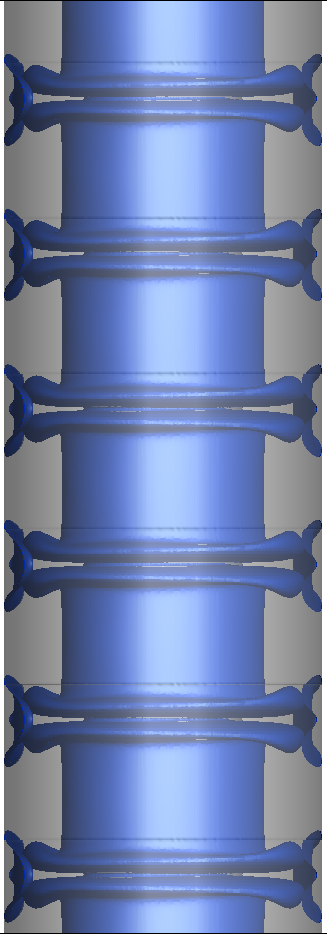}
        \caption{Re = 200}
        \label{}
    \end{subfigure}
    \begin{subfigure}{0.2\textwidth}
        \centering
        \includegraphics[height=50mm]{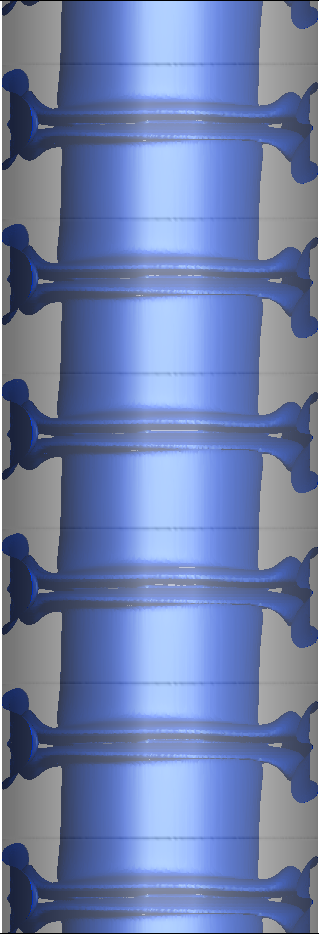}
        \caption{Re = 300}
        \label{fig:iso_eta_minor_axis_d}
    \end{subfigure}
    \caption{Isosurfaces of vorticity magnitude for $\eta = 2.0$, viewed from $y-z$ plane (view on the minor axis section of ellipse)}
    \label{fig:iso_eta_minor_axis}
\end{figure}

\subsection{Q criterion}
Q criterion is a mathematical tool used in fluid dynamics to identify regions prone to turbulence. A positive and high value of Q criterion, helps to visualise regions of high vorticity and shear, where vortex stretching is dominant. These regions are more likely to be turbulent compared to a negative Q criterion value indicating vortex compression. It is commonly used to visualise the location and extent of turbulent regions in a fluid flow. The value of Q is defined as the second invariant of the velocity gradient tensor given by:
\begin{equation}
    Q = \frac{1}{2}(||\Omega||^2 - ||S||^2)
\end{equation}
where $S$ is the strain rate tensor given as $S = \frac{1}{2}\left[\left(\partial u_i / \partial x_j\right)+\left(\partial u_j / \partial x_i\right)\right]$ and $\Omega$ is the rotation rate or vorticity tensor given as $\Omega = \frac{1}{2}\left[\left(\partial u_i / \partial x_j\right)-\left(\partial u_j / \partial x_i\right)\right]$

\subsubsection{Isosurfaces of Q criterion}

\begin{figure}[H]
    \centering
    \begin{subfigure}{0.24\textwidth}
        \centering
        \includegraphics[height=50mm]{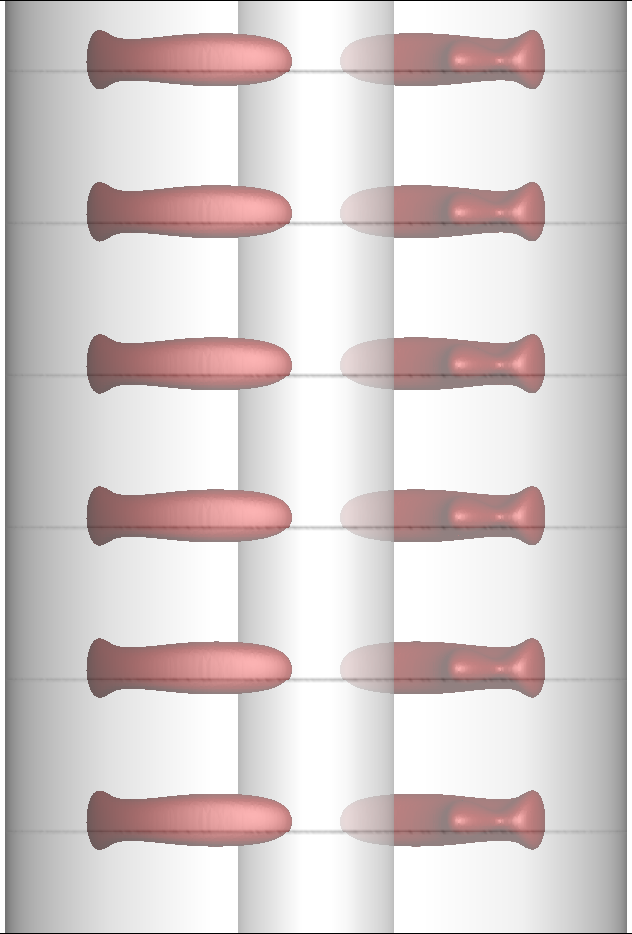}
        \caption{Re = 75}
        \label{}
    \end{subfigure}
    \hfill
    \begin{subfigure}{0.24\textwidth}
        \centering
        \includegraphics[height=50mm]{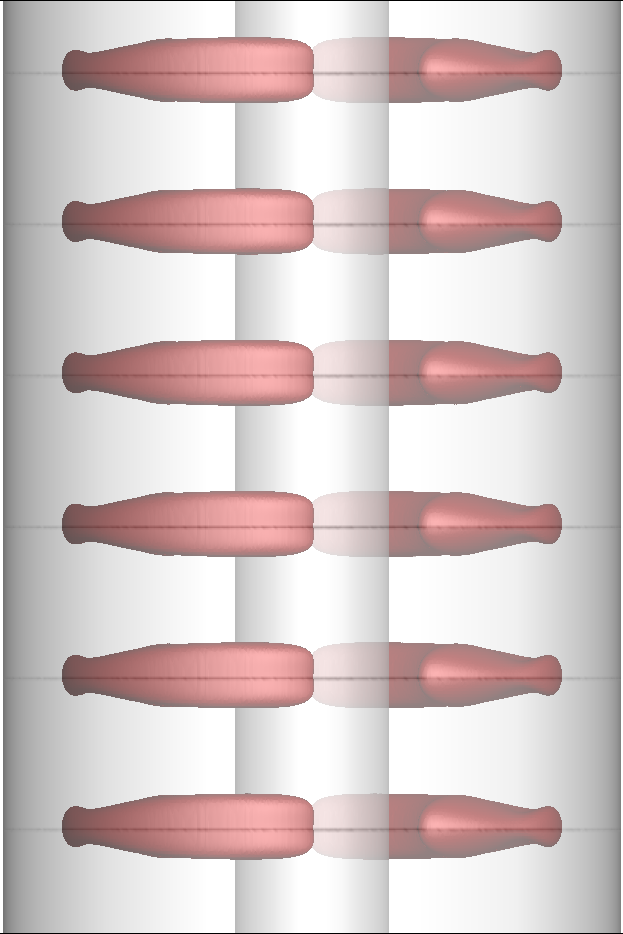}
        \caption{Re = 100}
        \label{}
    \end{subfigure}
    \hfill
    \begin{subfigure}{0.24\textwidth}
        \centering
        \includegraphics[height=50mm]{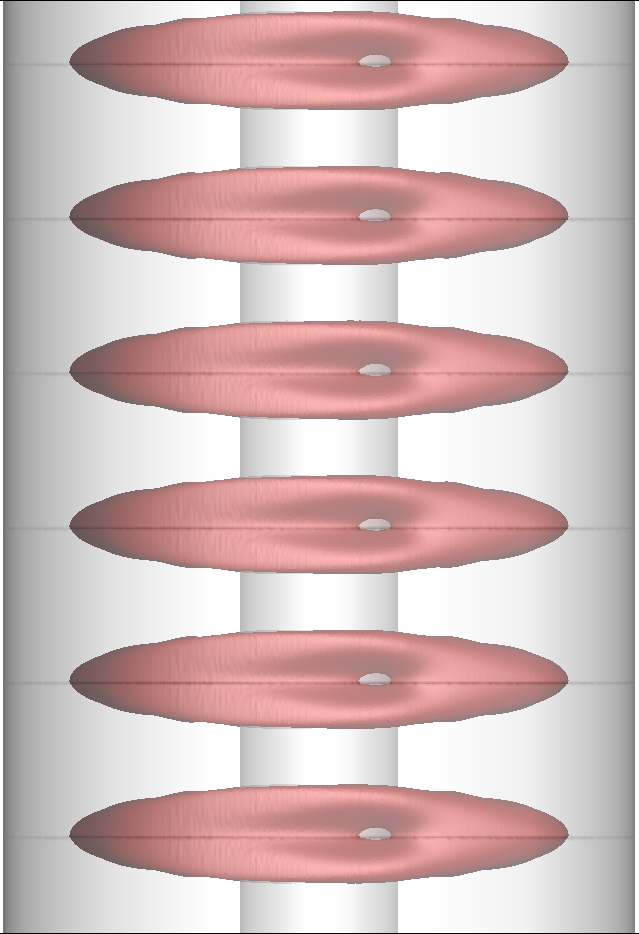}
        \caption{Re = 200}
        \label{}
    \end{subfigure}
    \hfill
    \begin{subfigure}{0.24\textwidth}
        \centering
        \includegraphics[height=50mm]{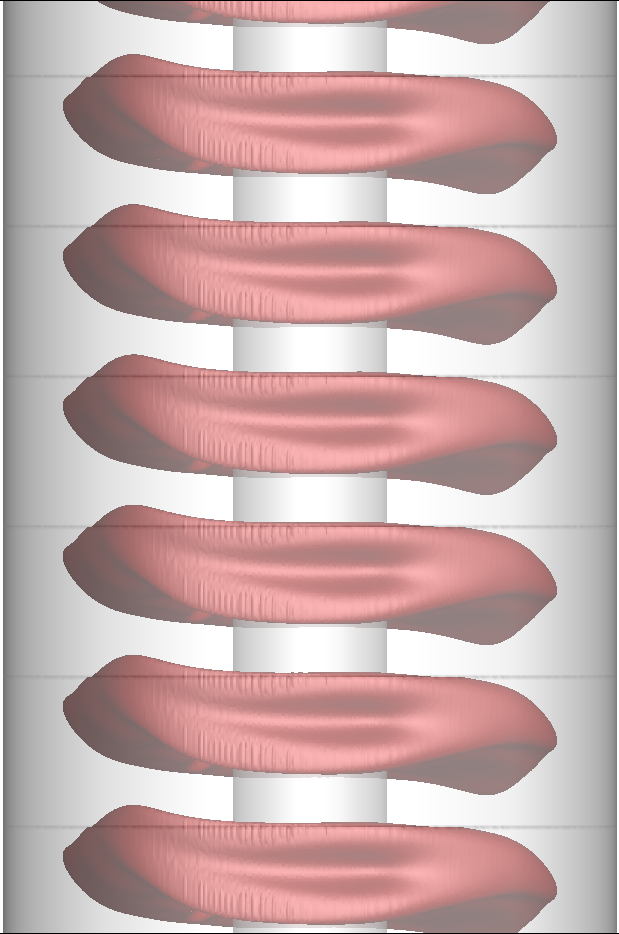}
        \caption{Re = 300}
        \label{}
    \end{subfigure}
    \caption{Isosurfaces of $Q = 1.0$, viewed from $x-z$ plane (view on the major axis section of ellipse)}
    \label{fig:iso_q_major_axis}
\end{figure}

\begin{figure}[H]
    \centering
    \begin{subfigure}{0.2\textwidth}
        \centering
        \includegraphics[height=50mm]{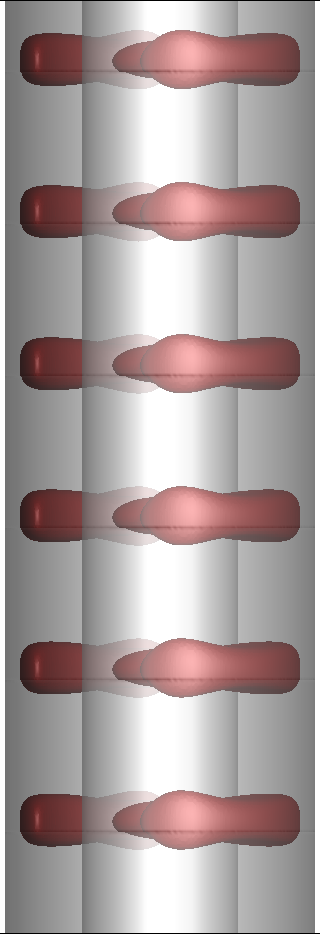}
        \caption{Re = 75}
        \label{}
    \end{subfigure}
    \begin{subfigure}{0.2\textwidth}
        \centering
        \includegraphics[height=50mm]{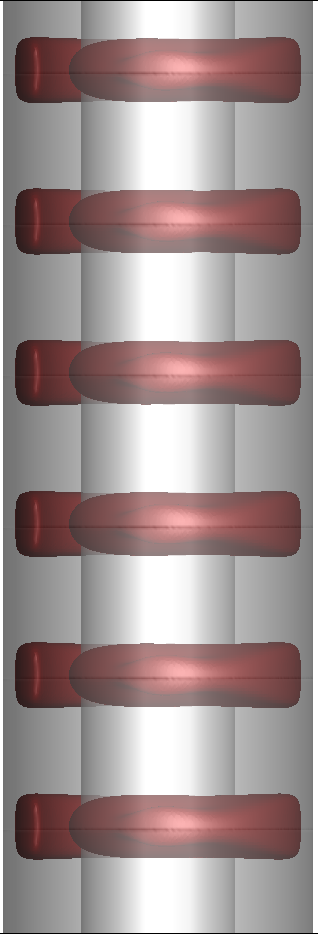}
        \caption{Re = 100}
        \label{}
    \end{subfigure}
    \begin{subfigure}{0.2\textwidth}
        \centering
        \includegraphics[height=50mm]{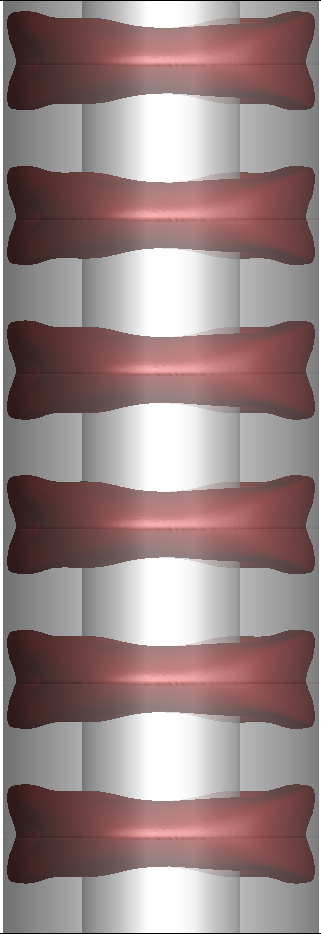}
        \caption{Re = 200}
        \label{}
    \end{subfigure}
    \begin{subfigure}{0.2\textwidth}
        \centering
        \includegraphics[height=50mm]{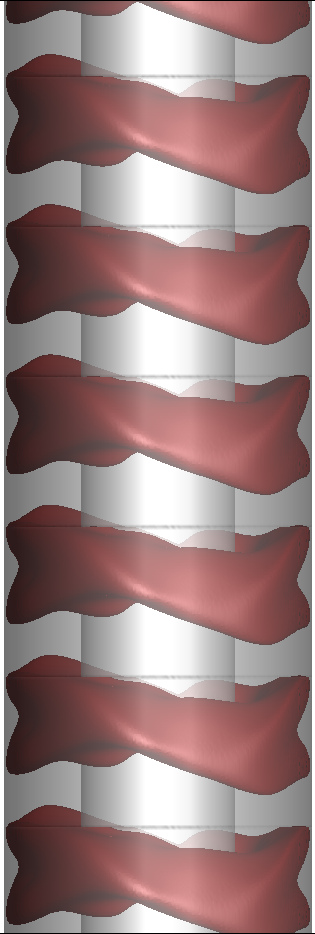}
        \caption{Re = 300}
        \label{}
    \end{subfigure}
    \caption{Isosurfaces of $Q = 1.0$, viewed from $y-z$ plane (view on the minor axis section of ellipse)}
    \label{fig:iso_q_minor_axis}
\end{figure}

The Q criteria surfaces in \cref{fig:iso_q_major_axis,fig:iso_q_minor_axis} indicate the tendency of flow to become turbulent. The region where the flow is more prone to turbulence is the region between two counter rotating Taylor cells. At $Re = 300$, where the wavy Taylor cells are formed, the Q surface is also tilted and wavy in nature. This actually indicates that the region between Taylor cells, where the axial velocity peaks, is where the flow tends to be turbulent before any other portion of the flow regime.

\subsection{Temperature distributions}

We finally present results of heat transfer from the inner wall. The inner wall is heated at a non-dimensional temperature of $1.0$ and the outer wall is maintained at $T=0$. The heat transfer rates in the form of the Nusselt number were earlier presented in \cref{fig:NuvsReplot}.

\subsubsection{Contours of temperature in r-z planes}

\begin{figure}[H]
     \centering
     \begin{subfigure}[b]{0.45\textwidth}
         \centering
         \includegraphics[height=50mm]{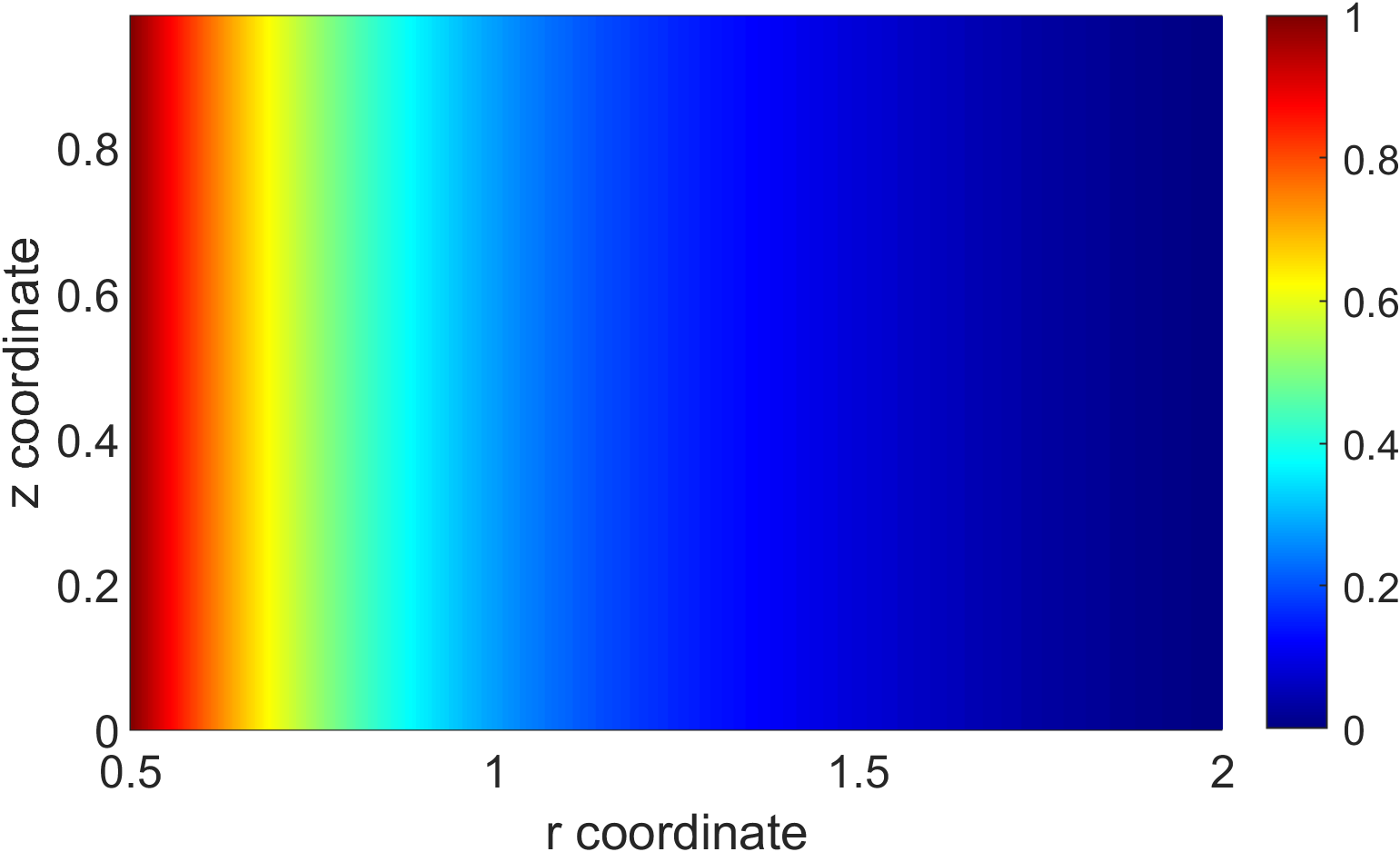}
         \caption{Re = 70}
         \label{}
     \end{subfigure}
     \hfill
     \begin{subfigure}[b]{0.45\textwidth}
         \centering
         \includegraphics[height=50mm]{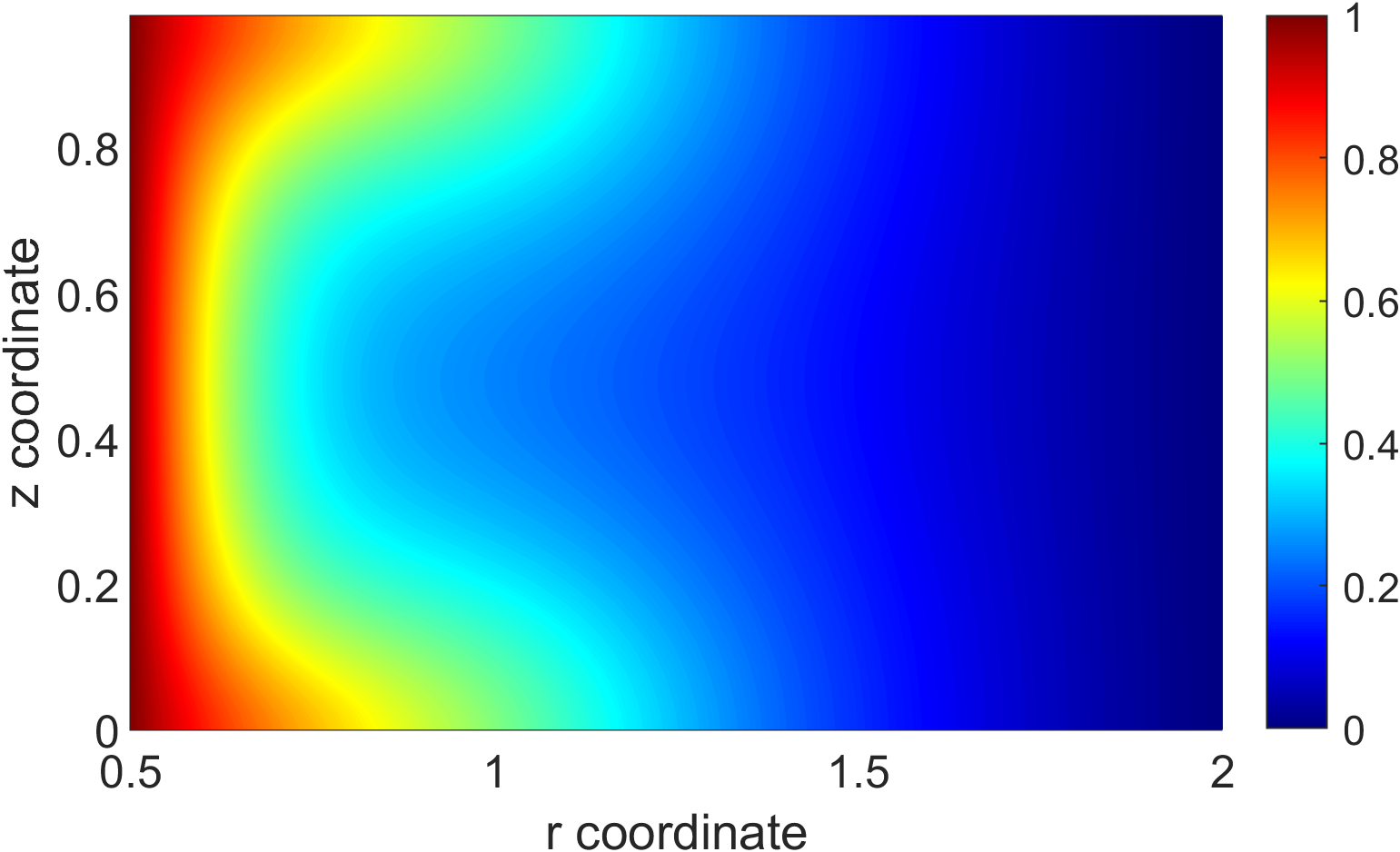}
         \caption{Re = 75}
         \label{}
     \end{subfigure}
     \hfill
     \par \bigskip
     \begin{subfigure}[b]{0.45\textwidth}
         \centering
         \includegraphics[height=50mm]{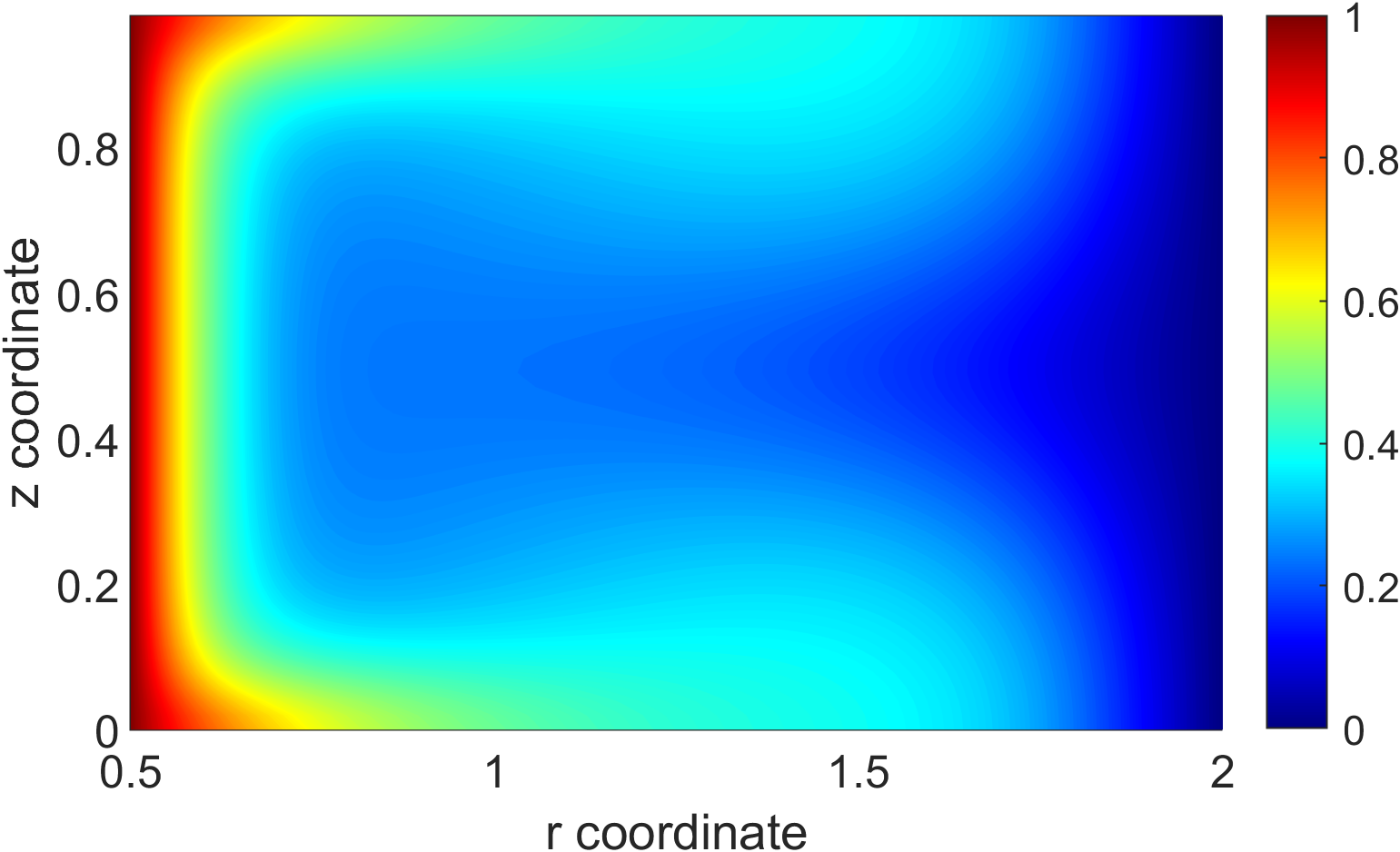}
         \caption{Re = 200}
         \label{}
     \end{subfigure}
     \hfill
     \begin{subfigure}[b]{0.45\textwidth}
         \centering
         \includegraphics[height=50mm]{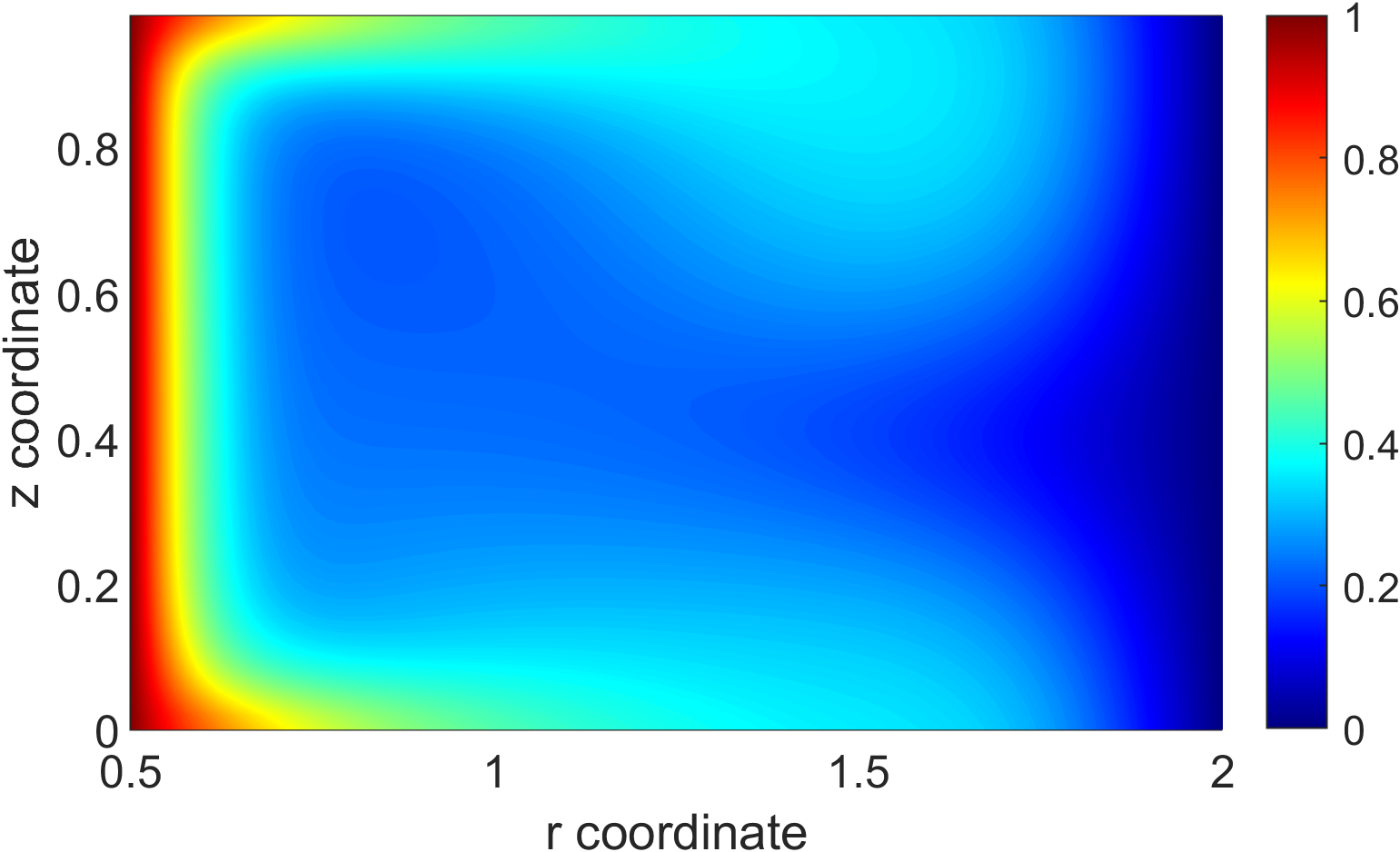}
         \caption{Re = 300}
         \label{}
     \end{subfigure}
     \caption{Contours of temperature in the plane along semi major axis of ellipse}
        \label{fig:T_rz_section_deg0}
\end{figure}

\begin{figure}[H]
     \centering
     \begin{subfigure}[b]{0.19\textwidth}
         \centering
         \includegraphics[height=50mm]{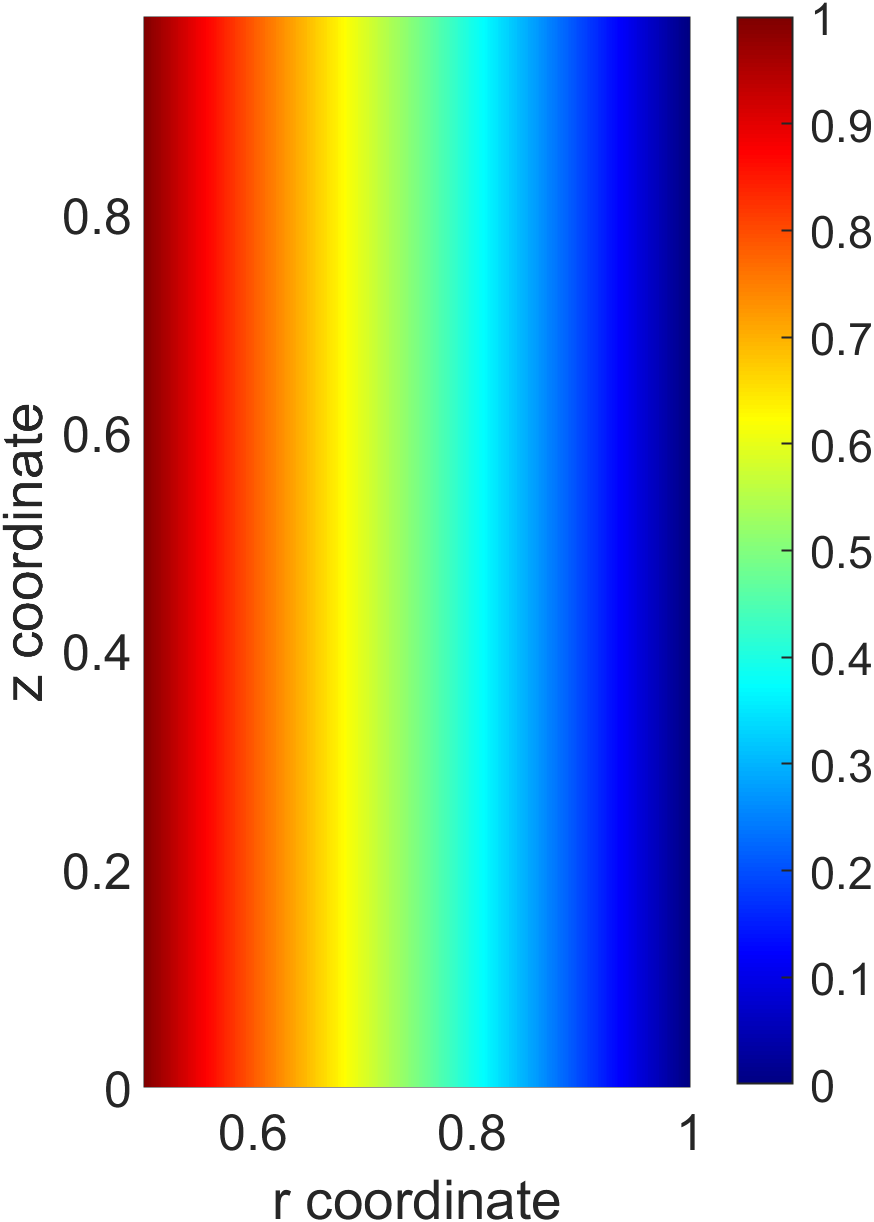}
         \caption{Re = 70}
         \label{}
     \end{subfigure}
     \hfill
     \begin{subfigure}[b]{0.19\textwidth}  
         \centering
         \includegraphics[height=50mm]{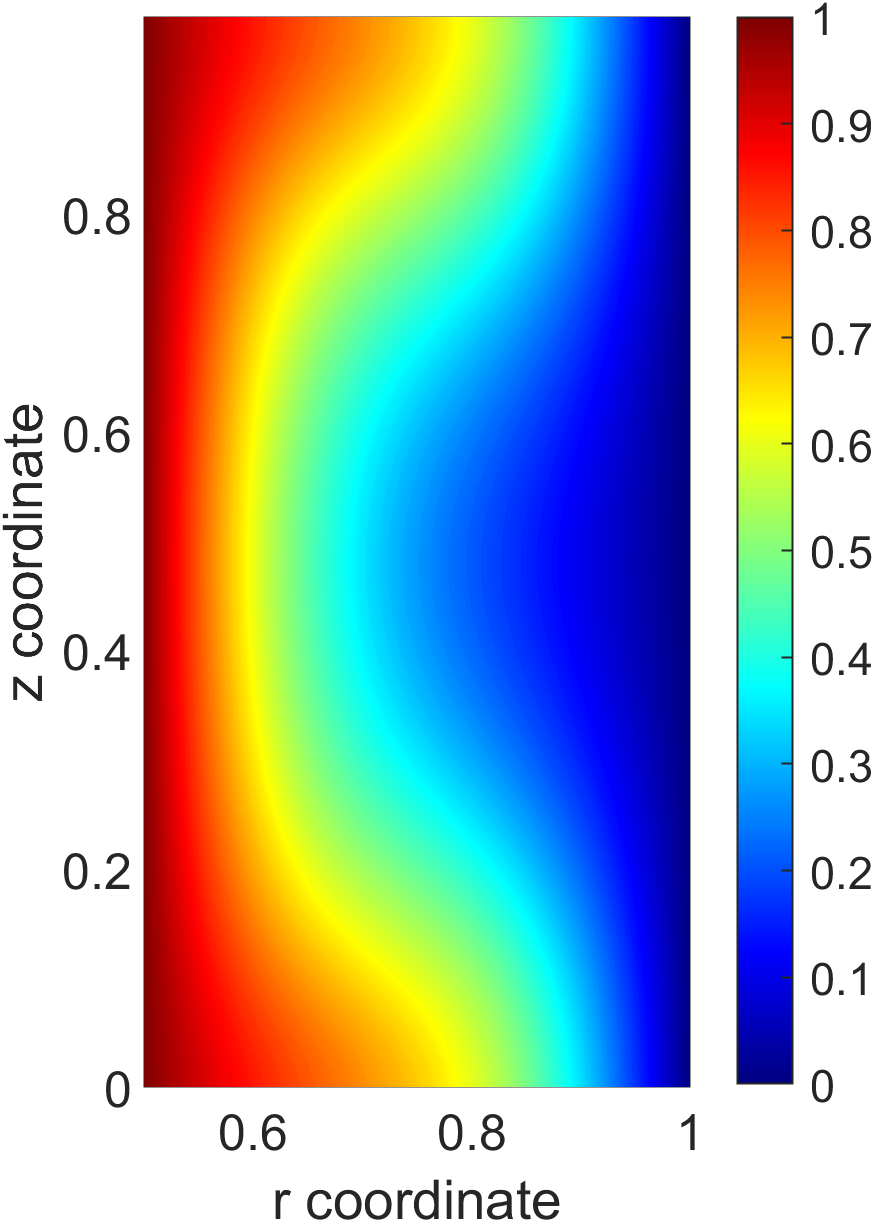}
         \caption{Re = 75}
         \label{}
     \end{subfigure}
     \hfill
     \begin{subfigure}[b]{0.19\textwidth}
         \centering
         \includegraphics[height=50mm]{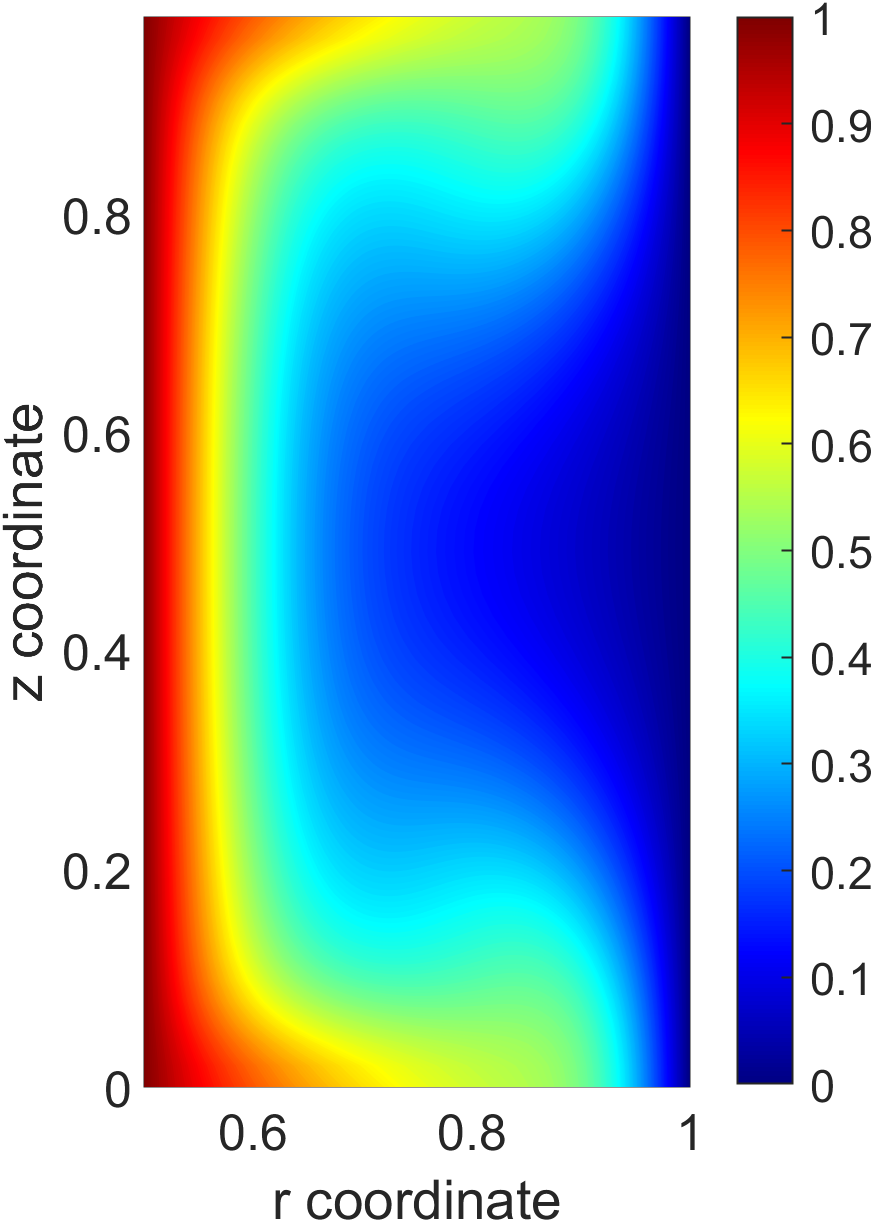}
         \caption{Re = 200}
         \label{}
     \end{subfigure}
     \hfill
     \begin{subfigure}[b]{0.19\textwidth}
         \centering
         \includegraphics[height=50mm]{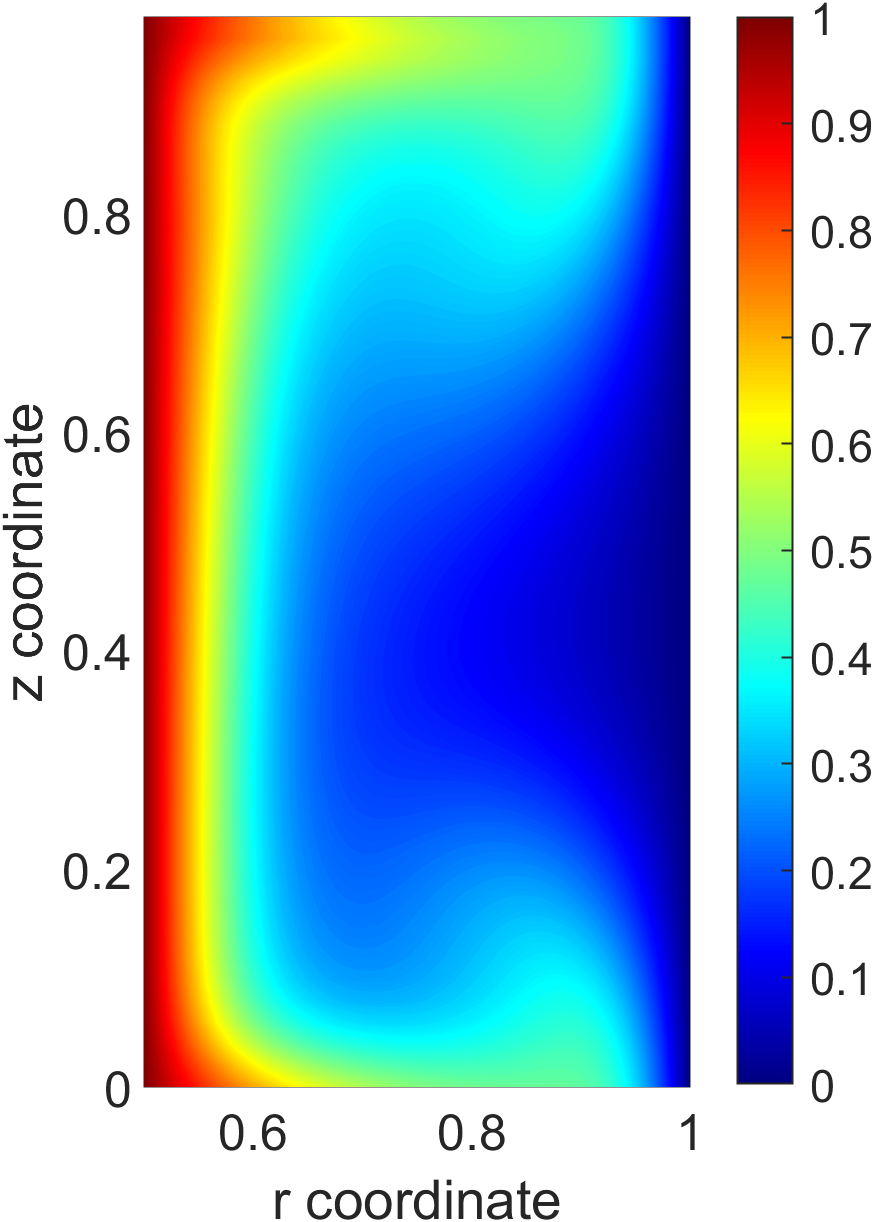}
         \caption{Re = 300}
         \label{}
     \end{subfigure}
     \caption{Contours of temperature in the plane along semi minor axis of ellipse}
        \label{fig:T_rz_section_deg90}
\end{figure}

The temperature contours for supercritical Reynolds number show the effect of fluid motion in the Taylor cells. The hot fluid from the top and bottom circulations is carried from inner wall to the outer wall and returns the colder fluid from outer wall to the inner cylinder. This is seen to increase with Reynolds number as the strength of the circulation increases. The same pattern is seen in the minor gap plane. Similar to all the other field variables that we have discussed, the contours shown in \cref{fig:T_rz_section_deg0,fig:T_rz_section_deg90} has a symmetry about the plane $z=0.5$ for Reynolds number lesser than 300. The lose of symmetry that we see at the Reynolds number of 300 is due to the wavy nature of Taylor cells at this regime.

\subsubsection{Isosurfaces of temperature}

Isosurfaces of temperature are in the form of bellows for supercritical cases with Reynolds number less than 300. Since temperature is solved as a scalar transport equation, it follows the fluid flow. i.e, the bulge on the bellows occur at the region between the Taylor cells where it meets the outer cylinder, since the flow is radially outward in this region. Similarly the trough of the bellow shape occurs at the section between the Taylor cells where the flow is radially inward meeting the inner cylinder. The surface for $Re = 300$ has the slanted nature as the Taylor cells have a spiral nature. 

\begin{figure}[H]
    \centering
    \begin{subfigure}{0.24\textwidth}
        \centering
        \includegraphics[height=50mm]{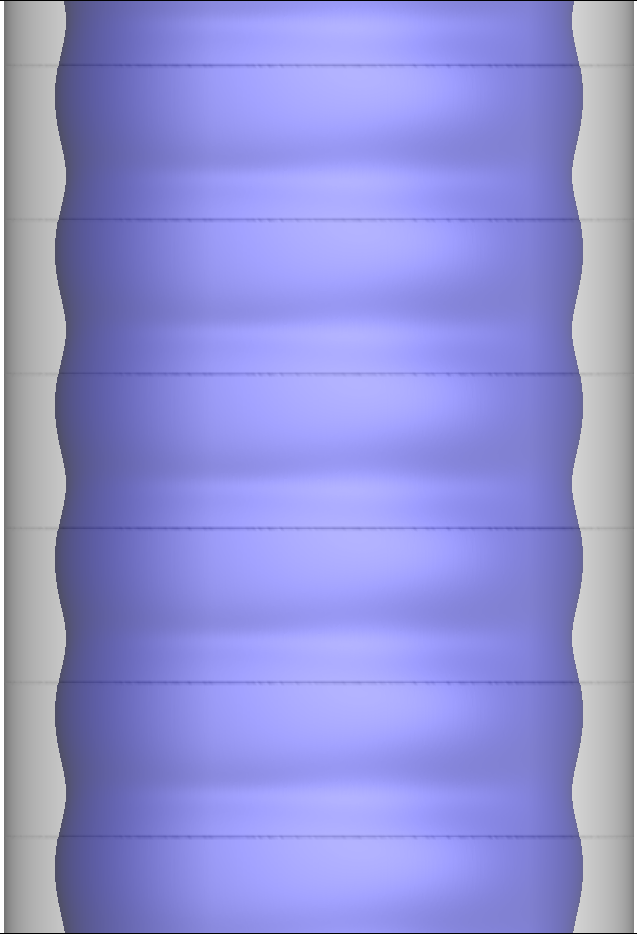}
        \caption{Re = 75}
        \label{}
    \end{subfigure}
    \hfill
    \begin{subfigure}{0.24\textwidth}
        \centering
        \includegraphics[height=50mm]{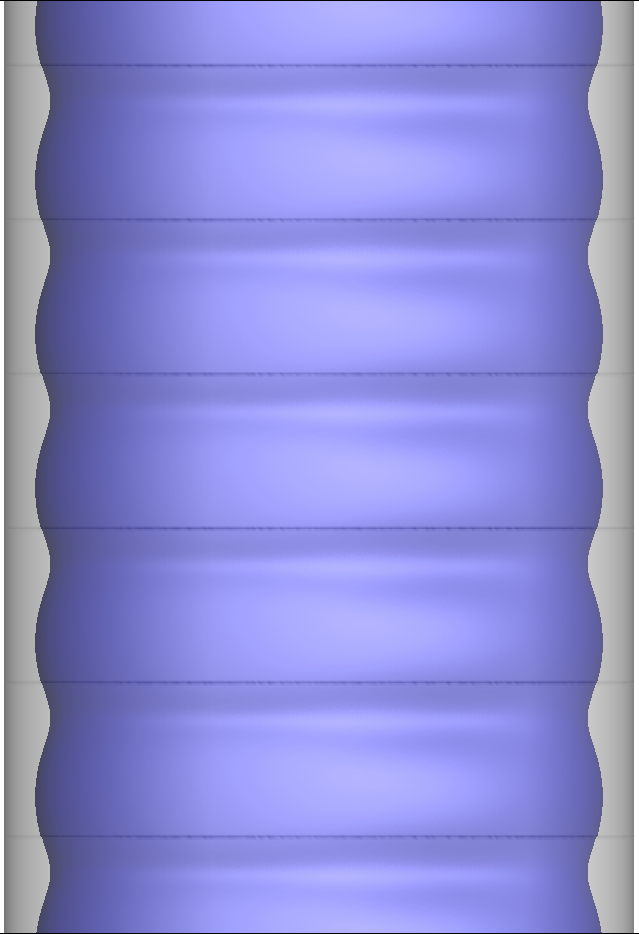}
        \caption{Re = 100}
        \label{}
    \end{subfigure}
    \hfill
    \begin{subfigure}{0.24\textwidth}
        \centering
        \includegraphics[height=50mm]{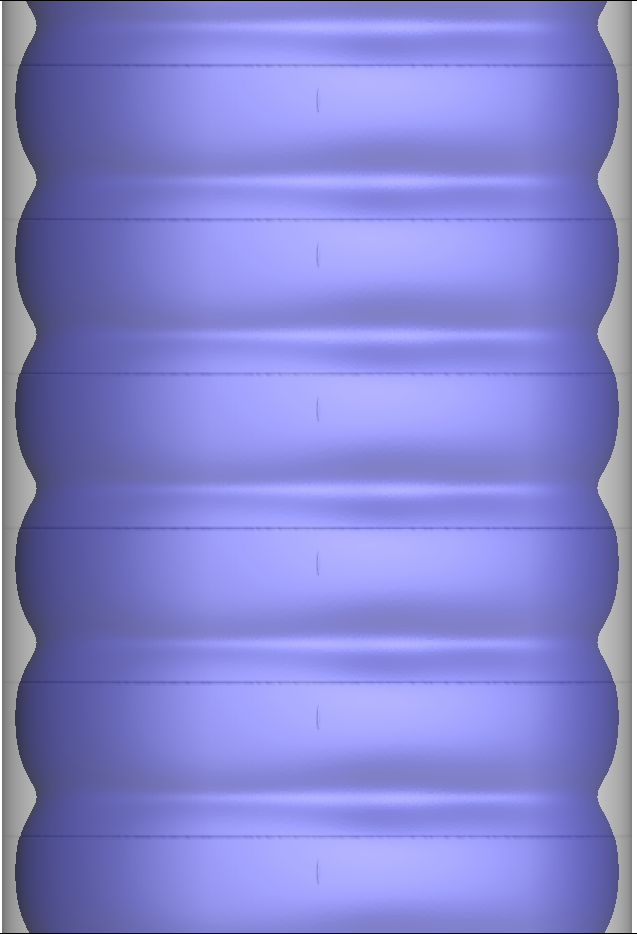}
        \caption{Re = 200}
        \label{}
    \end{subfigure}
    \hfill
    \begin{subfigure}{0.24\textwidth}
        \centering
        \includegraphics[height=50mm]{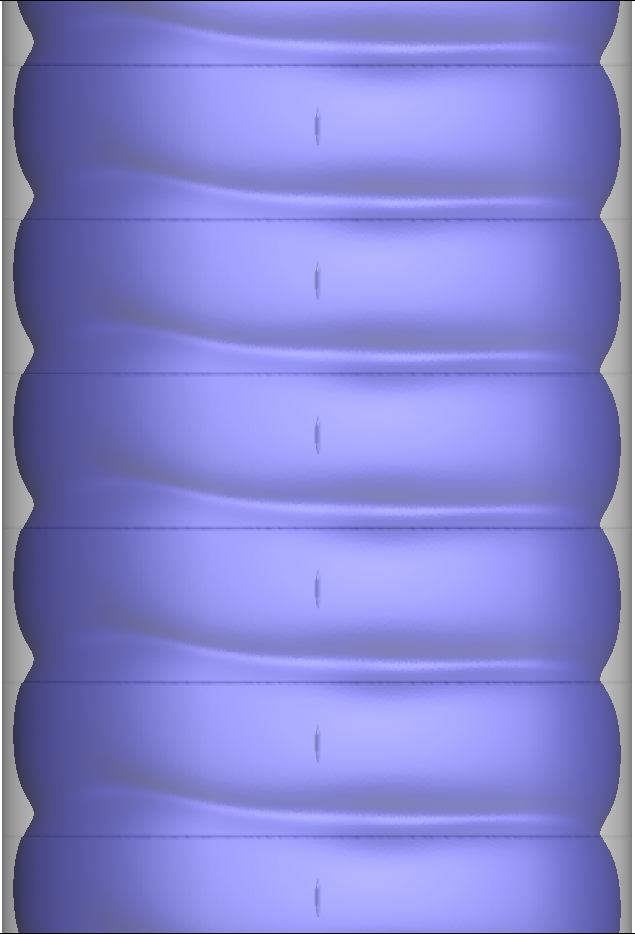}
        \caption{Re = 300}
        \label{}
    \end{subfigure}
    \caption{Isosurfaces of temperature for $T = 0.1$, viewed from $x-z$ plane (view on the major axis section of ellipse)}
    \label{fig:iso_T_major_axis}
\end{figure}

\begin{figure}[H]
    \centering
    \begin{subfigure}{0.2\textwidth}
        \centering
        \includegraphics[height=50mm]{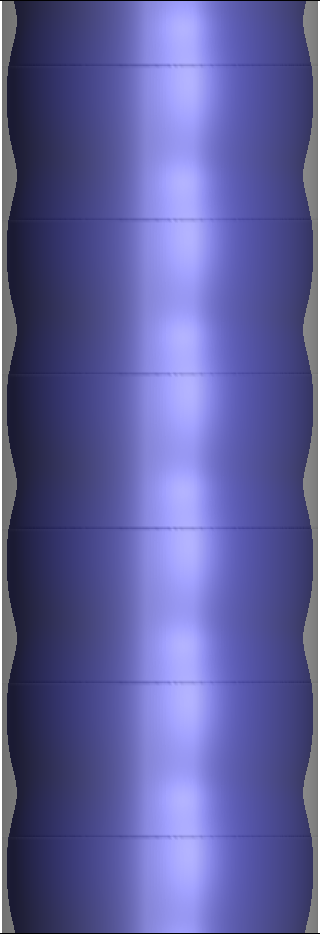}
        \caption{Re = 75}
        \label{}
    \end{subfigure}
    \begin{subfigure}{0.2\textwidth}
        \centering
        \includegraphics[height=50mm]{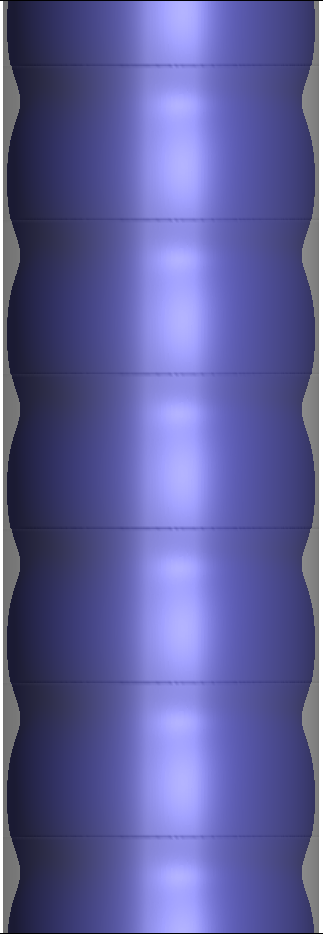}
        \caption{Re = 100}
        \label{}
    \end{subfigure}
    \begin{subfigure}{0.2\textwidth}
        \centering
        \includegraphics[height=50mm]{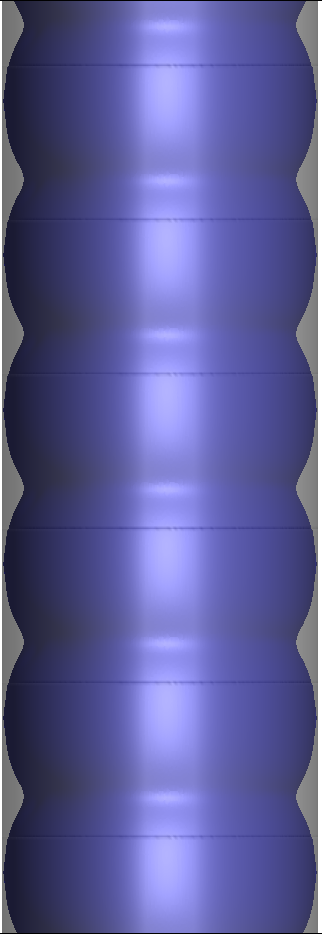}
        \caption{Re = 200}
        \label{}
    \end{subfigure}
    \begin{subfigure}{0.2\textwidth}
        \centering
        \includegraphics[height=50mm]{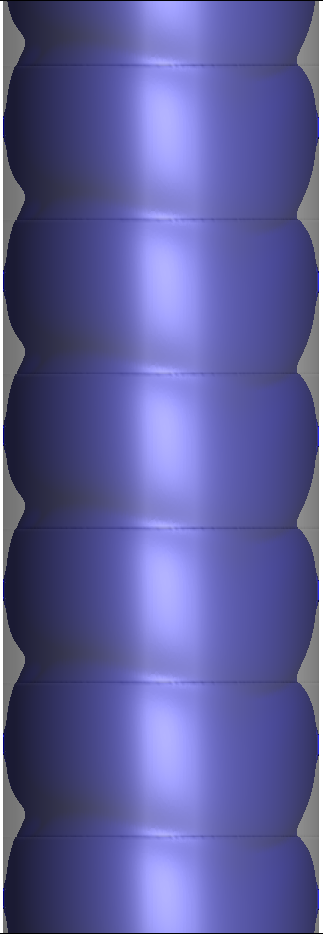}
        \caption{Re = 300}
        \label{}
    \end{subfigure}
    \caption{Isosurfaces of temperature for $T = 0.1$, viewed from $y-z$ plane (view on the minor axis section of ellipse)}
    \label{fig:iso_T_minor_axis}
\end{figure}

\section{Summary}
\label{sec:summary}
In this paper, we have applied a novel Fourier-spectral meshless technique to study the development and characteristics of Taylor cells in an ellipse (aspect ratio of 2) with an inner circular cylinder rotating concentrically. The meshless technique solves the Navier-Stokes equations in Cartesian coordinates at scattered locations using stencils derived through Radial Basis Function interpolations. We have assumed periodicity in the axial direction, allowing a Fourier-spectral expansion along the axis of the cylinders. The Navier-Stokes equations are first converted to the Fourier space and the resulting equations for wave numbers are solved by the meshless method. The distributions of velocity, pressure, temperature, and vorticity are presented for subcritical and supercritical regions. The torque on the inner cylinder and the critical Reynolds number for formation of the Taylor cells are found to be nearly the same as the concentric circular cylinder results in literature. However, the structure of the Taylor cells is different in the ellipse because of the compression and expansion of the Taylor vortices as they pass between the major and minor gaps. At low Reynolds numbers, the Taylor cells are concentrically structured forming a square shape in the minor gaps and a rectangular shape in the major gaps respectively, as a result of the stretch and squeeze between the gaps. However, these structured cells slowly deform when the Reynolds number reaches 300, indicating a transition to a wavy pattern. We present isosurfaces of the axial velocity, pressure, vorticity, and temperature which indicate the shape of the Taylor cells. These surfaces of iso-values appear consistent with the squeezing and expansions of the cells between the gaps.

\section*{Data Availability}
The data that support the findings of this study are available from the corresponding author upon reasonable request.

\section*{Author Declarations}
The authors have no conflicts to disclose.

\bibliography{References}

\end{document}